\documentclass[letterpaper,12pt]{article}

\usepackage{fancyhdr}
\usepackage{amsmath}
\usepackage{amsbsy}
\usepackage{amssymb}
\usepackage{amsthm}
\usepackage{amscd}
\usepackage{amsfonts}
\usepackage{supertabular}
\usepackage{graphicx}
\usepackage{verbatim}
\usepackage{subfigure}
\usepackage{epsfig}
\usepackage{xspace}
\usepackage{euscript}
\usepackage{alltt}
\usepackage{boxedminipage}
\usepackage{float}
\usepackage{times}
\usepackage[colorlinks]{hyperref}
\usepackage[round,authoryear]{natbib}
\usepackage{setspace}
\usepackage{amsmath, amssymb, graphics, setspace}
\usepackage{algorithm}
\usepackage{booktabs}
\usepackage{algorithmic}
\usepackage{lscape}
\usepackage{wrapfig}
\usepackage{subfigure}%
\usepackage[hang,small,bf]{caption}
\usepackage{color}
\usepackage{authblk}
\usepackage[pagewise,displaymath]{lineno}
\usepackage{morefloats}
\usepackage{epstopdf} 
\usepackage{gensymb}

\def\Put(#1,#2)#3{\leavevmode\makebox(0,0){\put(#1,#2){#3}}}

\begin{document}

%\linenumbers

\title{Quantification of lubrication and particle size distribution effects on tensile strength and stiffness of tablets}

\author[a]{Sonia M. Razavi}
\author[b]{Marcial Gonzalez\thanks{Corresponding author at: School of Mechanical Engineering, Purdue University, West Lafayette, IN 47907, USA. Tel.:~+1~765 494 0904. Fax: +1 765 496 7537 \\ \indent E-mail address: marcial-gonzalez@purdue.edu (M. Gonzalez)}}
\author[a]{Alberto M. Cuiti\~{n}o}
\affil[a]{\small Department of Mechanical and Aerospace Engineering, Rutgers, The State University of New Jersey, Piscataway, NJ 08854, USA}
\affil[b]{\small School of Mechanical Engineering, Purdue University, West Lafayette, IN 47907, USA}

\maketitle

\begin{abstract}

We adopt a Quality by Design (QbD) paradigm to better control the mechanical properties of tablets. To this end, the effect of particle size distribution, lubricant concentration, and mixing time on the tensile strength and elastic modulus of tablets is studied. Two grades of lactose, monohydrate and spray-dried, are selected. Tablets are compressed to different relative densities ranging from $0.8$ to $0.94$ using an instrumented compaction simulator. We propose a general model, which predicts the elastic modulus and tensile strength envelope that a specific powder can obtain based on its lubrication sensitivity for different particle size distributions. This is possible by introducing a new dimensionless parameter in the existing tensile strength and elastic modulus relationships with relative density. A wide range of lubrication conditions is explored and a predictable model is callibrated. The mechanical properties of lactose monohydrate tablets are noticeably dependent on particle size, unlike spray-dried lactose where little to almost no sensitivity to particle size is observed. The model is designed in a general fashion that can capture mechanical quality attributes in response to different lubrication conditions and particle size, and it can be extended to powders than undergo different deformation mechanisms, complex mixtures, and doubly convex tablets. Therefore, the model can be used to map the achievable design space of any given formulation.

\end{abstract}


\setlength\parindent{0pt}

\section{Introduction}

Lubricants are one of the key ingredients in the pharmaceutical formulations to improve flowability, increase bulk powder density, and reduce die wall friction and ejection forces \citep{lachman1976theory, moody1981tablet,hirai1982effect,dansereau1987effect,miller1988pharmaceutical,sheskey1995effects,podczeck1996influence,faqih2007effect}.  Magnesium stearate (MgSt) is the most frequently used lubricant \citep{shangraw1993survey}; typically added to the formulation in small amounts ($0.25\%-1.0\%$ (w/w)) \citep{lindberg1972evaluation,miller1988pharmaceutical}. It has been shown that MgSt can adversely affect the physical and chemical properties of tablets \citep{johansson1984granular,wang2010lubrication}. Hypothetically, MgSt forms a layer on the host particles weakening the interparticle bonding \citep{bolhuis1975film,de1978bonding, hussain1990secondary}. The lubricant type and concentration, type of mixer and its operation method, and mixing time are all important processing variables that affect the powder compactibility, interparticle bonding and thus, final mechanical properties of tablets \citep{asker1975investigation,shah1977mechanism,bossert1980effect,dansereau1987effect,
otsuka1993effects,kikuta1994effect,otsuka2004effects}. However, the deformation mechanism of host particles also play a role \citep{bolhuis1975film}. For example, brittle materials that undergo fragmentation are said to be unaffected by MgSt due to the creation of unexposed surfaces during compression \citep{de1978bonding,jarosz1984effect}. In contrast, plastically deformable powders are significantly impacted by lubricant mixing \citep{doelker1987comparative,bolhuis1995materials,mitrevej1996spray}. \citet{mollan1996effects} ascribed the reduction in the total work of compaction by increasing the lubricant concentration to decreased particle cohesiveness. \citet{zuurman1999effect} argued that the decrease in tablet strength of pharmaceutical powders such as microcrystalline cellulose mixed with MgSt is caused by a more extensive relaxation of the lubricated tablets corresponding to a weaker interparticle bonding.

Over the past decade there has been growing interest in quantifying what the powder experiences in mixing with lubricant to enable a more robust prediction of tablet quality attributes. The blender parameters were translated to a more relevant and fundamental variables, strain and shear rate, using a modified Couette shear cell to better quantify lubrication effect \citep{mehrotra2007influence,llusa2010measuring}. Shear rate is proportional to the energy input rate per unit mass and total strain is proportional to the total energy input per unit mass. \citet{narang2010effect} derived a dimensionless equation to quantify total shear imparted by the force feeder on the granulation in terms of a shear number, which provides guidance to the scale-up and interchangeability of tablet presses. \citet{kushner2010scale} proposed an empirical model, which can describe the impact of both formulation and process parameters on the extent of lubrication in a pharmaceutical powder blend. 

Particle size distribution (PSD) also plays an important role on the compaction and tablet properties \citep{roberts1986effect}. A decrease in particle size of the powdered material has been shown to increase tablet porosity \citep{mckenna1982effect,de1986studies}. Smaller particles are inclined to be more cohesive since the interparticle cohesive forces are comparable to the weight of the particles making them more compressible \citep{castellanos2005relationship,faqih2006experimental}. Reduction in particle size typically results in an increase in the mechanical strength of tablets \citep{shotton1961strength,hersey1967effect,mckenna1982effect,ragnarsson1985force}. This is attributed to a greater packing density after the particle rearrangement and an increase in the surface area available for interparticulate attractions \citep{rhinesmetal,vromans1985studies,de1986studies}. Attempts were made to correlate specific surface area to the mechanical strength of tablets and a linear relationship was found for different types of lactose \citep{vromans1985studies,de1986studies}. However, \citet{nystrom1993bonding} suggested that the intermolecular forces are the dominating mechanism in the compactibility of powders and only in some cases the available surface area could be used to establish a model to correlate with mechanical strength of tablets. On the contrary, sodium chloride tablets have been reported to become stronger as their particle size increased associated to more bonding between particles through solid bridges \citep{alderborn1982studies}.

\citet{katikaneni1995consolidation} investigated the tableting properties and predominant consolidation mechanism of ethylcellulose as lubricant concentration and particle size varied individually. The concurrent effect of lubrication and particle size on mechanical properties of pharmaceutical tablets during and after compaction has also been explored. \citet{van1987effect} was the first to show that tablet properties change after the same MgSt mixing time for different particle sizes of Avicel PH 102. In more recent years, \citet{almaya2008effect} examined the effect of particle size on lubricant sensitivity for different types of materials. They concluded that for MCC (a plastically deforming material) particle size impacts tablet strength only in the presence of lubricant. For starch (a viscoelastic material) tablet strength is affected by the particle size with or without added lubricant. Finally, for dibasic calcium phosphate dihydrate (a brittle material) particle size has no effect on tablet strength with or without the lubricant. Nevertheless, there is no previous work that goes beyond the qualitative predictions.

The primary goal of the present study is to adopt a Quality by Design (QbD) paradigm to better control the mechanical properties of tablets. We aim to quantify the lubrication effect combined with the particle size on the tensile strength and elastic modulus of tablets. To this end, the envelope of mechanical quality attributes of two grades of lactose, namely lactose $\alpha$-monohydrate (LM) and spray-dried lactose (SDL), caused by different PSD and lubrication conditions was explored. Tablets were compressed to different relative densities ranging from $0.8$ to $0.94$ using an instrumented compaction simulator. We propose a general model for predicting the elastic modulus and tensile strength spectrum that a specific powder can obtain based on its lubrication sensitivity for different PSDs. This was possible by introducing a new dimensionless parameter in the existing tensile strength and elastic modulus relationships that is a non-linear function of the PSD, lubricant concentration and its mixing time with the host particles. A wide range of lubrication conditions was explored and the model exhibited a good predictability. The mechanical properties of LM tablets were noticeably dependent on particle size, unlike SDL where little to almost no sensitivity to initial particle size was observed. The model is designed in a general fashion that can capture mechanical quality attributes in response to different lubrication conditions and initial particle size.

\section{Material and methods}

The materials used in this study include $\alpha$-lactose monohydrate (Foremost Farms, Wisconsin, USA), Spray-dried Fast-Flo lactose monohydrate N.F. (Foremost Farms, Wisconsin, USA) and magnesium stearate N.F. non-Bovine (Mallinckrodt, Missouri, USA) as lubricant.

The true density of lactose monohydrate (LM), spray-dried lactose (SDL), and magnesium stearate (MgSt) powders was measured using an AccuPyc Pycnometer (Accupyc II $1340$, Micromeritics) with helium as density medium. The powders were dried at $50^{\circ}\text{C}$ for $24$ hours before the test. 

Each powder was sieved through a vibrational sieve shaker (Octagon $2000$, Endecotts Ltd., England) into different particle size distributions. LM was divided into three particle size fractions $0$-$75$, $75$-$106$, and $106$-$150$ $\mu$m. SDL was divided into four particle size fractions $0$-$75$, $75$-$106$, $106$-$150$, and $150$-$212$ $\mu$m. The sieve shaker was operated at amplitude of $8$. The as-received powder was poured in the top pan of the clamped sieve stack. The powders on the lower pan (corresponding to $0$-$75$ $\mu$m) were collected at an interval of approximately $15$~min. This procedure was repeated until the powder in the lower pan was a negligible amount. 

The particle size distribution (PSD) was measured using a Beckman Coulter LS $13~320$ laser diffraction particle size analyzer to ensure if the desired distribution was achieved.  MgSt was pre-sieved through a $\#50$ mesh ($300$ $\mu$m opening) prior to mixing with powders using a laboratory scale resonant acoustic mixer (labRAM) (Resodyn Acoustic Mixers, Butte, Montana, USA). The mixing intensity ($0-100$\%) is the parameter that can be controlled in the LabRAM, which determines the amplitude of the mechanical vibration, translating into acceleration values ($0-100$ g's) depending on the load mass \citep{osorio2015evaluation}. In all the experimental work presented here the acceleration of $40~$g was used.  In other words, for each blending condition, based on the powder mass and powder properties, the mixing intensity was adjusted to give the same acceleration. Regardless, the variations in powder mass were kept minimal. Overall, MgSt concentration and mixing time varied from $0.25$\% to $2$\% and $30$~sec to $2400$~sec, respectively, aiming to produce tablets with a wide range of mechanical properties. Samples were stored in airtight plastic bags until used. 

The samples were compacted using a Presster tablet press simulator (The Metropolitan Computing Corporation of East Hanover, NJ) equipped with an $8$~mm flat round face, B-type tooling. A Fette $1200$ tablet press with $250$~mm compaction roll diameter was emulated at a constant speed of $25$~rpm. A dwell time of $26$~ms, corresponding to a production speed of $36,000$ tablets per hour, was used. Compression force and punch displacement are measured via strain gauges placed on the compression roll pins and a linear variable displacement transducer connected to each punch, respectively. No pre-compression force was applied. The total number of tablets per case varied from $8$ to $17$. All the compacted tablets were stored at ambient room temperature and inside a sealed plastic bag and kept for at least $24$ hours prior to any characterization.

The mass of tablets was measured with a precision balance ($\pm0.001$~g, Adventurer Ohaus). The thickness and diameter of tablets were measured using a MultiTest~$50$ (MT$50$) tablet hardness tester (Sotax, Allschwil, Switzerland). From these measurements, the relative density of the tablets was calculated
\begin{equation}
\bar{\rho} = \frac{\frac{4m}{{\pi}D^2t}}{\rho_{t}}
\label{eq:RD}
\end{equation}
where $m$, $D$, and $t$ are the mass, diameter, and thickness of the tablet and $\rho_{t}$ is the true density of the blend determined by
\begin{equation}
\frac{1}{\rho_{t}} = \sum_{i=1}^{2}  \frac{n_{i}}{\rho_{t,i}}
\label{eq:truedensity}
\end{equation}
$n$ and $\rho_{t}$ represent the concentration on a mass basis and true density of each ingredient.

Ultrasound (US) measurements were made using a pulser/receiver unit (Panametrics, $5077$PR), a pair of longitudinal wave contact transducers (Panametrics, V$606$-RB) with a central frequency and diameter of $2.25$MHz and $13$mm, respectively, a digitizing oscilloscope (Tektronix TDS$3052$), and a computer controlling the data acquisition. Parafilm tape was used, suggested by \citet{hakulinen2008ultrasound}, to improve the contact between transducers and tablets. From the acquired data, the time of flight (TOF) was obtained using the first peak of the received US signal. The speed of sound, SOS, was calculated as follows:
\begin{equation}
\text{SOS} = \frac{t}{\text{TOF}},
\end{equation}
A delay time of $0.8$$\mu$s was measured for the US setup independent of the material used. In subsequent measurements this delay time was subtracted from the measured TOF. 

The elastic modulus, E, of each tablet was then calculated from the SOS assuming the material is isotropic \citep{akseli2009non,razavi2016toward}:
\begin{equation}
\text{E} = \text{SOS}^2\rho_{b}.
\label{eq:SOS}
\end{equation}
The tablets were diametrically compressed using an MT$50$ tablet hardness tester. The tensile strength of tablets, $\sigma_{t}$, was computed using Hertz solution \citep{fell1970determination}
\begin{equation}
\sigma_{t} = \frac{2F}{\pi Dt},
\label{eq:hertz}
\end{equation}
where $F$ is the breaking force.

\section{Results and discussion}

The average true density of LM, SDL, and MgSt was measured to be $1.555$, $1.546$, and $1.040$~gcm$^{-3}$, respectively. According to \citep{yohannes2015role}, the true density of lactose particles was found to be independent of the PSD. Thus, the density measurements were only conducted on the as-received samples and the changes in the true density of powders with different PSD were assumed negligible. A total of $19$ blends of LM and $23$ blends of SDL were prepared varying in MgSt concentration ($c_{l}$), mixing time ($t_{m}$), and PSD, listed in Table \ref{Table:DOE}. 

\begin{figure}[H]
\centering
\begin{tabular}{cc}
\subfigure[]
{
\includegraphics[scale=0.4]{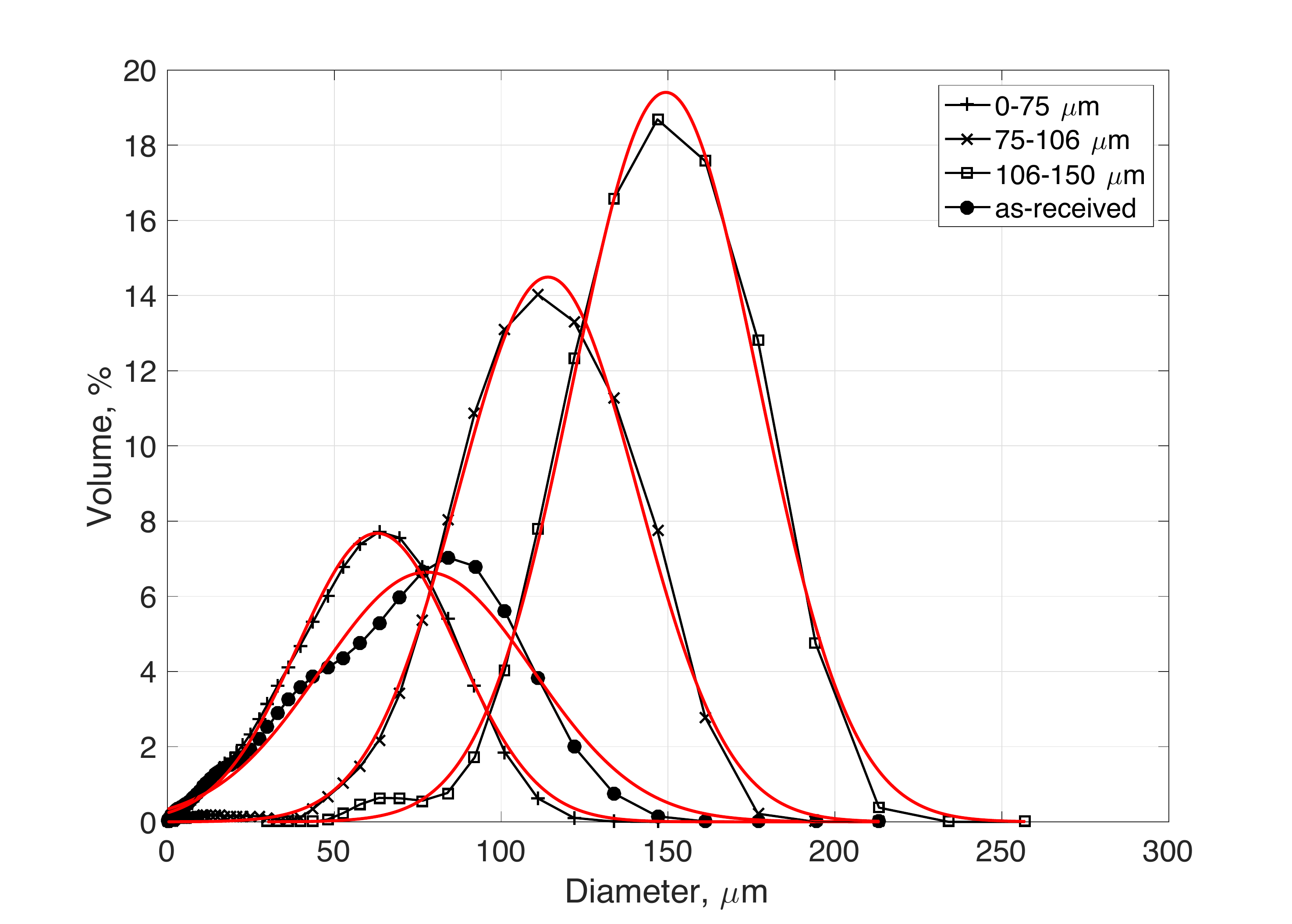}
\label{Fig:LM-PSD}
}
\\
\subfigure[]
{
\includegraphics[scale=0.4]{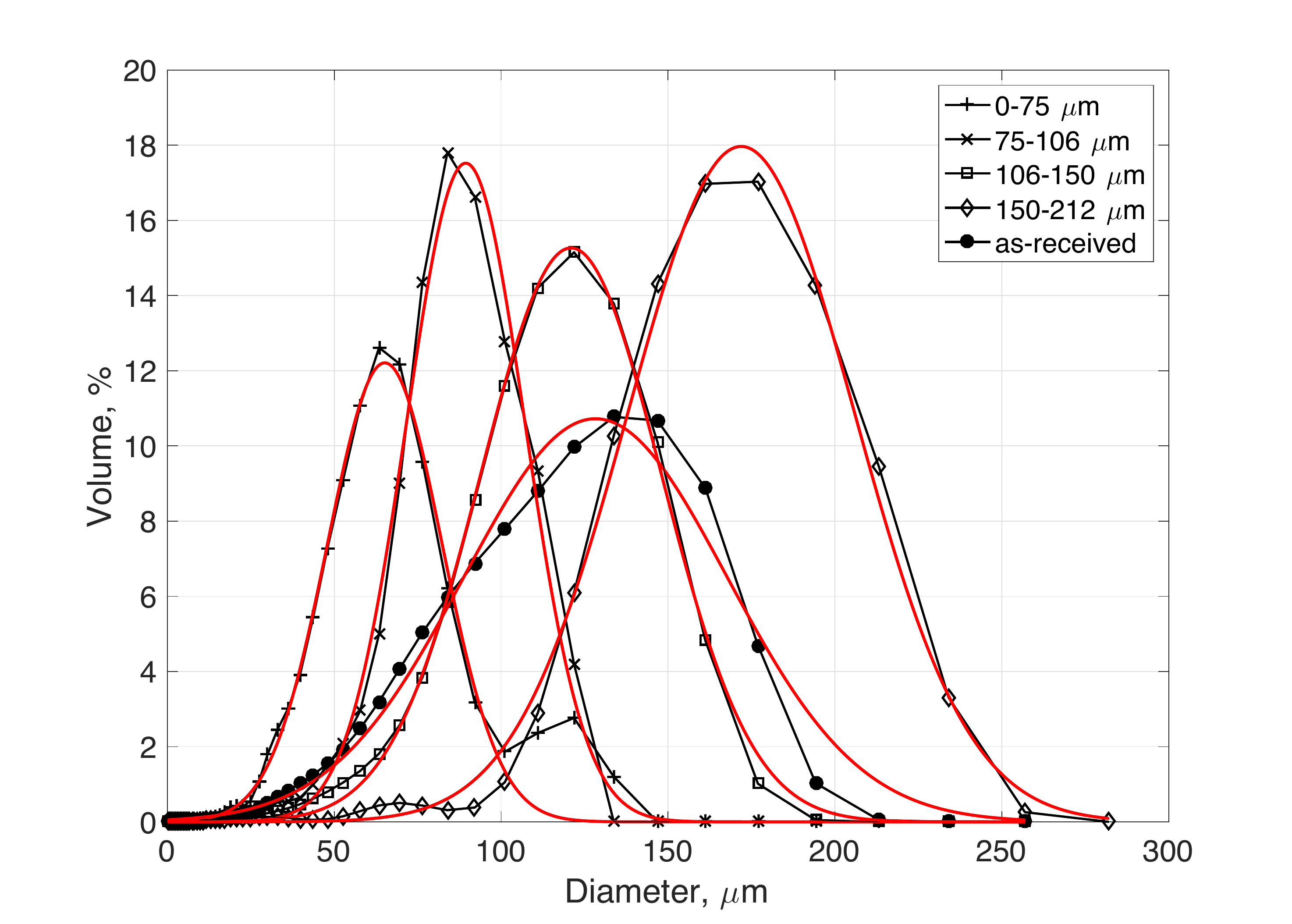}
\label{Fig:SDL-PSD}
}
\end{tabular} 
\caption{Particle size distribution of (a) lactose monohydrate and (b) spray-dried lactose. Red lines show the Gaussian fitting for each distribution.} 
\label{Fig:PSD}
\end{figure}

Figs.~\ref{Fig:LM-PSD} and \ref{Fig:SDL-PSD} show PSD measured for the sieved and as-received samples of LM and SDL, respectively. A Gaussian distribution was fitted to all the curves using Matlab 2016a \citep{MATLAB:2016} and the mean ($\mu$) and standard deviation ($\sigma$) for each PSD are reported in Table \ref{Table:DOE}. There is a clear difference between the as-received powders. As-received LM contains mainly of fine particles, whereas as-received SDL has larger particles. This justfies the addition of the fourth PSD ($150-212$ $\mu$m) selected for SDL. For each PSD sample, there are particles smaller (except for $0-75$ $\mu$m) and larger than the target PSD. Possible reasons were mentioned in \citep{yohannes2015role}. Optical microscopy on samples with different PSD showed insignificant change in particle shape. Thus, particle shape was assumed to be a constant in this study.

Although PSD (in particular, $d_{10}$) was shown to be increased by the RAM mixing time \citep{osorio2016effect}, our goal is to be able to correlate the mechanical strength of tablets to the initial properties of the powder. Thus, the PSD measurement was only conducted on the unlubricated samples.

\subsection{Effect of particle size on lubricant sensitivity on compaction properties}

Compaction pressure is calculated by dividing the compaction force over the cross sectional area of the tooling used. The in-die relative density is determined by Eq. (\ref{eq:RD}), but $t$ varies as the gap between the upper and lower punches changes during loading and unloading. During loading the punches get closer and the thickness of the powder bed decreases until it reaches its minimum, where the maximum compaction pressure is applied. When the force is released, the unloading stage starts, some of the energy is recovered and the tablet mostly expands axially. The total work input during the compaction process is the area under the loading curve of a force-displacement profile. In this study, instead of force-displacement, we used compaction pressure versus in-die relative density profiles, which allows for comparison between samples of different sizes.

\begin{figure}[H]
\centering
\begin{tabular}{cc}
\subfigure[]
{
\hspace{-1cm}
\includegraphics[scale=0.27]{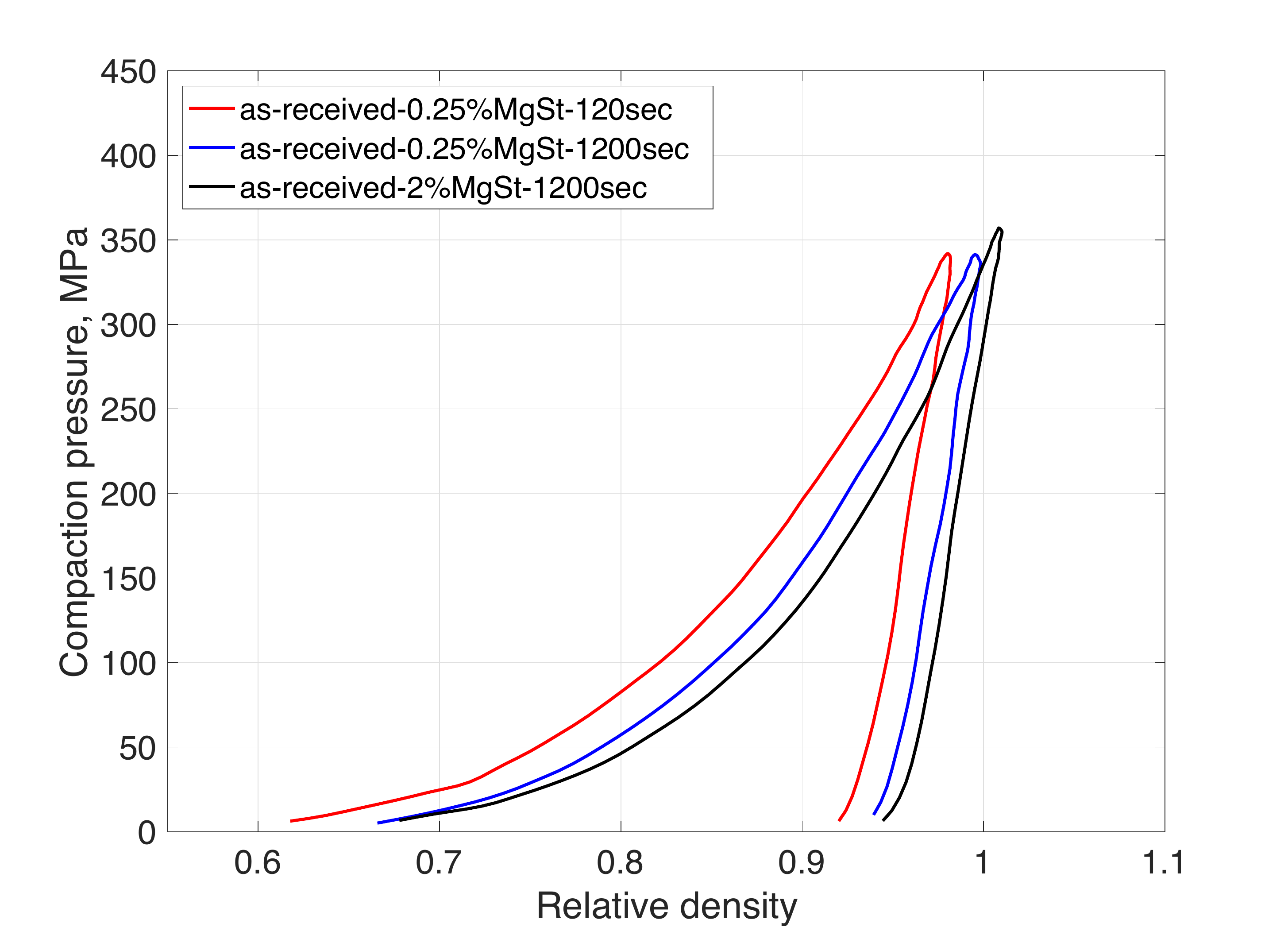}
\label{Fig:CP-AR-LM}
}
\hspace{-1cm}
\subfigure[]
{
\includegraphics[scale=0.27]{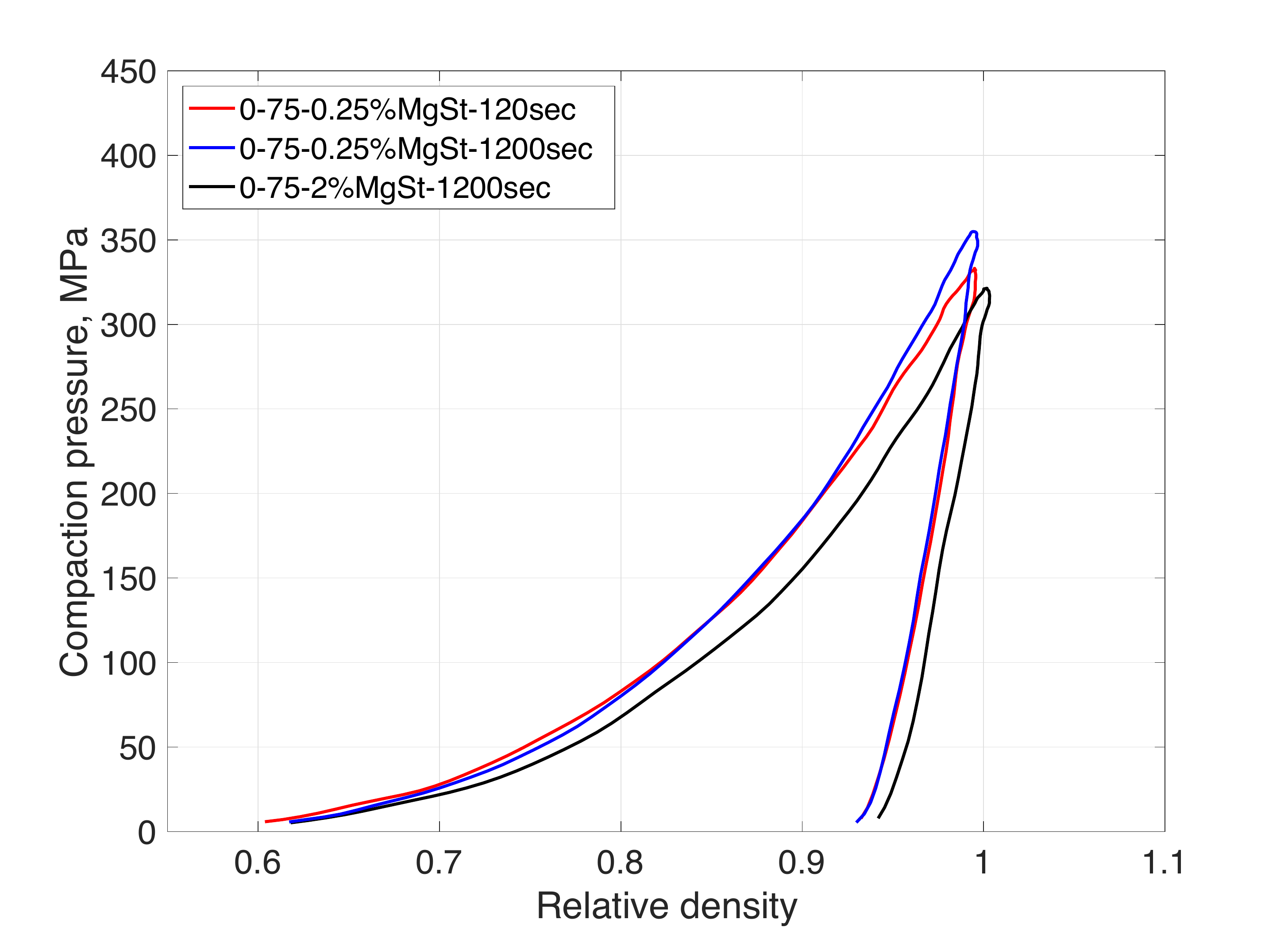}
\label{Fig:CP-0-75-LM}
}
\\
\subfigure[]
{
\hspace{-1cm}
\includegraphics[scale=0.27]{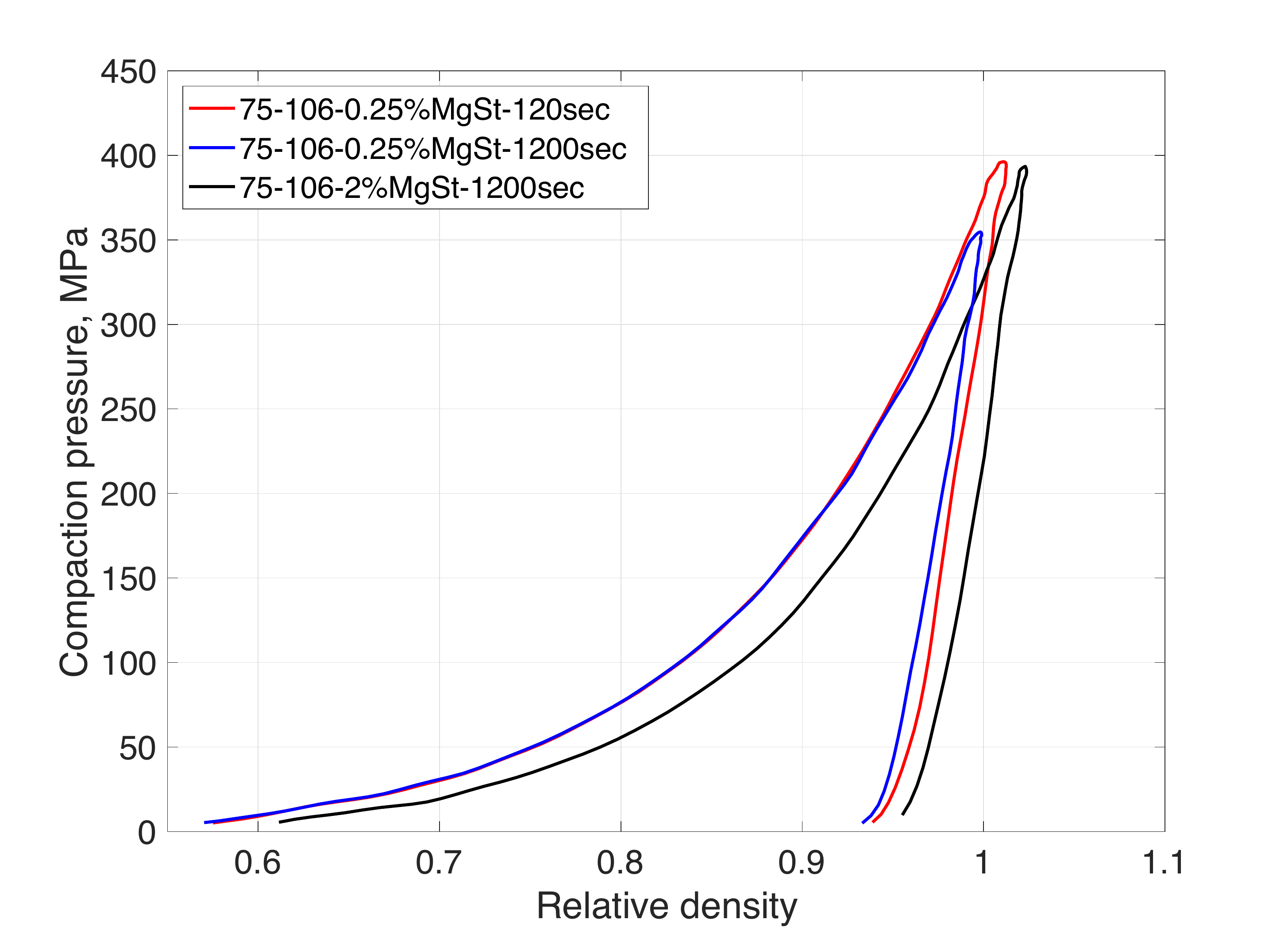}
\label{Fig:CP-75-106-LM}
}
\hspace{-1cm}
\subfigure[]
{
\includegraphics[scale=0.27]{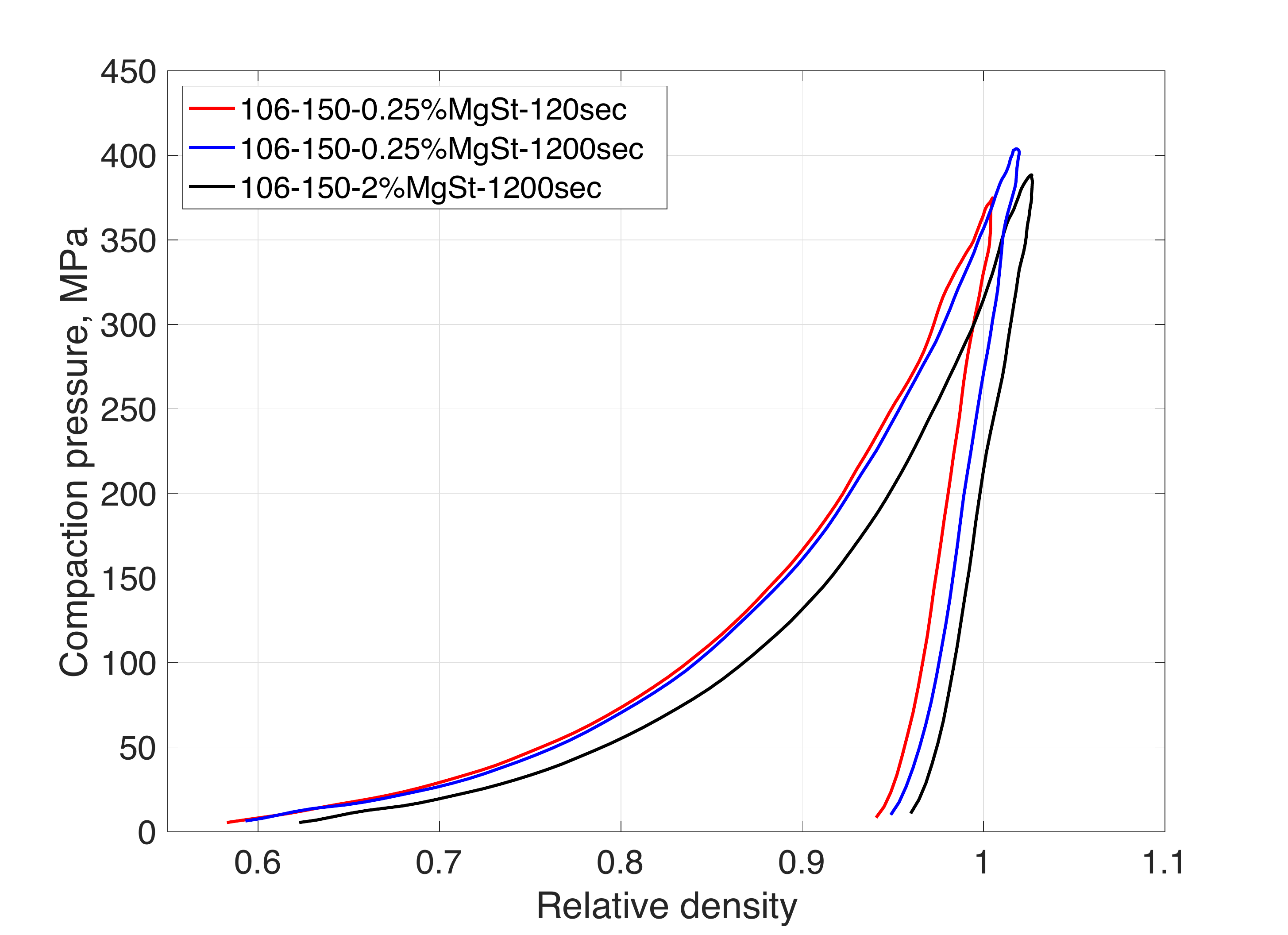}
\label{Fig:CP-106-150-LM}
}
\end{tabular} 
\caption{Lubrication effect on compaction pressure vs. in-die relative density of lactose monohydrate for different particle size distributions: (a) as-recieved, (b) 0-75 $\mu$m, (c) 75-106 $\mu$m, and (d) 106-150 $\mu$m.} 
\label{Fig:CP-LM}
\end{figure}

Fig.~\ref{Fig:CP-LM} compares the axial compaction pressure versus in-die relative density profiles of three similar lubrication conditions for as-received and sieved LM powders. It has been reported that better packing can be achieved during die filling, in particles exposed to higher shear strain levels causing higher initial relative densities (i.e., the die-fill relative density) \citep{mehrotra2007influence,pawar2016effect}. In this study, this phenomenon was noticeably observed in the as-received LM. For the sieved samples, by keeping the MgSt concentration constant, the mixing time did hardly affect the loading and unloading path as a function of relative density. On the other hand, in all the cases, the increase in MgSt concentration results in a different compaction profile, moving toward the right-hand side. This shift is partially caused by the decrease in the true density of the blend. Fig.~\ref{Fig:CP-0-75-LM} discernibly shows that the forces evolved during compression are slightly lower when more lubrication was used, i.e. less work was needed. It has been reported that with increase in lubricant level lower input work is expected attributing it to reduced particle cohesiveness and decreased frictional effects at the punch faces and die wall \citep{mollan1996effects, velasco1997force}. 

\begin{figure}[H]
\centering
\begin{tabular}{cc}
\subfigure[]
{
\hspace{-0.9cm}
\includegraphics[scale=0.275]{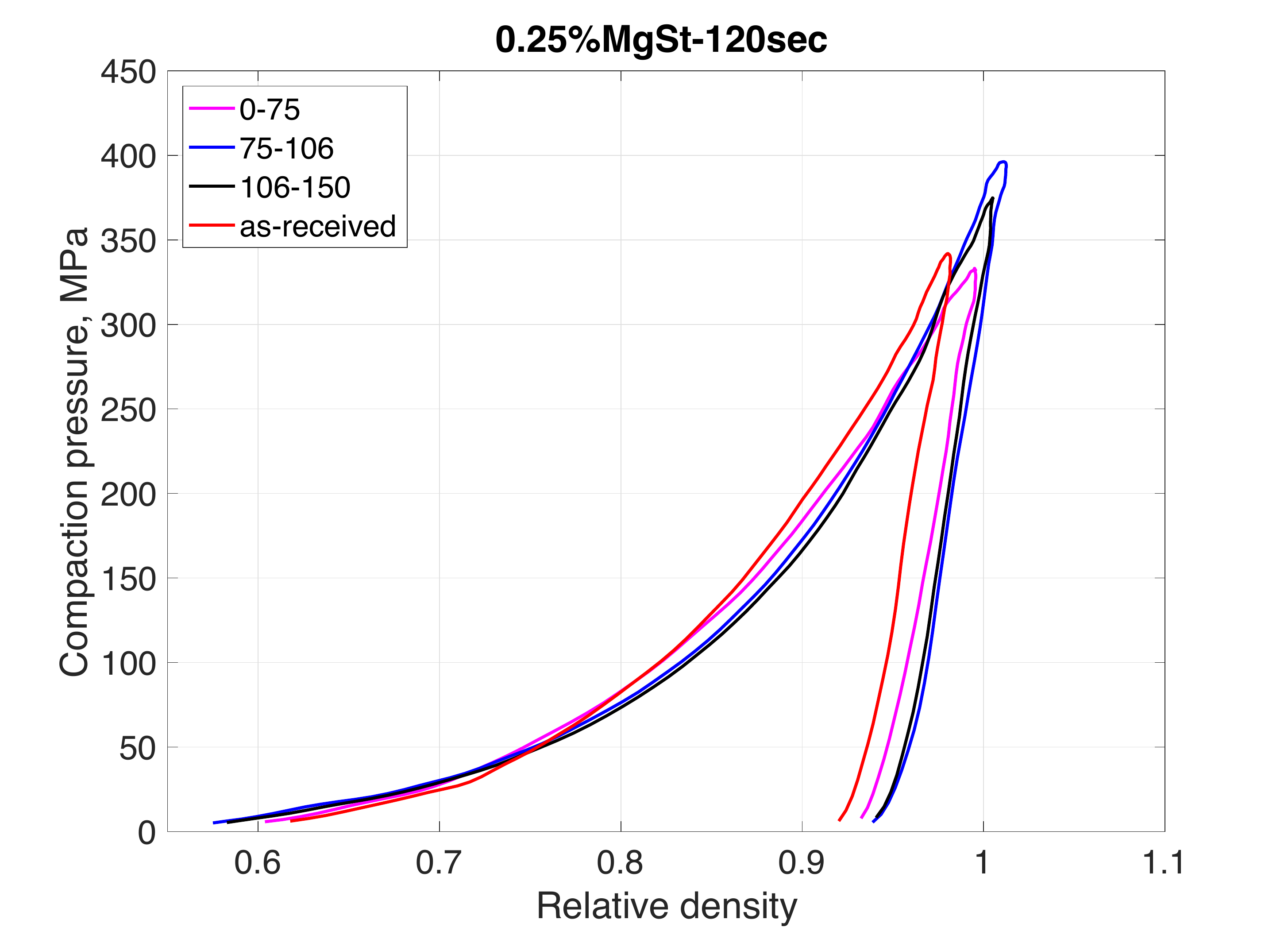}
\label{Fig:CP-025MgSt-120sec-LM}
}
\hspace{-1.1cm}
\subfigure[]
{
\includegraphics[scale=0.275]{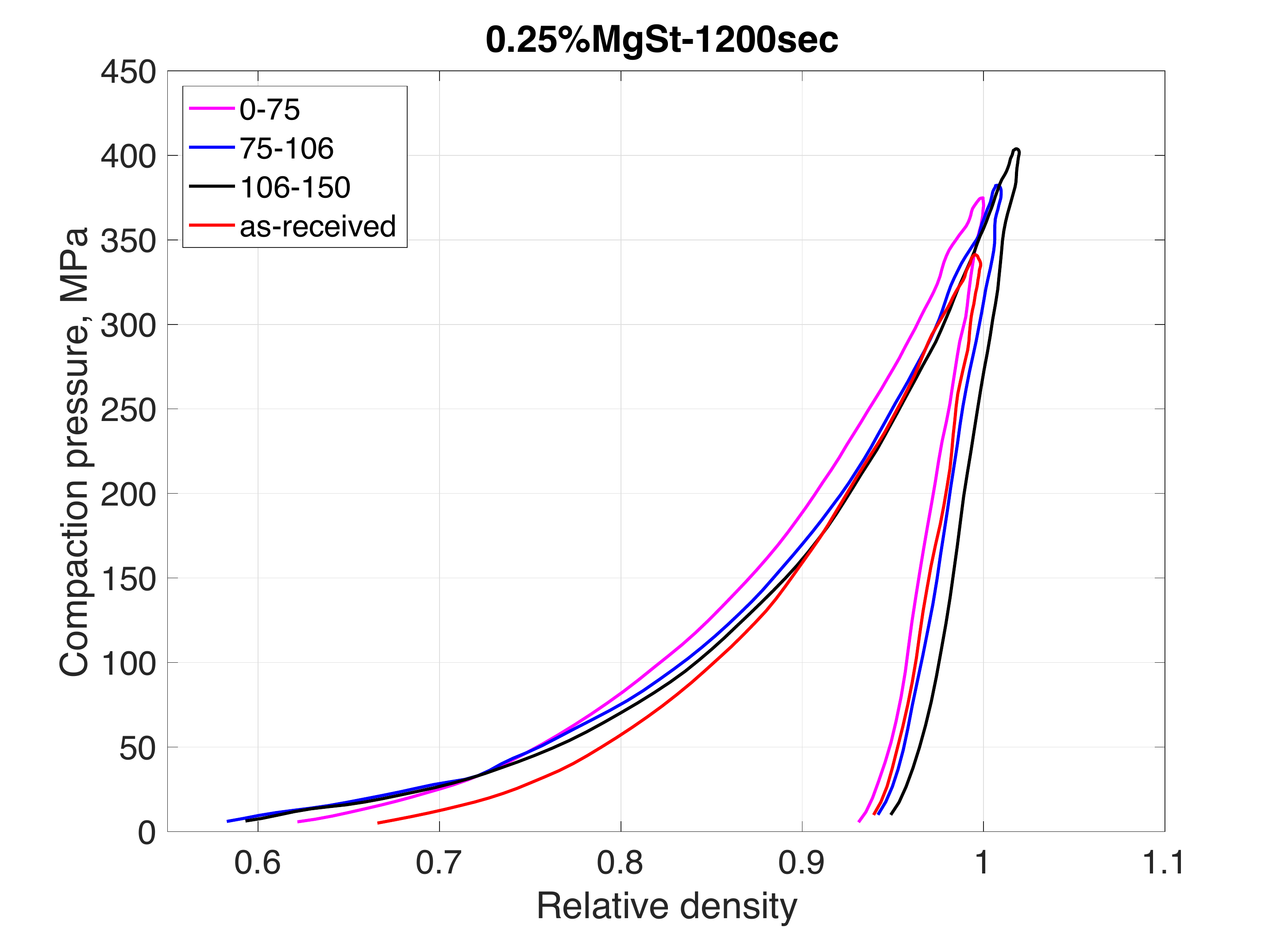}
\label{Fig:CP-025MgSt-1200sec-LM}
}
\hspace{-1cm}
\\
\subfigure[]
{
\includegraphics[scale=0.275]{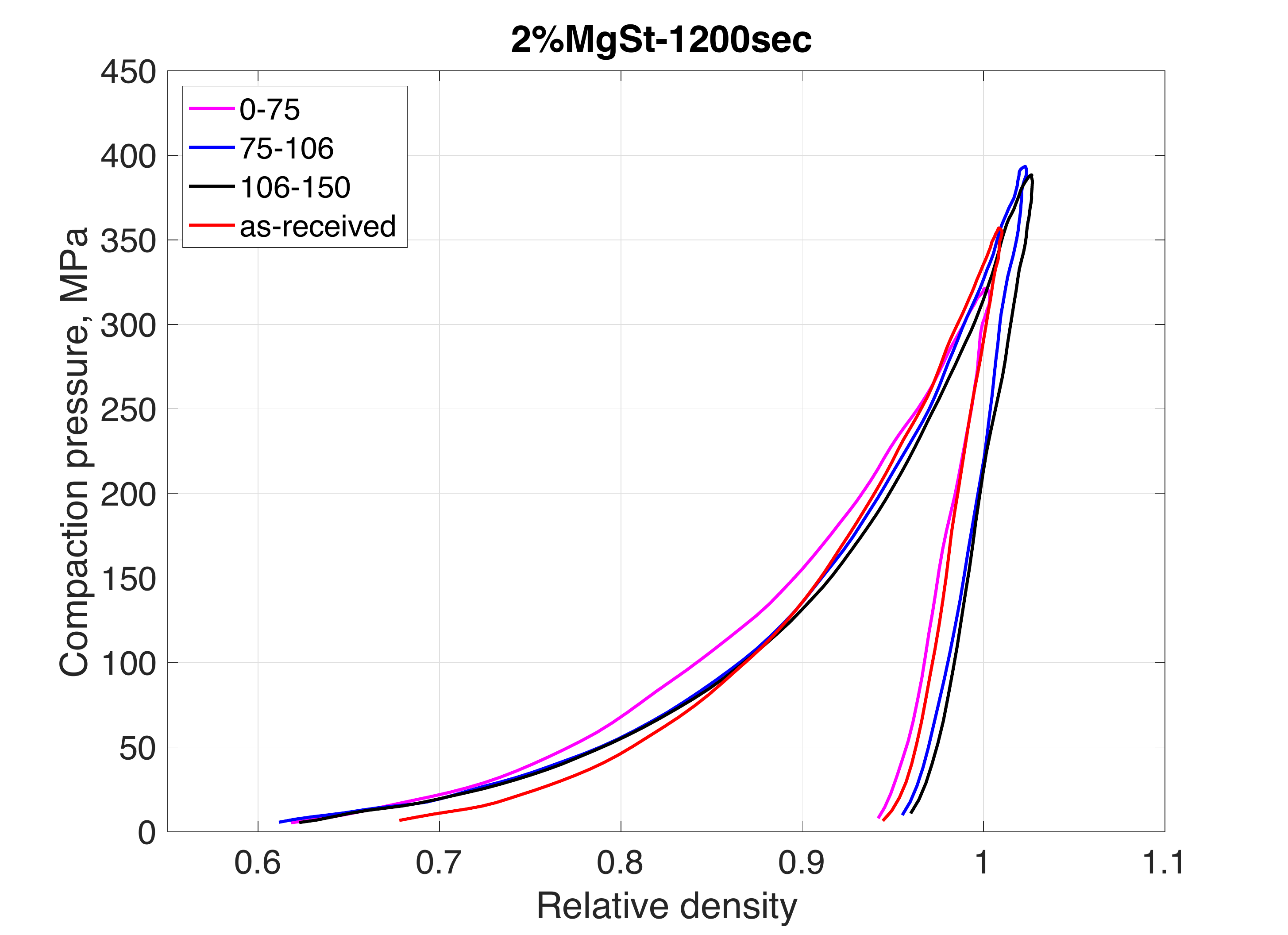}
\label{Fig:CP-2MgSt-1200sec-LM}
}
\end{tabular} 
\caption{Particle size effect on compaction pressure vs. in-die relative density of lactose monohydrate for different lubrication parameters: (a) 0.25$\%$MgSt-120sec, (b) 0.25$\%$MgSt-1200sec, and (c) 2$\%$MgSt-1200sec.} 
\label{Fig:CP-lubrication-LM}
\end{figure}

Fig.~\ref{Fig:CP-lubrication-LM} was plotted to better compare the compaction profiles for different PSD with the same lubrication history. There is no significant difference among the compaction profiles for each condition, indicating that PSD does not affect the deformation behavior of LM with the presence of lubrication. The high initial relative density in the as-received sample compared to the sieved samples, %especially in the high sheared cases ($t_{m}=1200~sec$), 
may be explained by its relatively larger standard deviation in PSD ($31.85~\mu$m) (Table \ref{Table:DOE}). The fine particles fill the voids in between the large particles increasing the bulk density of the powder \citep{yohannes2015role}. 

\begin{figure}[H]
\centering
\begin{tabular}{cc}
\subfigure[]
{
\includegraphics[scale=0.36]{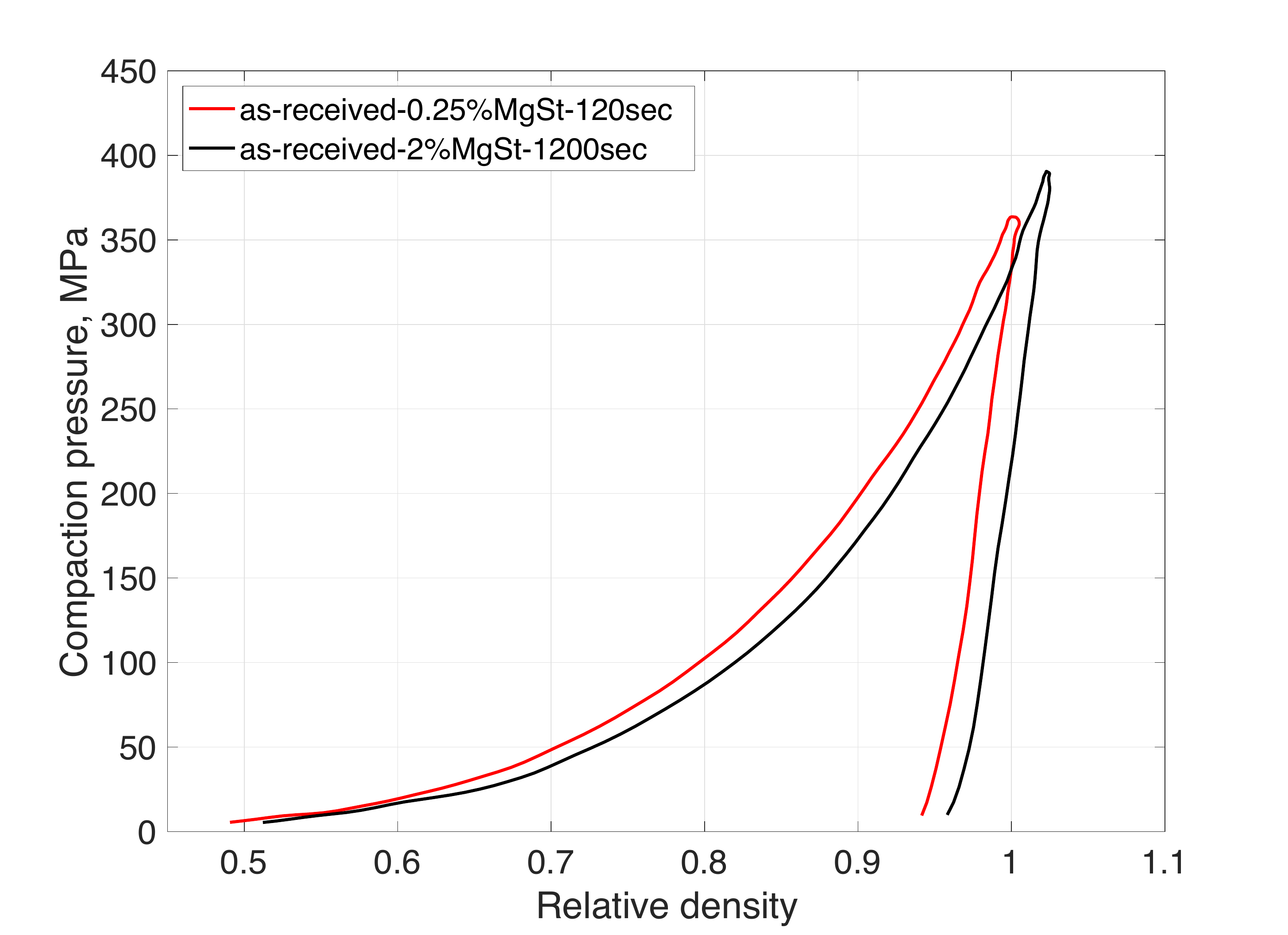}
\label{Fig:CP-AR-SD}
}
\\
\subfigure[]
{
\includegraphics[scale=0.36]{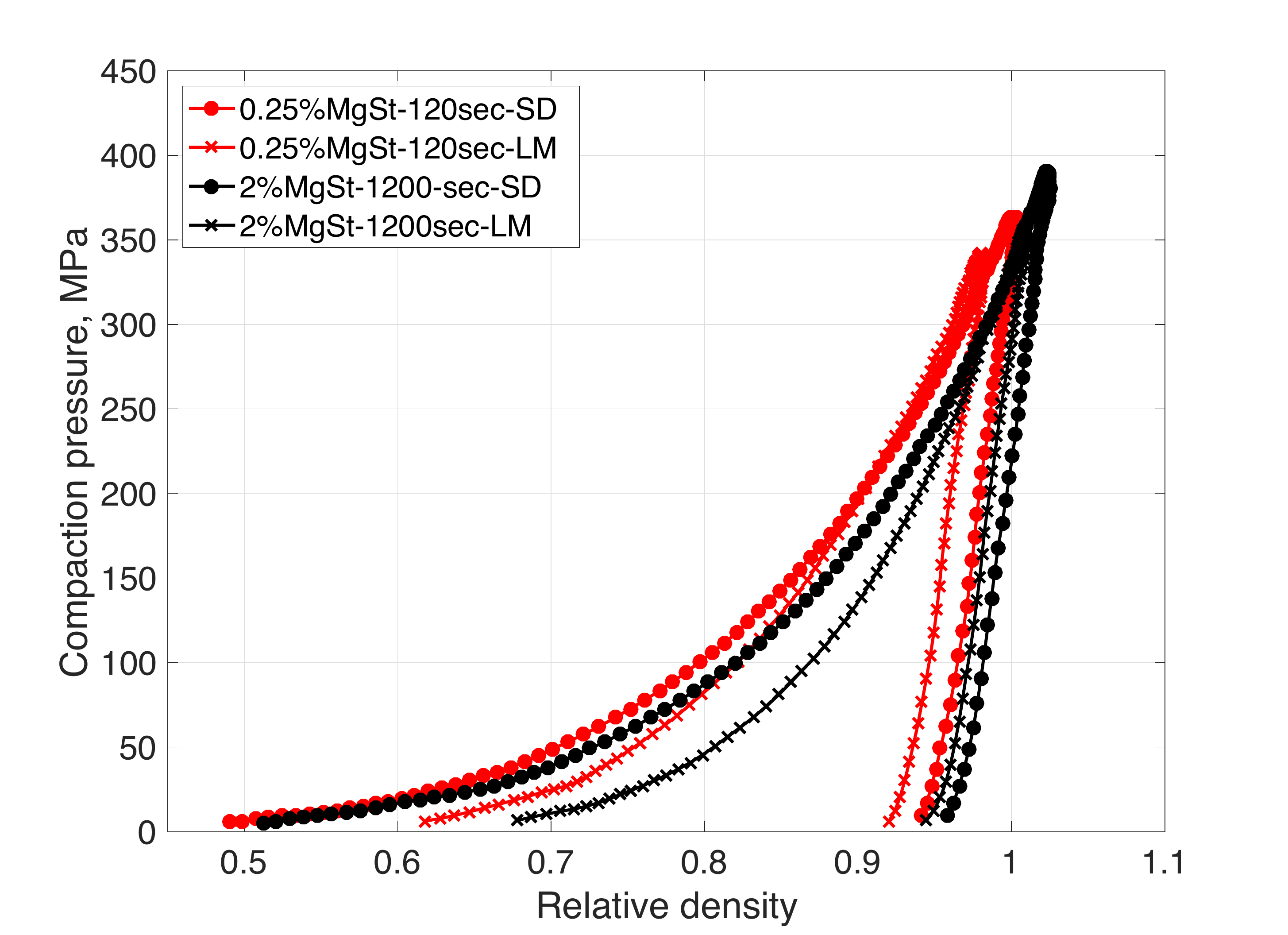}
\label{Fig:CP-LM-SD}
}
\end{tabular} 
\caption{(a) Compaction pressure vs. in-die relative density for as-received spray-dried lactose. (b) Comparison of compaction curves of as-received lactose monohydrate and spray-dried lactose for two extreme lubrication conditions.} 
\label{Fig:CP-AR-LM-SD}
\end{figure}

The compaction pressure- in-die relative density profiles of as-received SDL for two extreme lubrication conditions ascertained little lubricant sensitivity of SDL, as depicted in Fig.~\ref{Fig:CP-AR-SD}. Fig.~\ref{Fig:CP-LM-SD} compares the compaction behavior between the two as-received powders.  The total work input during the compaction process is higher for SDL. \citet{mollan1996effects} argued that as the total work input increases, the compacted powder is expected to be stronger as a result of higher energies used to form bonds between particles. If this hypothesis holds, we expect SDL tablets to be much stronger than LM tablets. However, we should be mindful that the energy of a tablet formation that affects the bond strength, according to the first law of thermodynamics, is associated with both the work done on the powder to form a tablet and the heat released by the system \citep{coffin1983determination}.

It is also interesting to mention that for the same lubrication condition the difference in the compaction profiles between the two powders almost disappears when it reaches its maximum. This happens at relative densities beyond $0.9$, where the area of true contact between particles is large.

\subsection{Effect of particle size on lubricant sensitivity on tensile strength and stiffness of tablets}

\begin{figure}[H]
\centering
\begin{tabular}{cc}
\vspace{-0.41cm}
\subfigure[]
{\hspace{-1cm}
\includegraphics[scale=0.25]{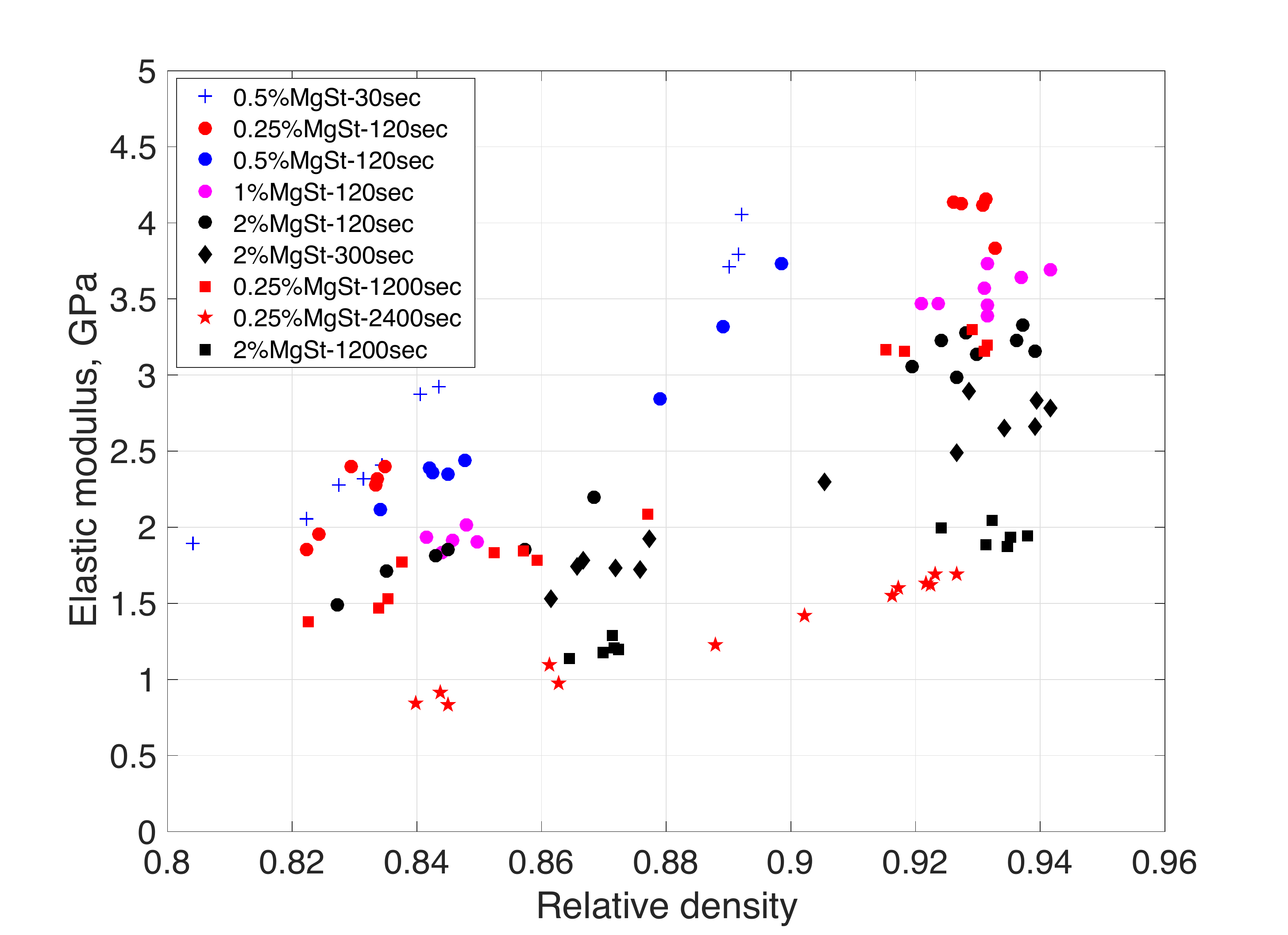}
\label{Fig:E-0-75-LM}
}
\hspace{-1cm}
\subfigure[]
{
\includegraphics[scale=0.25]{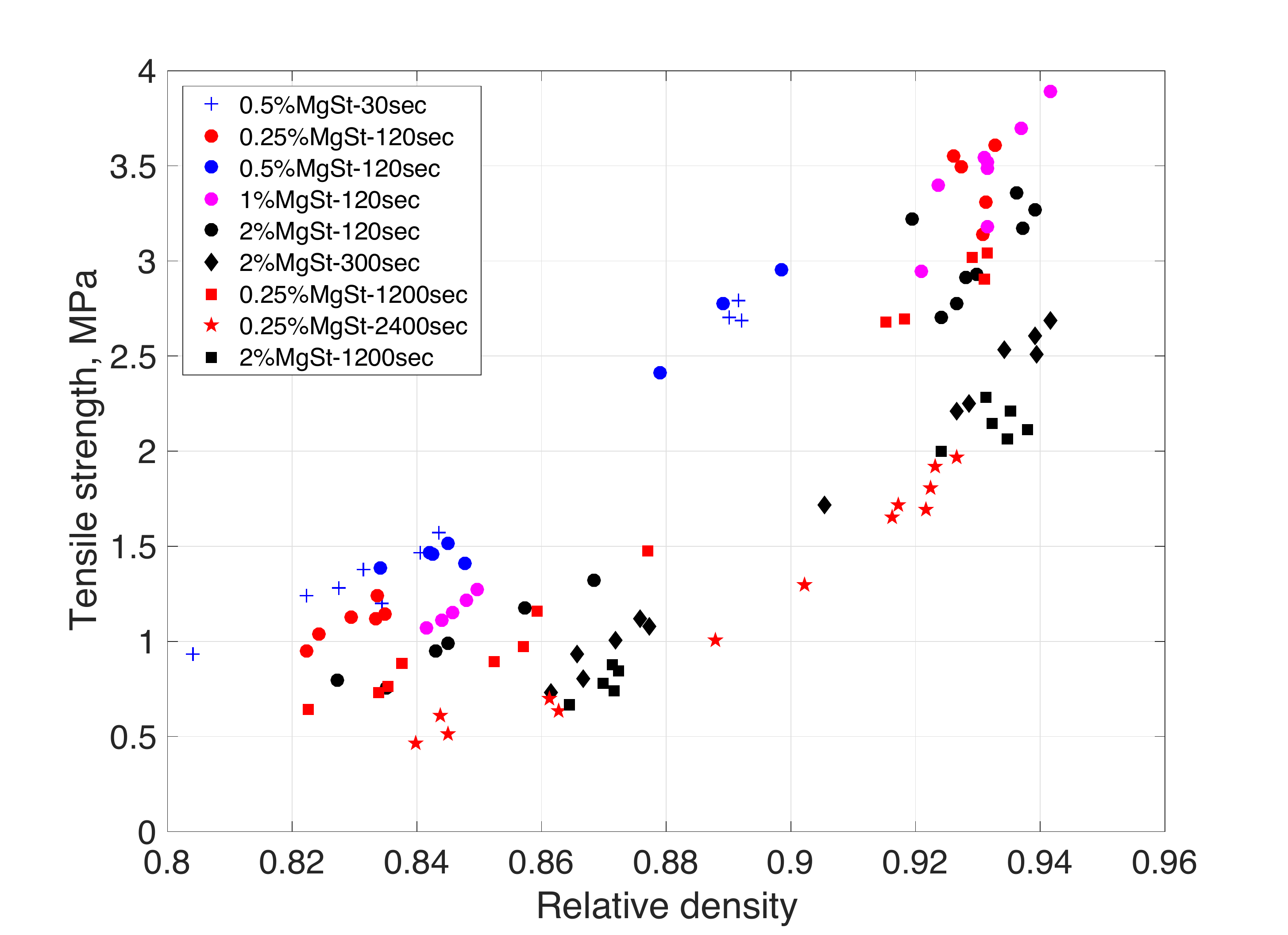}
\label{Fig:TS-0-75-LM}
}
\\
\vspace{-0.41cm}
\subfigure[]
{\hspace{-1cm}
\includegraphics[scale=0.25]{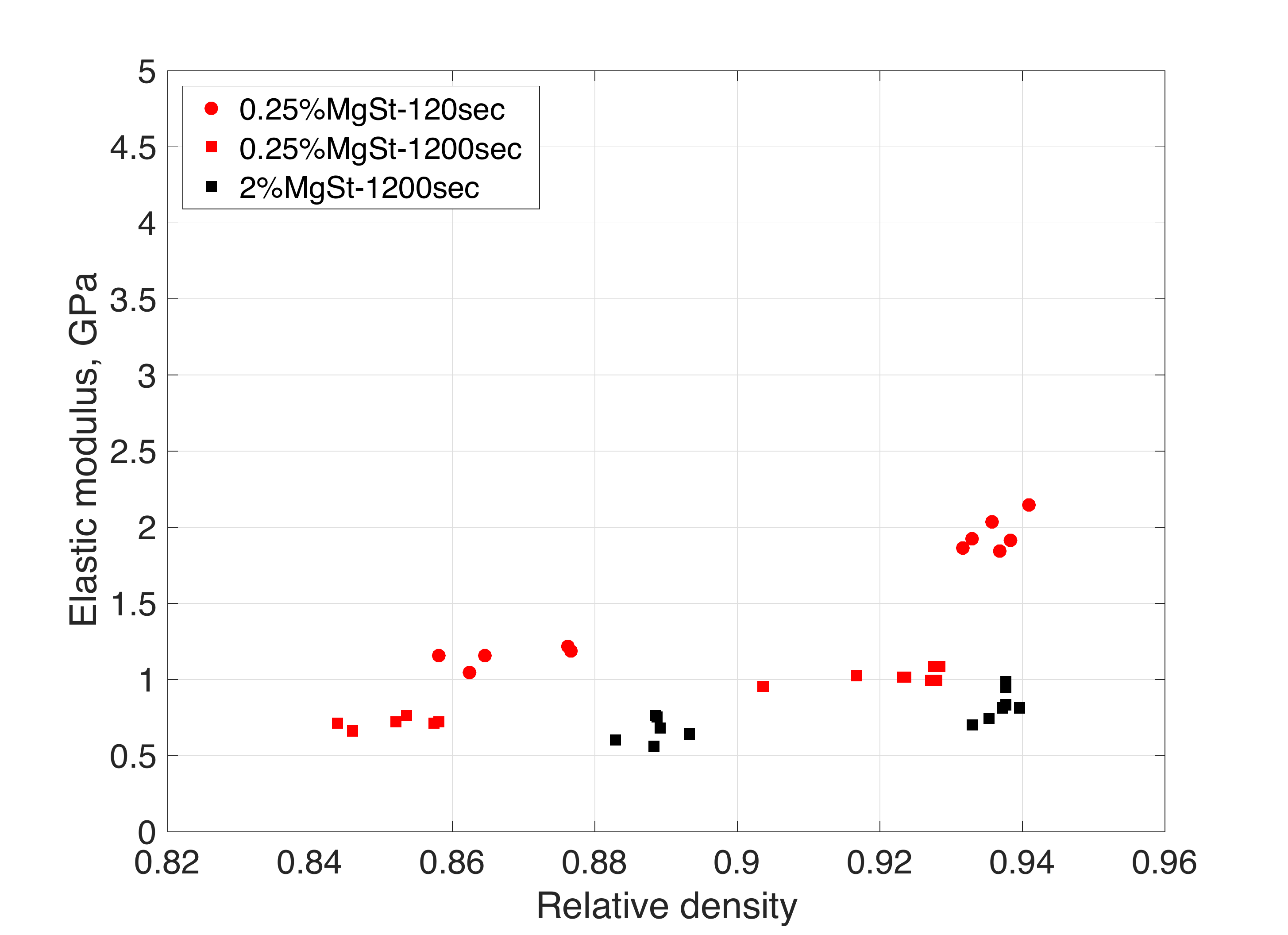}
\label{Fig:E-75-106-LM}
}
\hspace{-1cm}
\subfigure[]
{
\includegraphics[scale=0.25]{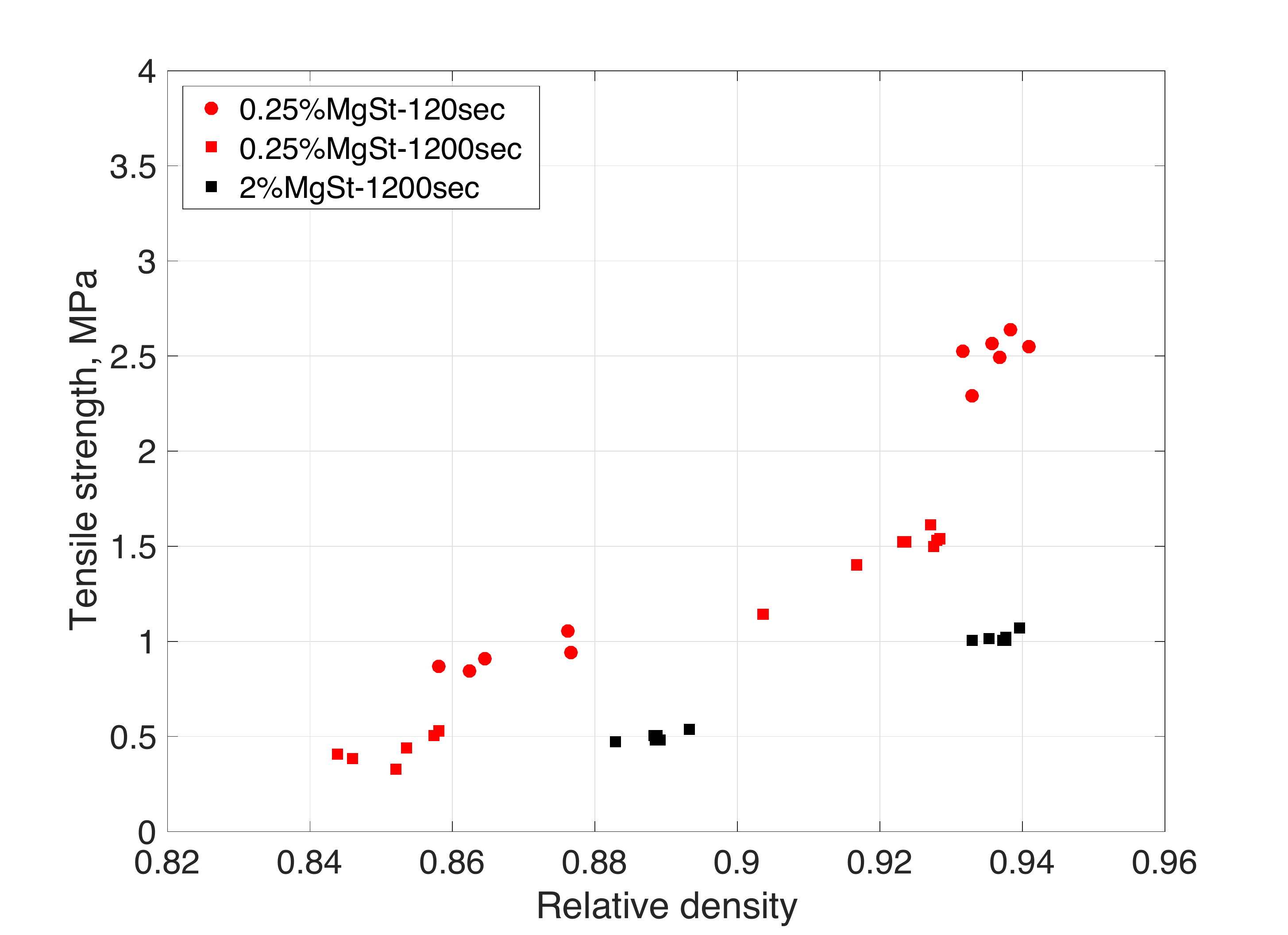}
\label{Fig:TS-75-106-LM}
}
\\
\vspace{-0.41cm}
\subfigure[]
{\hspace{-1cm}
\includegraphics[scale=0.25]{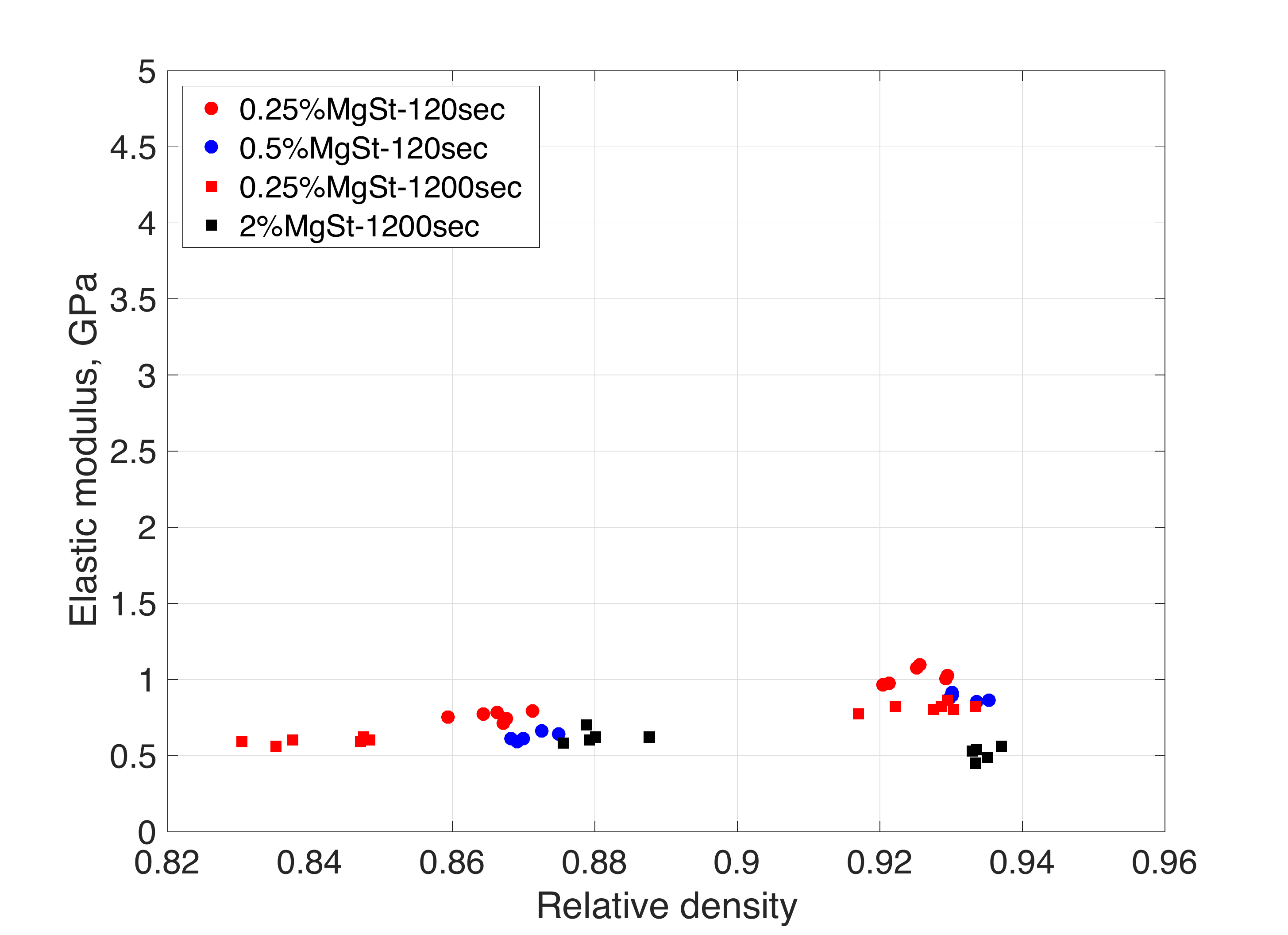}
\label{Fig:E-106-150-LM}
}
\hspace{-1cm}
\subfigure[]
{
\includegraphics[scale=0.25]{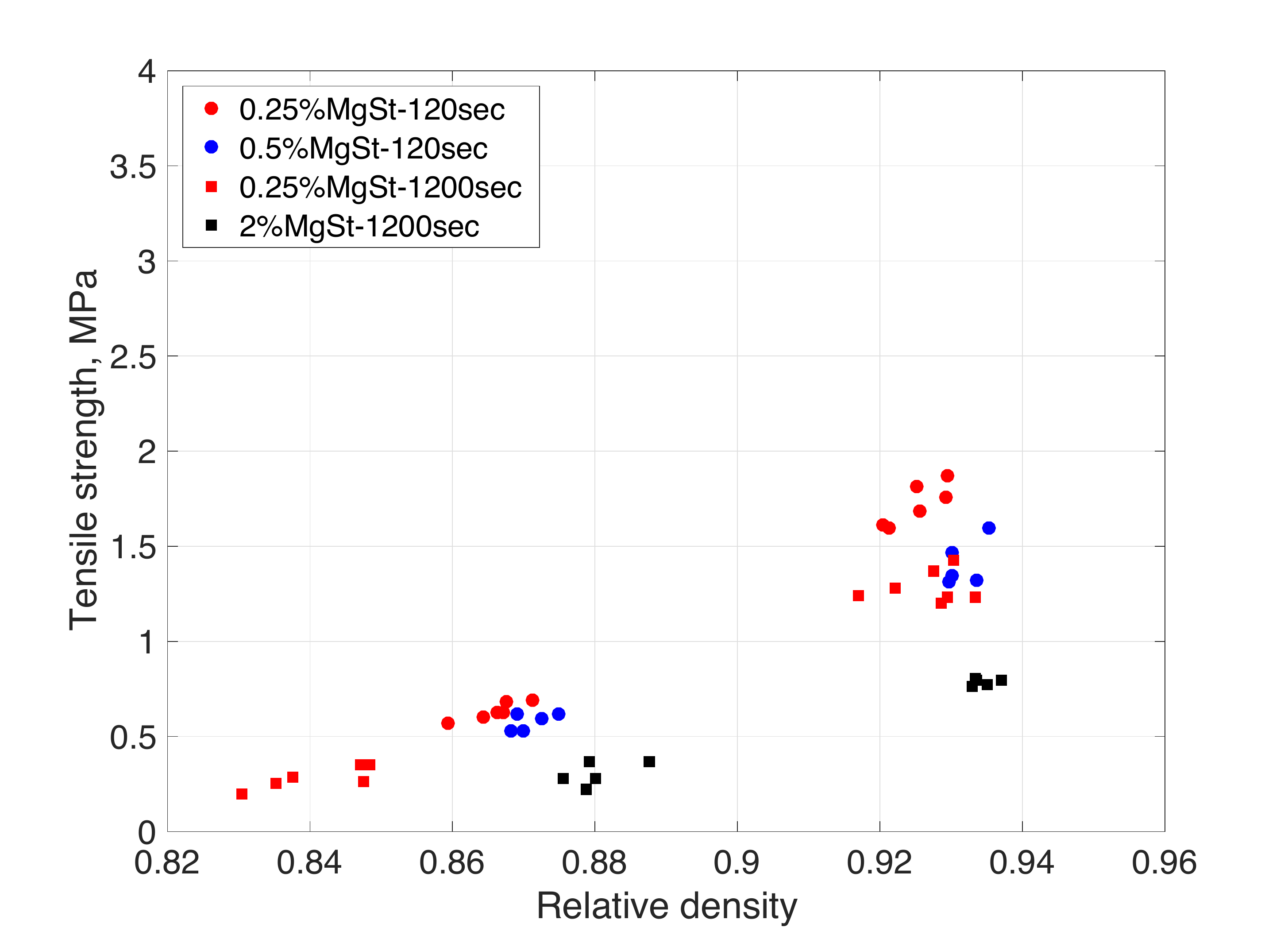}
\label{Fig:TS-106-150-LM}
}
\\
\vspace{-0.35cm}
\subfigure[]
{
\hspace{-1cm}
\includegraphics[scale=0.25]{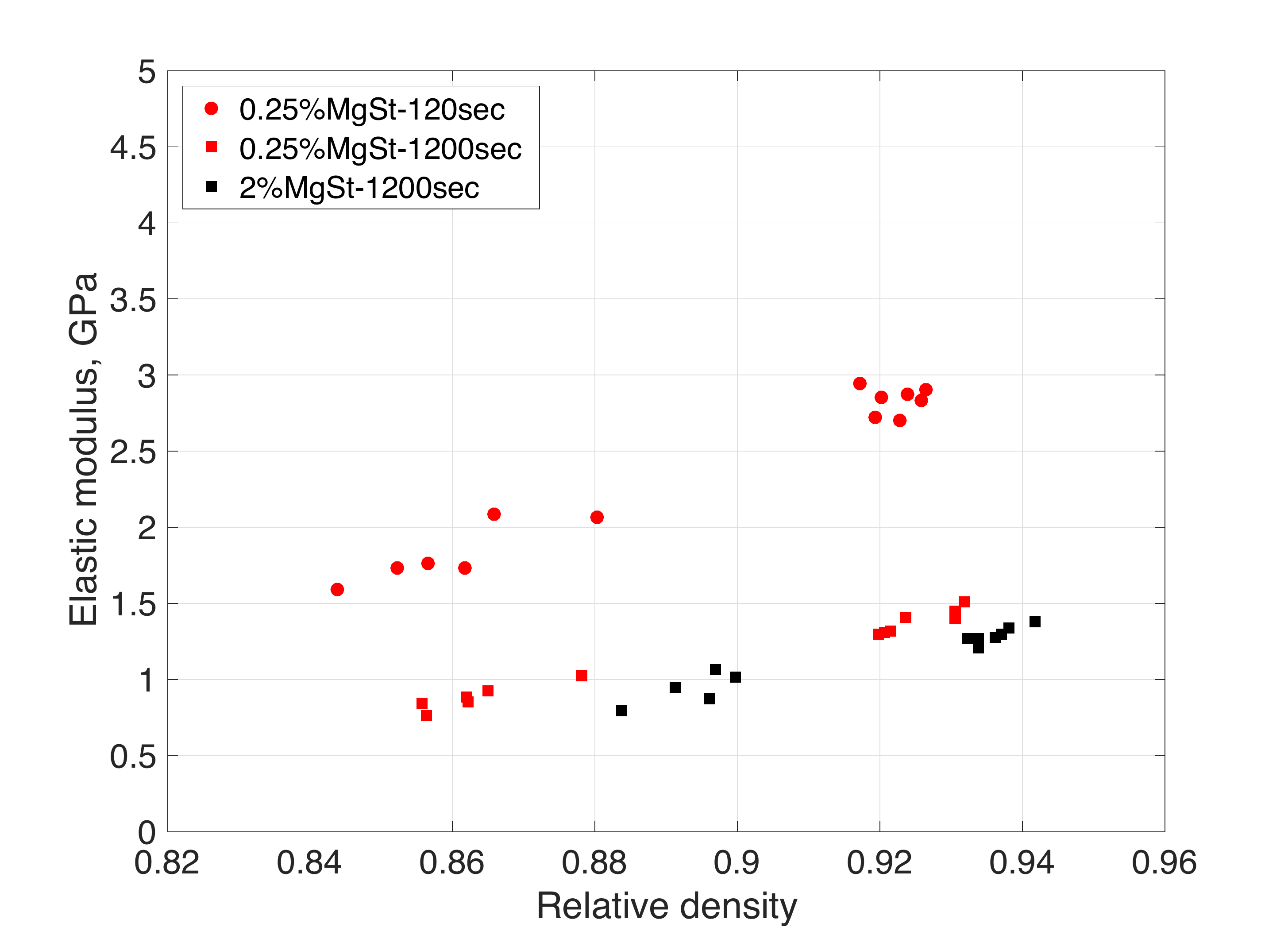}
\label{Fig:E-AR-LM}
}
\hspace{-1cm}
\subfigure[]
{
\includegraphics[scale=0.25]{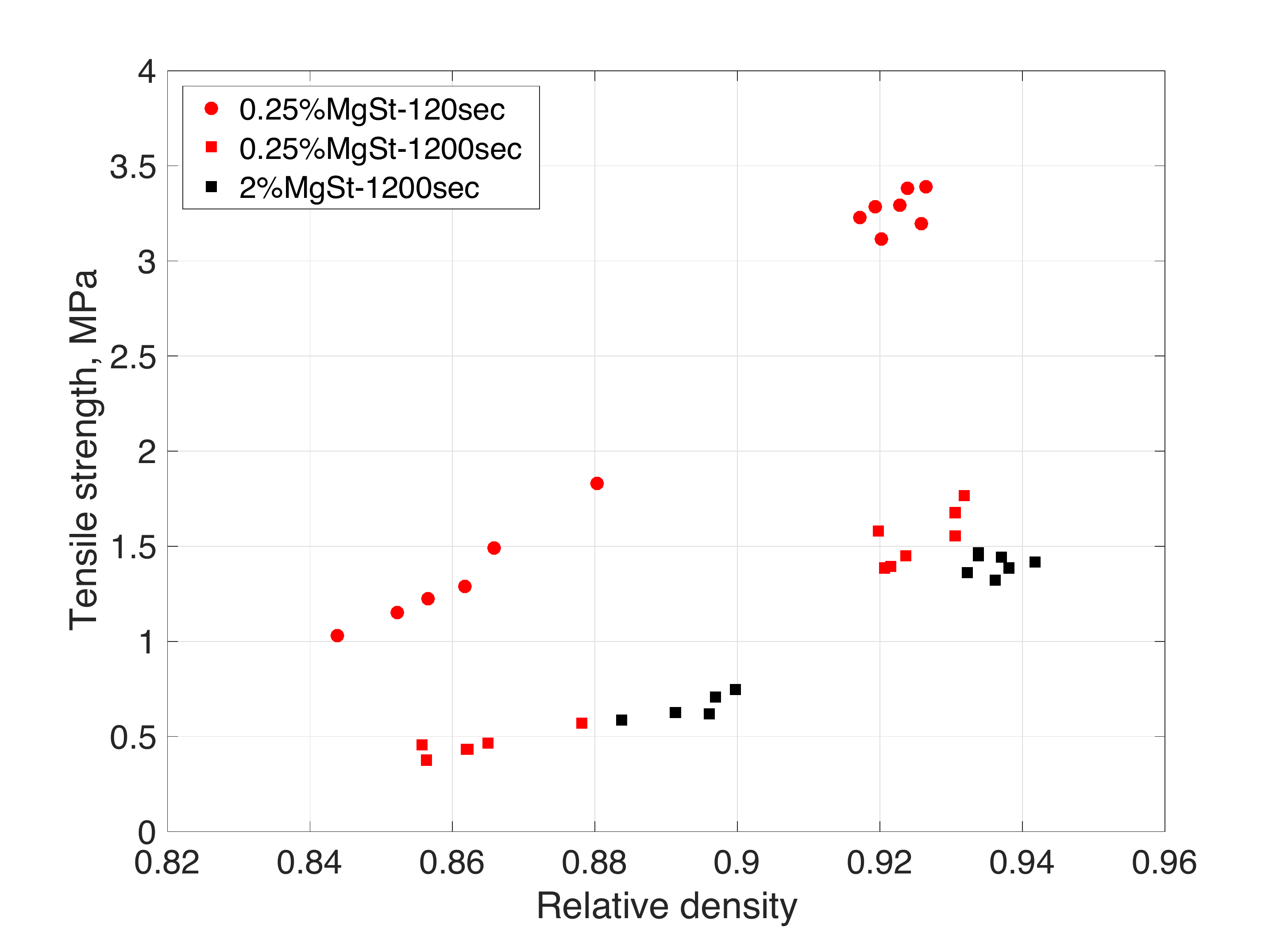}
\label{Fig:TS-AR-LM}
}
\end{tabular} 
\caption{Elastic modulus and tensile strength vs. out-of-die relative density for different PSDs of lactose monohydrate; (a, b) 0-75 $\mu$m, (c, d) 75-106 $\mu$m, (e, f) 106-150 $\mu$m, and (g, h) as-received.} 
\label{Fig:properties-LM}
\end{figure}

Elastic modulus and tensile strength versus out-of-die relative density for different PSDs of LM tablets are depicted in Fig.~\ref{Fig:properties-LM}. As was expected, both the elastic modulus and tensile strength decreased by adding more lubricant and/or mixing time. Although, the effect is more significant in smaller PSD compared to large PSD. This can be attributed to smaller particles having more available surface area to be covered by MgSt coating for the same tablet weight. For all the PSD levels, the lubrication affects the strength and stiffness level until it reaches a saturation regime, where the powder would no longer be affected with the addition of lubricant concentration or mixing time, in accordance with \citep{kikuta1994effect}. It should be noted that the lubricant concentration and mixing time do not affect the tensile strength and elastic modulus of tablets by the same rate. The results demonstrate an envelope for tensile strength and elastic modulus obtainable for tablets with relative densities ranging between $0.8$ and $0.94$ considering different PSDs and lubricant conditions.

\begin{figure}[H]
\centering
\begin{tabular}{cc}
\subfigure[]
{
\hspace{-1cm}
\includegraphics[scale=0.3]{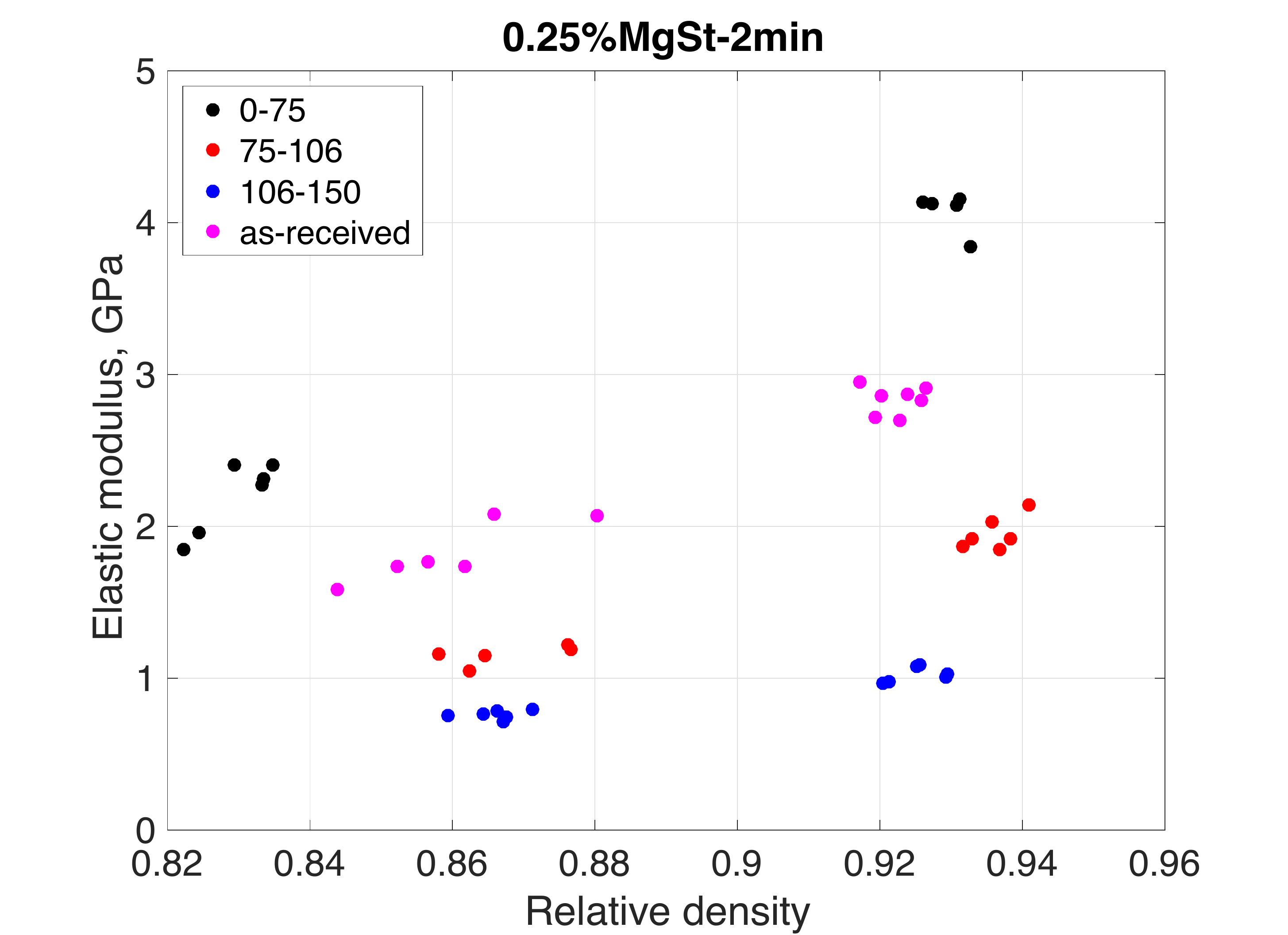}
\label{Fig:E-AR-LM}
}
\subfigure[]
{
\hspace{-1cm}
\includegraphics[scale=0.3]{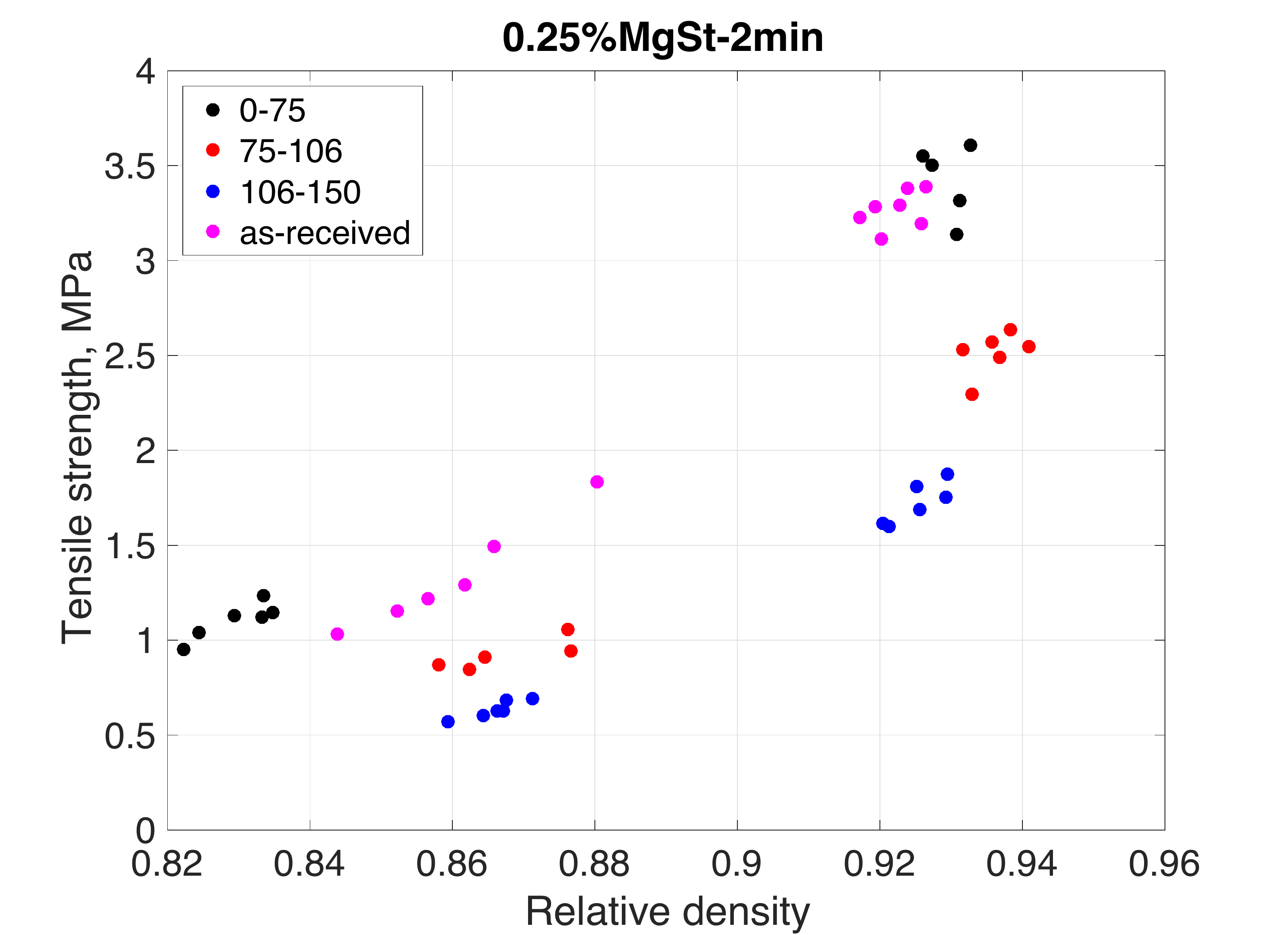}
\label{Fig:AR-150-LM}
}
\\
\subfigure[]
{
\hspace{-1cm}
\includegraphics[scale=0.3]{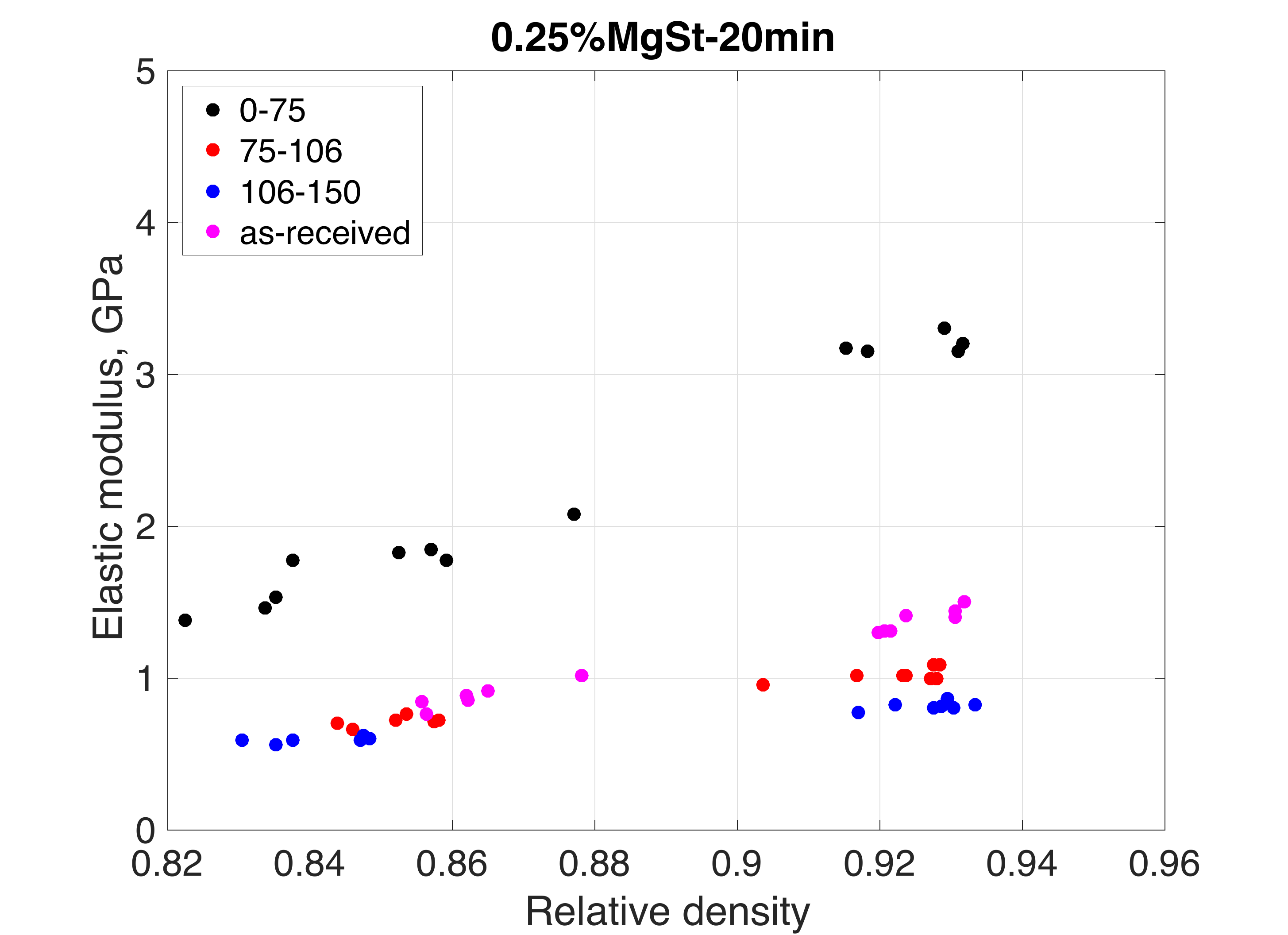}
\label{Fig:E-AR-LM}
}
\subfigure[]
{
\hspace{-1cm}
\includegraphics[scale=0.3]{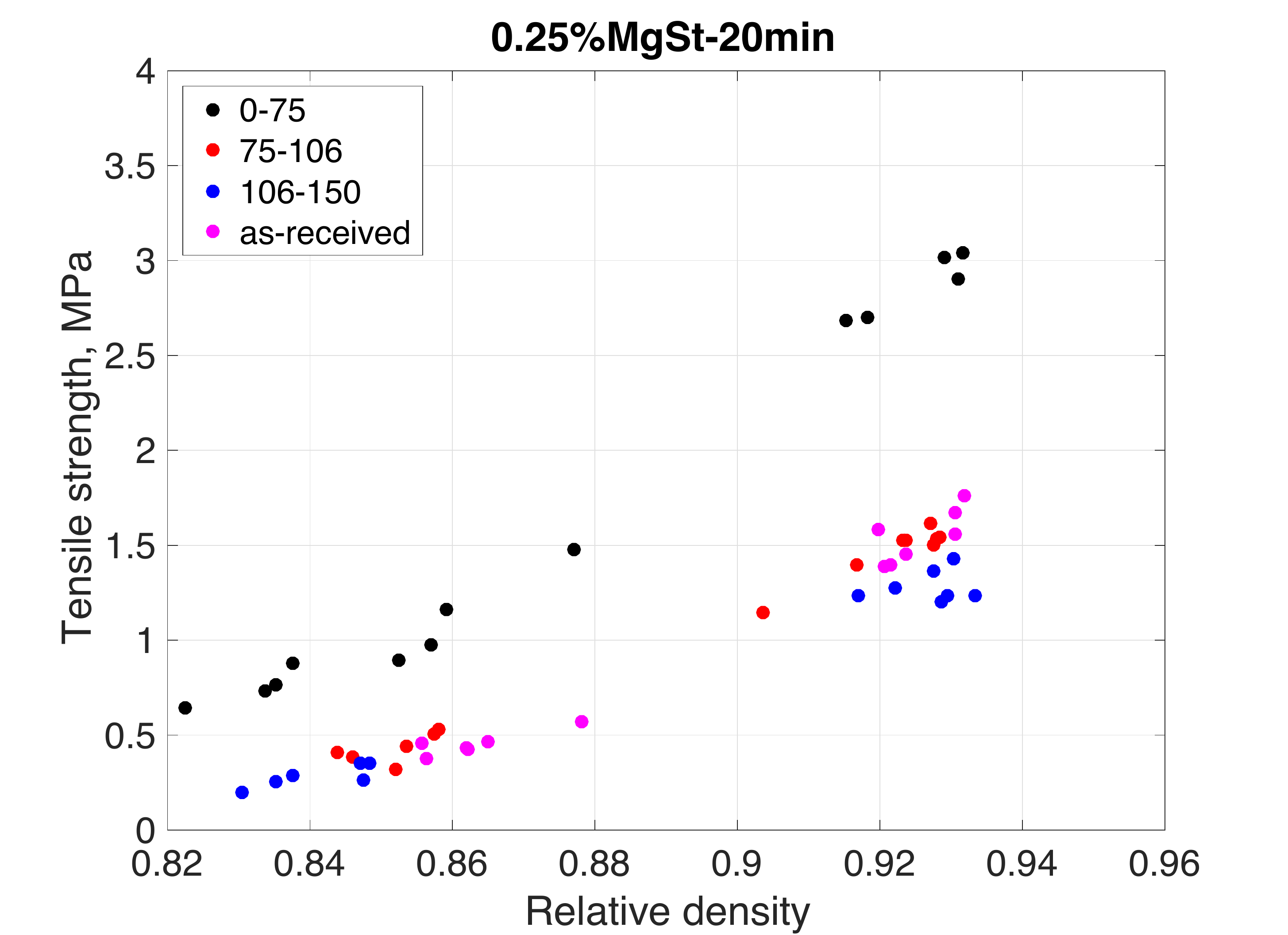}
\label{Fig:AR-150-LM}
}
\\
\subfigure[]
{
\hspace{-1cm}
\includegraphics[scale=0.3]{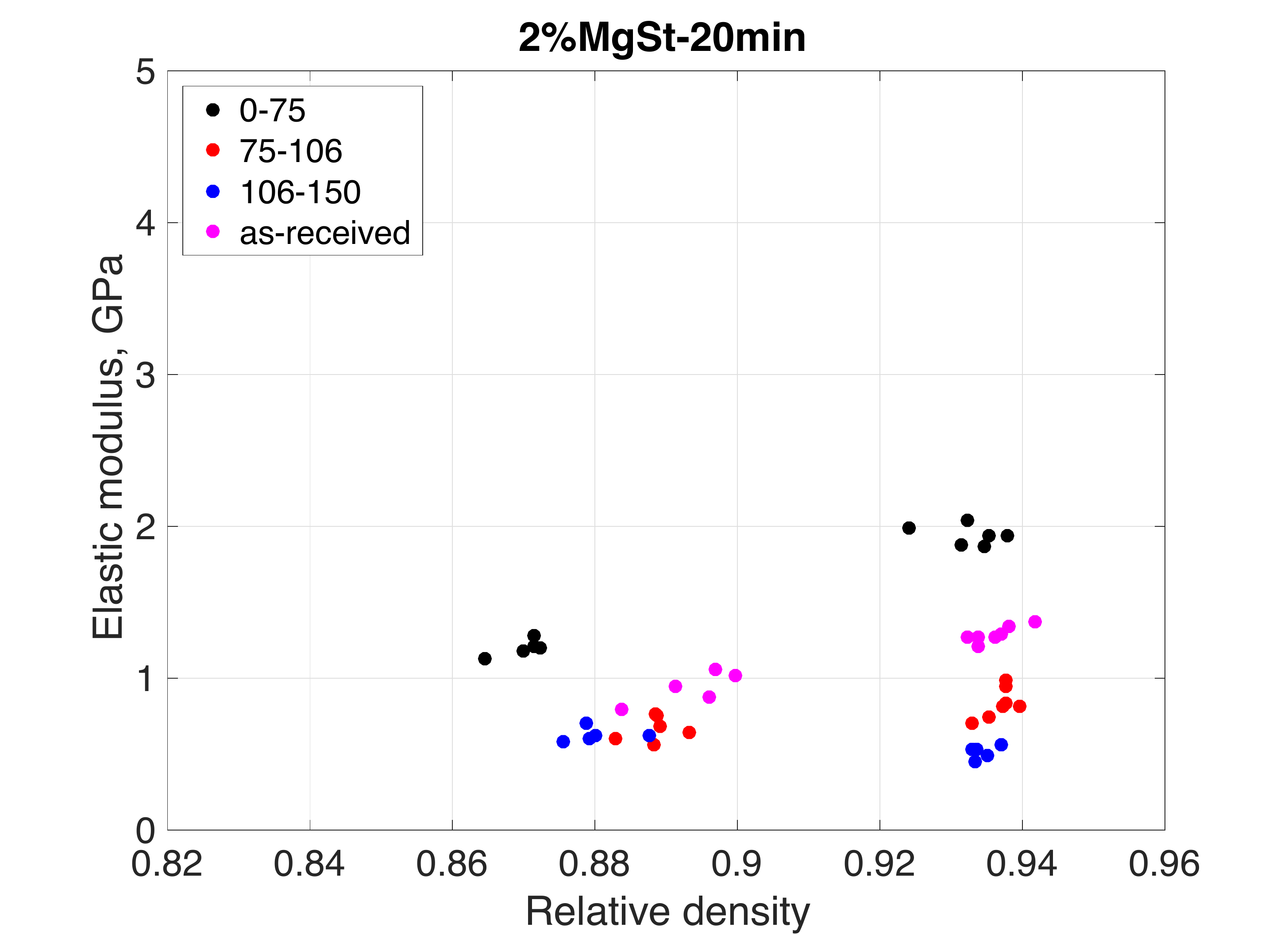}
\label{Fig:E-AR-LM}
}
\subfigure[]
{
\hspace{-1cm}
\includegraphics[scale=0.3]{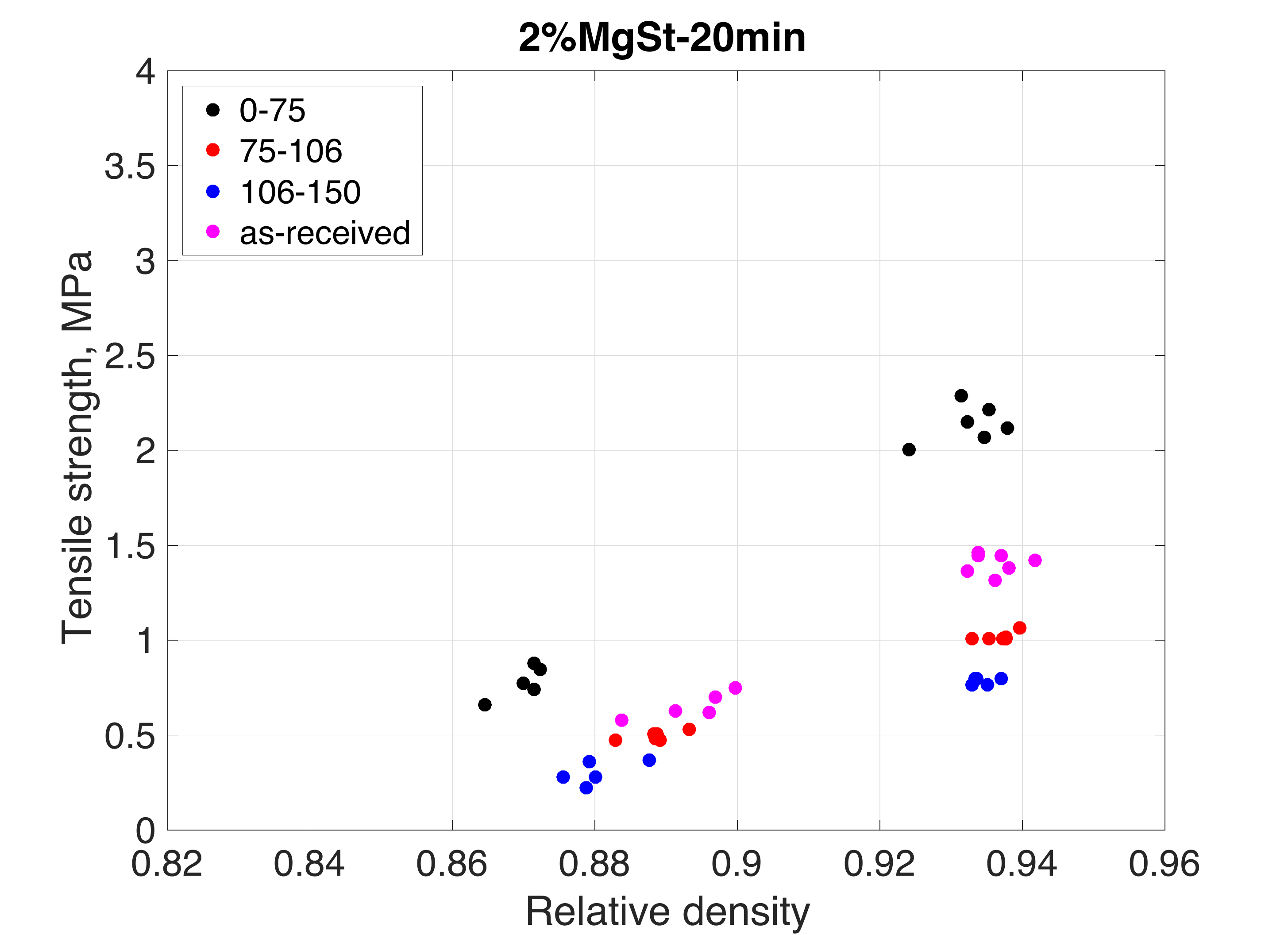}
\label{Fig:AR-150-LM}
}
\end{tabular} 
\caption{Partile size effect on elastic modulus and tensile strength of lactose monohydrate tablets at different lubrication conditions (a, b) 0.25\%MgSt-2min, (c, d) 0.25\%MgSt-20min, and (e, f) 2\%MgSt-20min.} 
\label{Fig:comparisons-LM}
\end{figure}

The initial particle size of LM affects the mechanical strength of tablets, as depicted in Fig.~\ref{Fig:comparisons-LM}. Large mean particle sizes exhibited faster response to tablet strength saturation by increasing lubricant concentration and/or mixing time. The reduction in strength of large particles ($106$-$150$ $\mu$m), even for the least lubricated tablets (case $13$), was to a degree that hardly particles formed a solid and the tablets were extremely weak. Obviously, in practical purposes this level of strength is not desirable but since our goal is to have a model based on lubrication sensitivity we created a vast collection of data.

The difference between each PSD is noticeable for the low lubrication condition (i.e.,  0.25\%MgSt-2min). This may be due to the more available surface area in smaller particles requiring more mixing time to fully be coated by MgSt.

The as-received tablets in the low lubrication condition show high elastic modulus and tensile strength because of having a large population of small particles. However, as the shear strain increases, the larger particles are overlubricated causing a significant drop in the elastic modulus and tensile strength.
Altogether, we have observed that the lubricant and PSD sensitivity of LM are more pronounced in the tablet properties than in its deformation behavior during compaction.

\begin{figure}[H]
\centering
\begin{tabular}{cc}
\subfigure[]
{
\hspace{-1cm}
\includegraphics[scale=0.27]{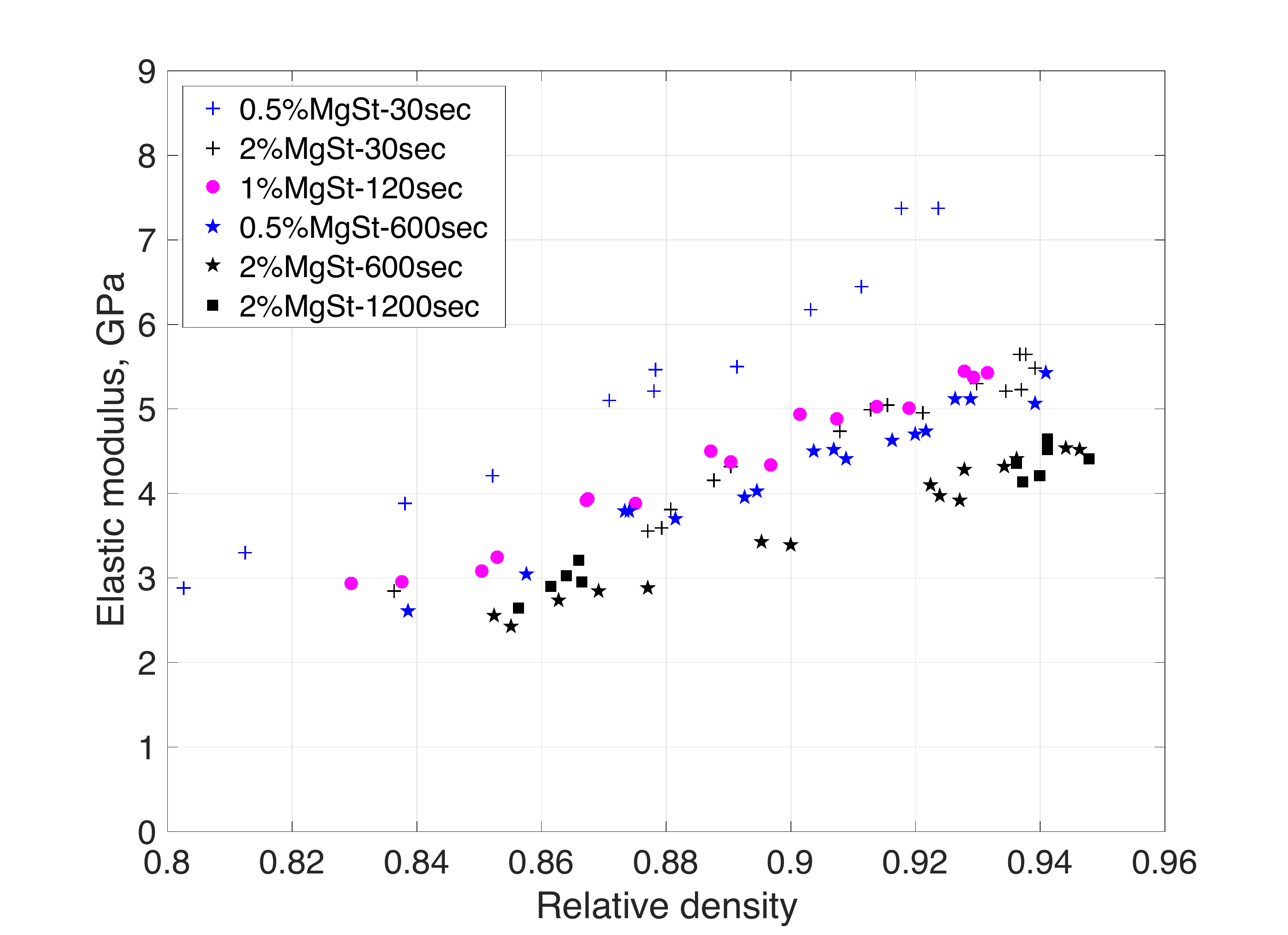}
\label{Fig:E-0-75-SD}
}
\subfigure[]
{
\hspace{-1cm}
\includegraphics[scale=0.27]{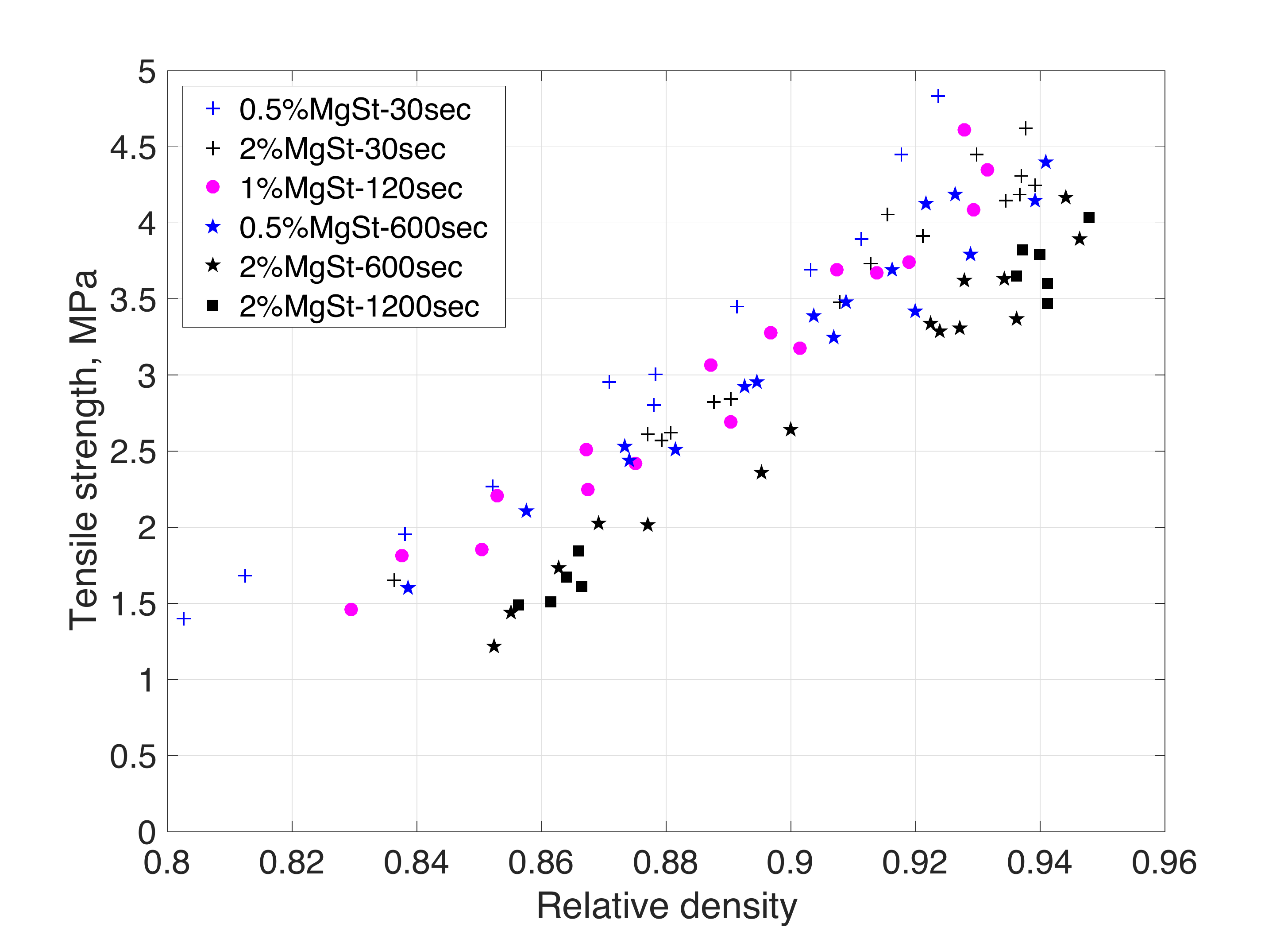}
\label{Fig:TS-0-75-SD}
}
\\\subfigure[]
{
\hspace{-1cm}
\includegraphics[scale=0.27]{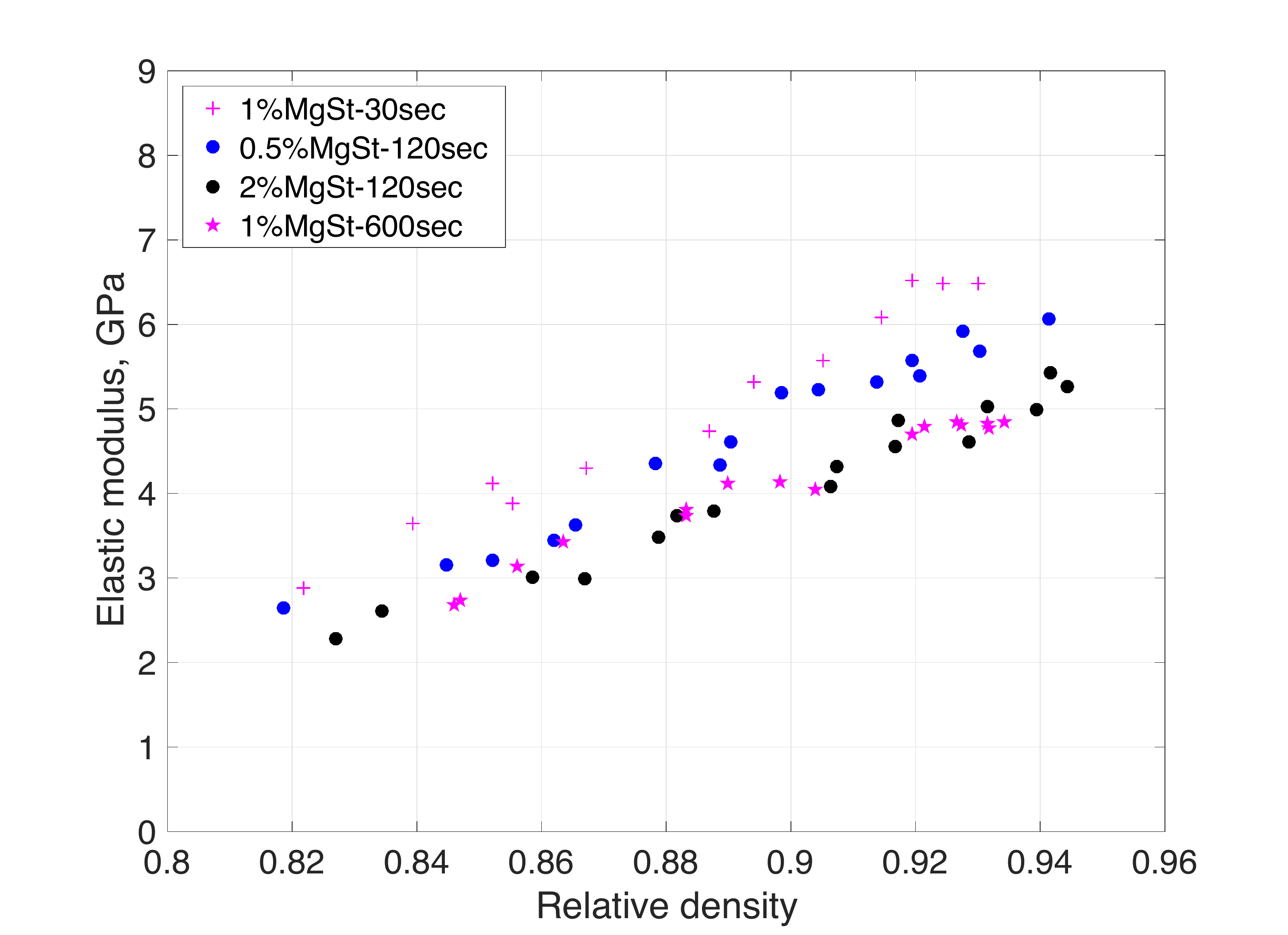}
\label{Fig:E-75-106-SD}
}
\subfigure[]
{
\hspace{-1cm}
\includegraphics[scale=0.27]{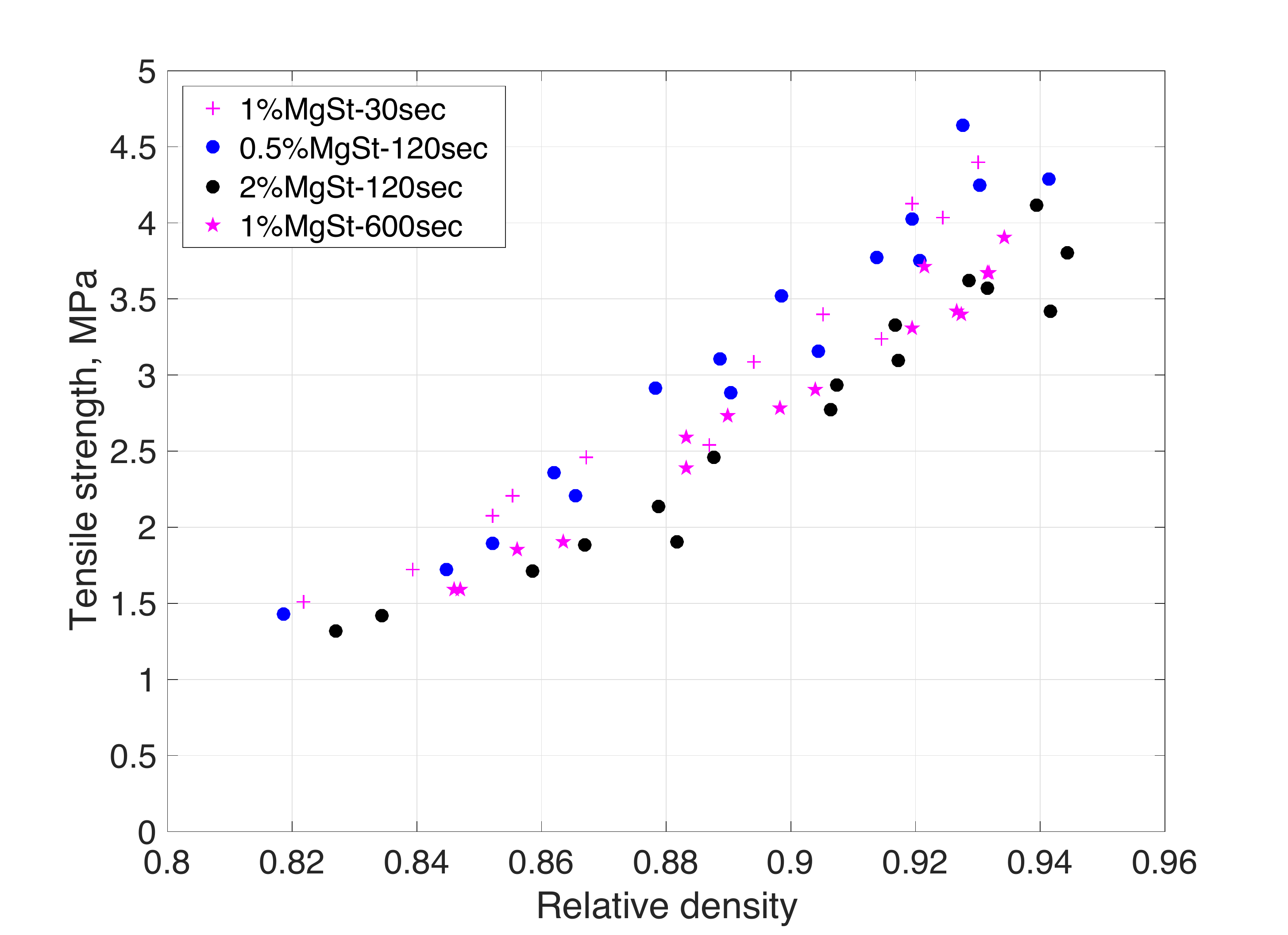}
\label{Fig:TS-75-106-SD}
}
\\\subfigure[]
{
\hspace{-1cm}
\includegraphics[scale=0.27]{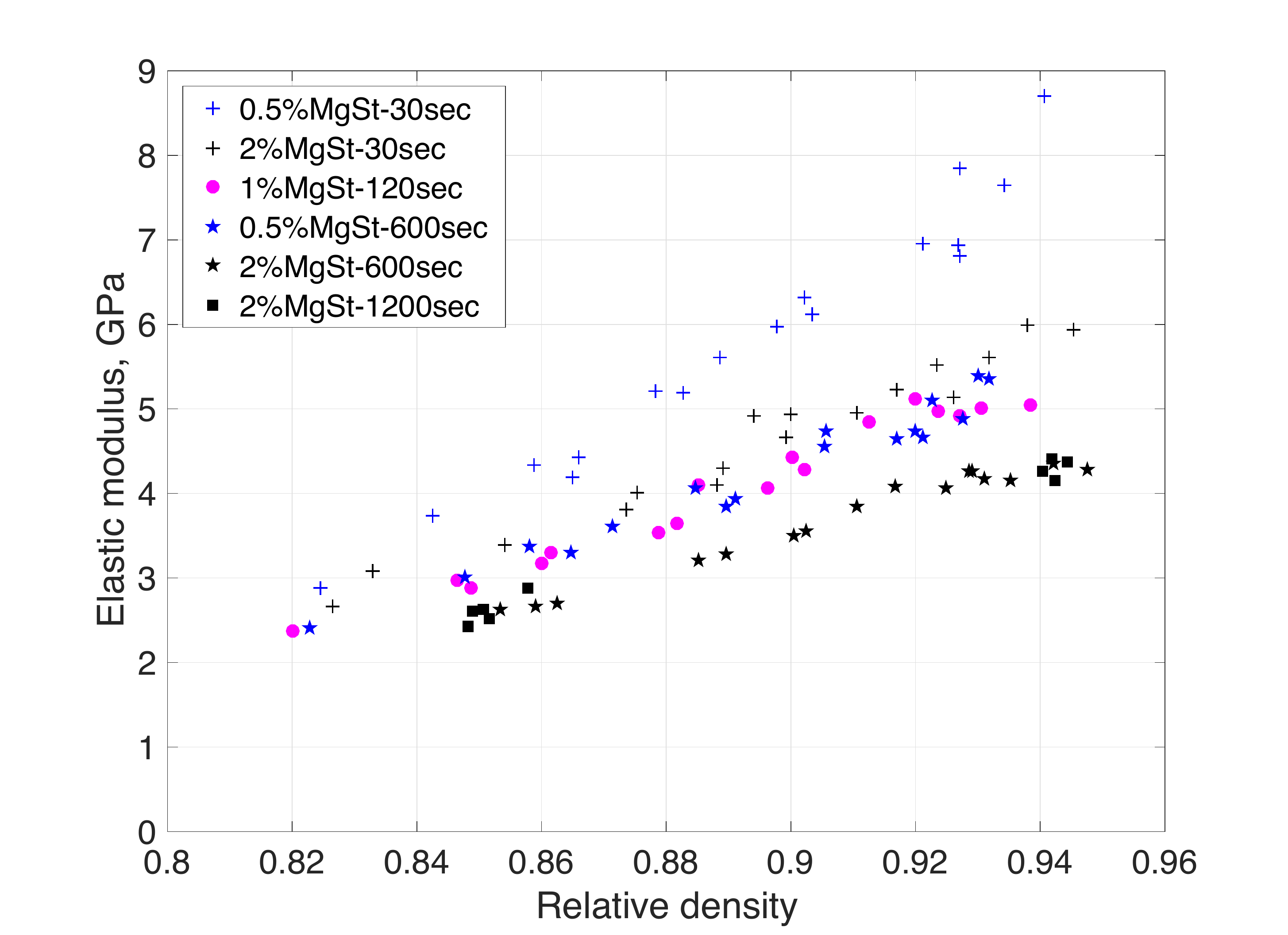}
\label{Fig:E-106-150-SD}
}
\subfigure[]
{
\hspace{-1cm}
\includegraphics[scale=0.27]{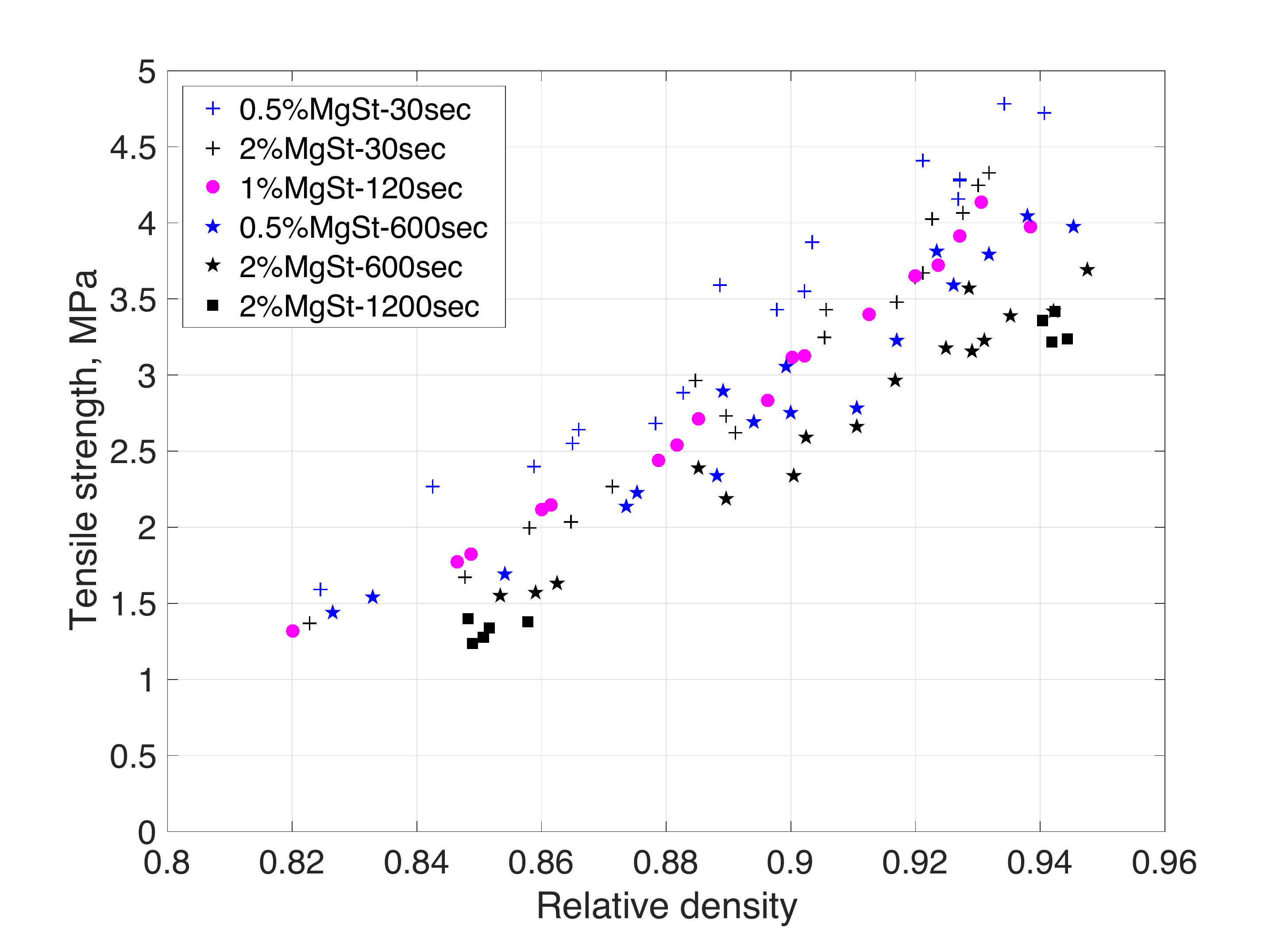}
\label{Fig:TS-106-150-SD}
}
\end{tabular} 
\caption{Elastic modulus and tensile strength vs. out-of-die relative density for different particle size distributions of spray-dried lactose; (a,b) 0-75 $\mu$m, (c,d) 75-106 $\mu$m, (e,f) 106-150 $\mu$m, (g,h) 150-212 $\mu$m, and (i,j) as-received.} 
\label{Fig:properties-SD}
\end{figure}

The elastic modulus and tensile strength of all the SDL tablets are plotted against their relative densities for different PSD levels in Fig.~\ref{Fig:properties-SD}. SDL shows a non-negligible sensitivity to lubrication and higher values of elastic modulus and tensile strength compared to LM tablets. This difference is more remarkable between the stiffness of tablets compared to the their tensile strength. 

Four different lubrication conditions were selected to compare the PSD effect on the tensile strength and elastic modulus of SDL tablets. According to Fig.~\ref{Fig:comparisons-SD}, unlike LM, particle size does not affect the tensile strength and elastic modulus of SDL tablets, taking into account different lubrication conditions.

All of the 42 cases (cf. Table \ref{Table:DOE}) were individually fitted to Eqs. (\ref{eq:Rossi}) \citep{rossi1968prediction} and ~(\ref{eq:leuenberger}) \citep{kuentz2000new}
\begin{equation}
\text{E} = \text{E}_{0}\left[1-\left(\frac{1}{1-{{\rho}_{c,\text{E}}}}\right)\left(1-\bar{\rho}\right)\right]
\label{eq:Rossi}
\end{equation}
\begin{equation}
\sigma_{t} = \sigma_{0}\left[1-\left(\frac{1-\bar{\rho}}{{1-\rho}_{c, \sigma_{t}}}\right)e^{(\bar{\rho}-\rho_{c, \sigma_{t}})}\right]
\label{eq:leuenberger}
\end{equation}
where $\text{E}_{0}$ and $\sigma_{0}$ are the elastic modulus and tensile strength at zero-porosity, respectively and ${{\rho}_{c,\text{E}}}$ and $\rho_{c, \sigma_{t}}$ are the relative density at which E and $\sigma_{t}$ go to zero, respectively. Table \ref{Table:TS-E-results} lists all the fitted parameters and $R^2$ values. A good fit was shown for all the cases ($R^2$ $>$ $0.93$), except for the elastic modulus fitting of cases $12$ and $16$, due to the implications of acquiring data from these tablets exhibiting low elasticity using the already selected settings on the ultrasound testing.

%%%%%%%%%%%%%%%%%%%%
\subsection {Quantitative model of lubrication and particle size distribution effects on tensile strength and stiffness of tablets}

Toward quantifying the PSD and lubricant sensitivity effects on mechanical properties of tablets, we introduce a parameter $C$ into the tensile strength and elastic modulus relationships with relative density (Eqs. (\ref{eq:Rossi}) and (\ref{eq:leuenberger})). For the sake of simplicity and generality, we assume a form for $C$ that captures the leading order term of the variables, that is
\begin{equation}
C = \frac{{c_{l}}^{x_{1}} ~{t_m}^{x_2} ~\mu^{x_3}}{x_4}
\label{Eq:2-parameter}
\end{equation}
where \{${x_{1}} $, ${x_{2}} $, ${x_{3}}$, ${x_{4}}$\} are the fitting coefficients. The coefficient $x_{4}$ serves as a scaling parameter that, in addition, makes $C$ dimensionless. $C$ will depend on the response variable (i.e., tensile strength and elastic modulus) and material properties. Thus, we define four distinct $C$ parameters, referred to as $C_{\text{E}, LM}$ and $C_{{\sigma}, LM}$ for lactose monohydrate and {red}$C_{\text{E}, SDL}$ and $C_{{\sigma}, SDL}$ for spray-dried lactose.

In order to introduce $C$ into Eqs.~(\ref{eq:Rossi}) and (\ref{eq:leuenberger}), we need to find a relationship between $C$ and $\text{E}_{0}$, $\sigma_{0}$, ${{\rho}_{c,\text{E}}}$, and $\rho_{c, \sigma_{t}}$. Figs.~\ref{Fig:properties-LM} and \ref{Fig:properties-SD} show that the data points converge as $\bar{\rho}$ decreases, in agreement with what authors observed in \citep{razavi2016toward}. Thus, to keep the optimization problem simpler and mathematically less complex, we assume constant ${{\rho}_{c,\text{E}}}$ and $\rho_{c, \sigma_{t}}$ for each powder, regardless of its PSD, lubricant concentration or/and mixing time. Thus, as a first order approximation only $\text{E}_{0}$ and $\sigma_{0}$ are considered as functions of $C$. We parameterize $\text{E}_{0}$ and ${\sigma}_{0}$ as follows
\begin{equation}
\text{E}_{0} = \frac{\text{E}_{0,\emptyset}-\text{E}_{0,\infty}}{1+C_{\text{E}}}+\text{E}_{0,\infty},~~~~ \text{where}\\ ~~
C_{\text{E}}=\frac{{c_{l}}^{b_{1}}~t_{m}^{b_{2}}~\mu^{b_{3}}}{b_{4}}
\label{Eq:E0-C}
\end{equation}
\begin{equation}
\sigma_{0} =\frac{\sigma_{0,\emptyset}-\sigma_{0,\infty}}{1+C_{{\sigma}}}+\sigma_{0,\infty},~~~~
\text{where} ~~C_{{\sigma}}=\frac{{c_{l}}^{d_{1}}~t_{m}^{d_{2}}~\mu^{d_{3}}}{d_{4}}
\label{Eq:sigma0-C}
\end{equation}
where \{${b_{1}} $, ${b_{2}} $, ${b_{3}}$, ${b_{4}} $\} and \{${d_{1}} $, ${d_{2}} $, ${d_{3}}$, ${d_{4}} $\} are the fitting parameters presented in function $C$ for $\text{E}_{0}$ and ${\sigma}_{0}$, respectively. ($\text{E}_{0,\emptyset}$ and $\text{E}_{0,\infty}$) and ($\sigma_{0,\emptyset}$ and $\sigma_{0,\infty}$) correspond to properties for when $C=0$ and $C=\infty$, respectively. Thus, $\text{E}_{0,\emptyset}$ and $\text{E}_{0,\infty}$ are the maximum and minimum values $\text{E}_{0}$ can obtain, which are determined by fitting the experimental data to the above equation. The same holds for $\sigma_{0}$. For most materials and lubricants, $\text{E}_{0,\infty}$ (or $\sigma_{0,\infty}$) will go to zero because the lubricants prevent the formation of solid bridges. However, it has been reported that materials, which experience significant fracture may develop some tablet strength even if they are fully lubricated \citep{kikuta1994effect}. Therefore, the minimum values for $\text{E}_{0}$ and  ${\sigma}_{0}$ does not need to be zero for  ``infinite" lubrication. It is worth to emphasize that the functionality of $\text{E}_{0}$ and ${\sigma}_{0}$ is in fact unknown. In this study, we parameterized these two functions only in the interest of simplicity and generality. Other functionalities may be explored, which falls outside the scope of this paper.

$\text{E}_{0}$ and $\sigma_{0}$ were predicted from experimental results by solving the following general optimization problems
\begin{linenomath*}
\begin{eqnarray*}
\min_{\scriptscriptstyle{\{{b_{1}},...,\text{E}_{0,\infty},\rho_{c, {\text{E}}}\}}}
\left[ 
\sum_{i \in \mathcal{P}} 
\left(\text{E}(\bar{\rho_i})-{\text{E}_{0}}_{\scriptscriptstyle{\{{b_{1}},...,{b_{4}},\text{E}_{0,\emptyset},\text{E}_{0,\infty}\}}}\left(1-\frac{1-\bar{\rho_i}}{1-\rho_{c, {\text{E}}}}\right)\right)^2 \right]^{1/2}
\end{eqnarray*}
\end{linenomath*}
\begin{linenomath*}
\begin{eqnarray*}
\min_{\scriptscriptstyle{\{{d_{1}},...,\sigma_{0,\infty},\rho_{c, {\sigma_{t}}}\}}}
\left[ 
\sum_{i \in \mathcal{Q}} 
\left( \sigma_{t}(\bar{\rho_i})-{\sigma_{0}}_{\scriptscriptstyle{\{{d_{1}},...,{d_{4}},\sigma_{0,\emptyset},\sigma_{0,\infty}\}}}\left[1-\left(\frac{1-\bar{\rho_i}}{1-\rho_{c, {\sigma_{t}}}}\right)e^{(\bar{\rho{i}}-\rho_{c, \sigma_{t}})}\right]\right)^2 
\right]^{1/2}
\end{eqnarray*}
\end{linenomath*}
where $\mathcal{P}$ and $\mathcal{Q}$ are a set of experimental points obtained from ultrasound and diametrical compression tests, respectively. In this study, $\mathcal{P}$ and $\mathcal{Q}$ consisted of $316$ points for SDL and $204$ points for LM, respectively. It is noted that case $16$ (see Table \ref{Table:TS-E-results}) was removed from the optimization. Case $5$ from LM dataset and cases $24$ and $33$ from SDL dataset were randomly taken out and adopted later to validate the model.

The solution to the optimization problems forced $\text{E}_{0,\infty}$ and $\sigma_{0,\infty}$ to go to zero for both materials. Thus, Eqs. (\ref{Eq:E0-C}) and (\ref{Eq:sigma0-C}) were reduced to
\begin{equation*}
\text{E}_{0} = \frac{\text{E}_{0,\emptyset}}{1+\frac{{c_{l}}^{b_{1}}~t_{m}^{b_{2}}~\mu^{b_{3}}}{b_{4}}}
\label{Eq:E0-C-final}
\end{equation*}
\begin{equation*}
\sigma_{0} = \frac{\sigma_{0,\emptyset}}{1+\frac{{c_{l}}^{d_{1}}~t_{m}^{d_{2}}~\mu^{d_{3}}}{d_{4}}}
\label{Eq:sigma0-C-final}
\end{equation*}

The optimal values together with the residual errors for the optimization problems are in Tables~\ref{Table:LM-C} and~\ref{Table:SDL-C}. For LM, $b_{3}$ and $d_{3}$ values indicate that changes in PSD result in more drastic changes in elastic modulus and tensile strength of tablets compared to the lubricant concentration and mixing time. On the other hand, $c_{l}$ seems to be the most influential variable on the mechanical properties of SDL (see, ${b_{1}}$ and ${d_{1}}$ values in Tables~\ref{Table:LM-C} and~\ref{Table:SDL-C}). $\text{E}_{0,\emptyset}$ of SDL resulted in an unrealistic prediction. Hence, we attempted to produce tablets of SDL with no lubrication, but the compaction was not successful due to extremely high frictional and ejection forces. Caution must be taken in interpreting $\text{E}_{0,\emptyset}$ and $\sigma_{0,\emptyset}$, since the proposed model does not consider other physical mechanisims that prevent the formation of a tablet.

In summary, our proposed strategy shows that elastic modulus (or, tensile strength) is inversely proportional to a non-linear function of material and blending properties and can be presented as follows
\begin{equation}
{\text{E}}=\frac{\text{E}_{0,\emptyset}}{1+\frac{{c_{l}}^{b_{1}}~t_{m}^{b_{2}}~\mu^{b_{3}}}{b_{4}}}\left(1-\frac{1-\bar{\rho}}{1-\rho_{c, {\text{E}}}}\right)
\end{equation}
\begin{equation}
{\sigma_{t}}=\frac{\sigma_{0,\emptyset}}{1+\frac{{c_{l}}^{d_{1}}~t_{m}^{d_{2}}~\mu^{d_{3}}}{d_{4}}}\left[1-\left(\frac{1-\bar{\rho}}{1-\rho_{c, {\sigma_{t}}}}\right)e^{(\bar{\rho}-\rho_{c, \sigma_{t}})}\right]
\end{equation}

The groundwork of our model was to relate the variables that contribute to parameter $C$, to $\text{E}_{0}$ and ${\sigma}_{0}$. Figs.~\ref{Fig:E0-sigma0-LM} and \ref{Fig:E0-sigma0-SDL} demonstrate the functionality of the proposed relationship. The model fitted the data well for both materials. The validation points were in good agreement with the predicted curves. It should be noted that our optimization problems were constructed to minimize the sum of squared residuals between the experimental and predicted values of elastic modulus and tensile strength of tablets. Thereby, the fitting coefficients presented in Tables \ref{Table:LM-C} and \ref{Table:SDL-C} are not the optimal values for predicting $\text{E}_{0}$ and ${\sigma}_{0}$. 

Figs.~\ref{Fig:LM-val} and \ref{Fig:SDL-val} compare the validation measurements with model predictions for elastic modulus and tensile strength of LM and SDL tablets, respectively (case $5$ for LM and cases $24$ and $33$ for SDL). The agreement between the validations and model predictions are very promising.

Figs.~\ref{Fig:E-E} and \ref{Fig:sigma-sigma} show the relationship between actual (measured) and predicted elastic modulus and tensile strength for all the LM and SDL tablets. Good correlations were observed between the predicted and actual values. For elastic modulus predictions, $R^2$ of $0.94$ for both powders was found and for tensile strength predictions $R^2$ was $0.96$ and $0.91$ for SDL and LM, respectively.

Overall, our proposed model can be successfully adopted to predict the mechanical strength of tablets capturing the lubricant sensitivity and PSD effects. Contour plots of elastic modulus and tensile strength as a function of $C$ and relative density for the materials studied are presented in Figs.~\ref{Fig:contour-LM} and \ref{Fig:contour-SDL}. Establishing such plots assist to scan through the design space and systematically optimize the formulation and process. The model can be expanded to include other blend properties or processing parameters effects. 

\section{Summary and conclusion}

We have proposed a general framework for predicting a wide range of elastic modulus and tensile strength that a specific powder can attain based on its lubricant sensitivity and considering different particle size distributions. This was possible by introducing a new dimensionless parameter in the existing tensile strength and elastic modulus models of porous materials. Specifically, we propose that the elastic modulus and tensile strength at zero-porosity are a function of MgSt concentration, mixing time and mean PSD, while the relative densities at zero tablet stiffness and strength are not. The model showed good predictability for two grades of lactose, namely monohydrate and spray-dried grades. Possible avenues for extension of the proposed model is studying the applicability to powders that undergo different deformation mechanisms, the generalization to ternary or more complex mixtures, and the integration with optimal relationships for tensile strength of doubly convex tablets (see, e.g., \cite{razavi2015}). Establishing such predictive models helps drug formulators and manufacturers to optimize lubricant concentration and mixing conditions according to the desired mechanical strength and stiffness of tablets. 

We have covered a wide, if not the widest reported in the literature, strength level of lactose monohydrate and spray-dried lactose tablets with respect to particle size distributions and lubrication conditions. Fig. \ref{Fig:E-TS-total} depicts the tensile strength as a function of elastic modulus of all the lactose monohydrate and spray-dried lactose tablets tested in this study. The data points show a pattern with identifiable zones which provide the achievable design space, i.e., an elliptic space for lactose monohydrate and a trapezoidal space for spray-dried lactose. For a certain formulation, mapping the achievable design space allows for optimizing the processing variables for the desirable table mechanical properties.

We close by discussing opportunities and future work. It is clear that the proposed nonlinear functions for the elastic modulus and tensile strength at zero-porosity are not the only functions of lubrication and particle size parameters that exhibit the experimentally observed behavior. The systematic investigation of these functions is worthwhile directions of future research. Furthermore, the elusidation of the mechanistic basis of these relationships and how lubrication conditions affect particle properties and solid brigde formation during compaction are desirable for fundamentaly understanding the achievable design space map. Thus, particle mechanics simulations capable of describing strength formation and evolution during the compaction process are desirable \cite{gonzalez2012nonlocal,gonzalez2016microstructure,yohannes2016evolution,yohannes2017discrete,gonzalez2017generalized}, if beyond the scope of this paper.

%%%%%%%%%%%%%%%%%%%%%%
\section*{Acknowledgements}
%%%%%%%%%%%%%%%%%%%%%%

The authors gratefully acknowledge the support received from the NSF ERC grant number EEC-0540855, ERC for Structured Organic Particulate Systems. M.G. also acknowledges support from the NSF under grant number CMMI-1538861.

%%%%%%%%
\bibliographystyle{apalike}
\bibliography{PSD}
%%%%%%%%

\pagebreak

\pagebreak

\pagebreak

\pagebreak

\pagebreak

\pagebreak

{\renewcommand{\thefigure}{(7 Cont.)}
\addtocounter{figure}{-1}
\begin{figure}[H]
\centering
\begin{tabular}{cc}
\renewcommand{\thesubfigure}{(g)}
\subfigure[]
{
\hspace{-1.1cm}
\includegraphics[scale=0.27]{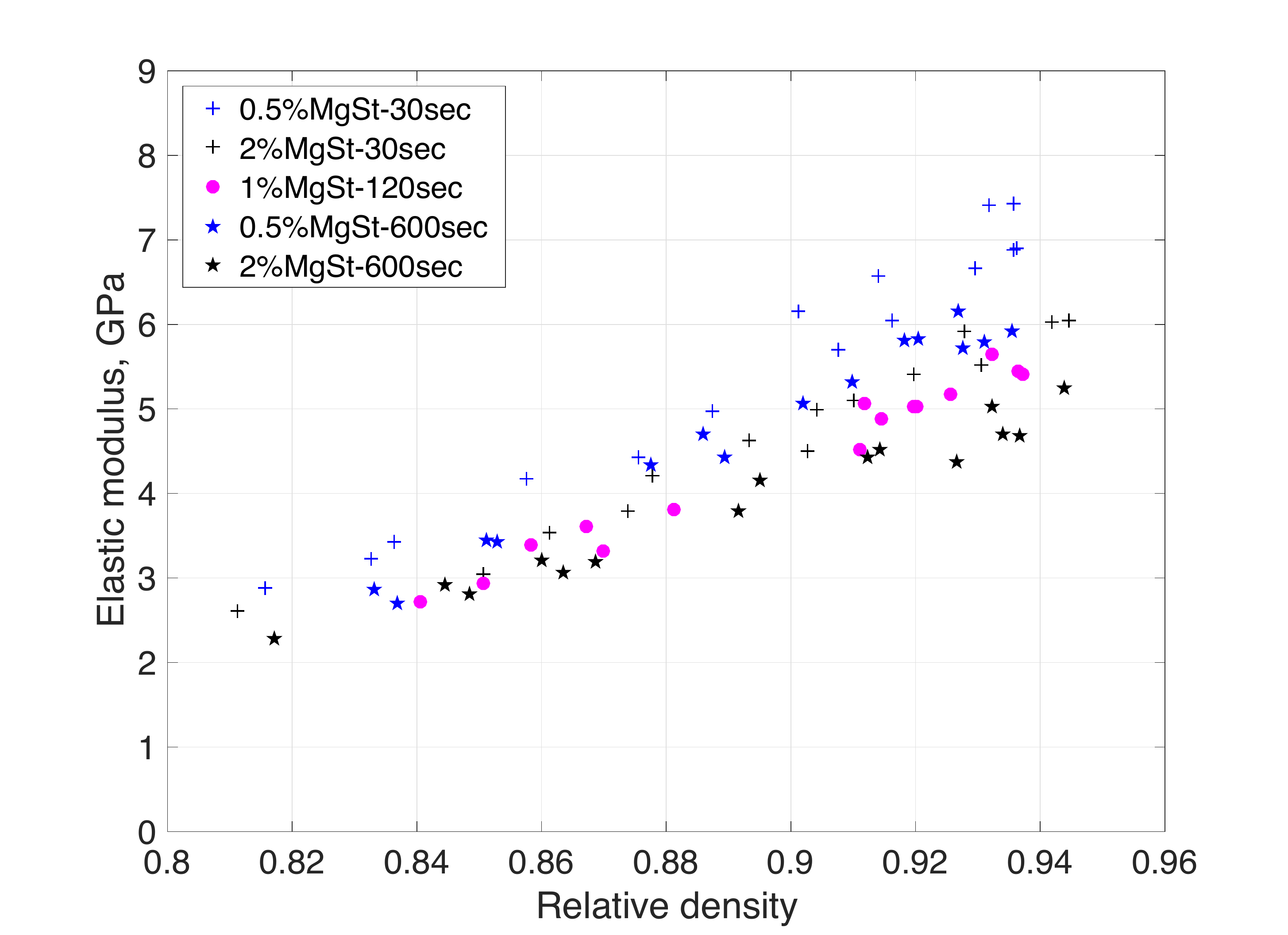}
\label{Fig:E-150-212-SD}
}
\renewcommand{\thesubfigure}{(h)}
\subfigure[]
{
\hspace{-1.1cm}
\includegraphics[scale=0.27]{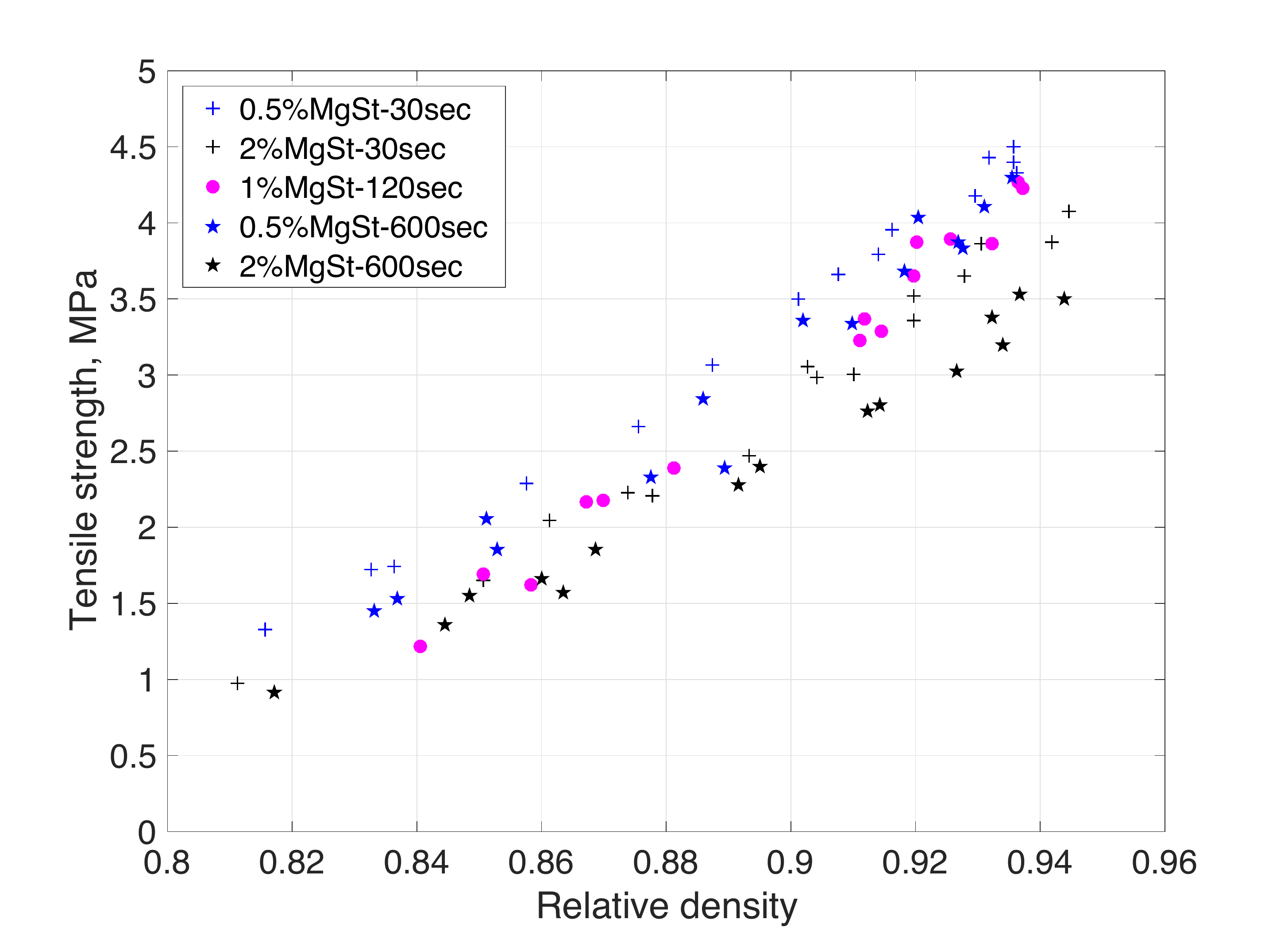}
\label{Fig:TS-150-212-SD}
}
\\
\renewcommand{\thesubfigure}{(i)}
\subfigure[]
{
\hspace{-1.1cm}
\includegraphics[scale=0.27]{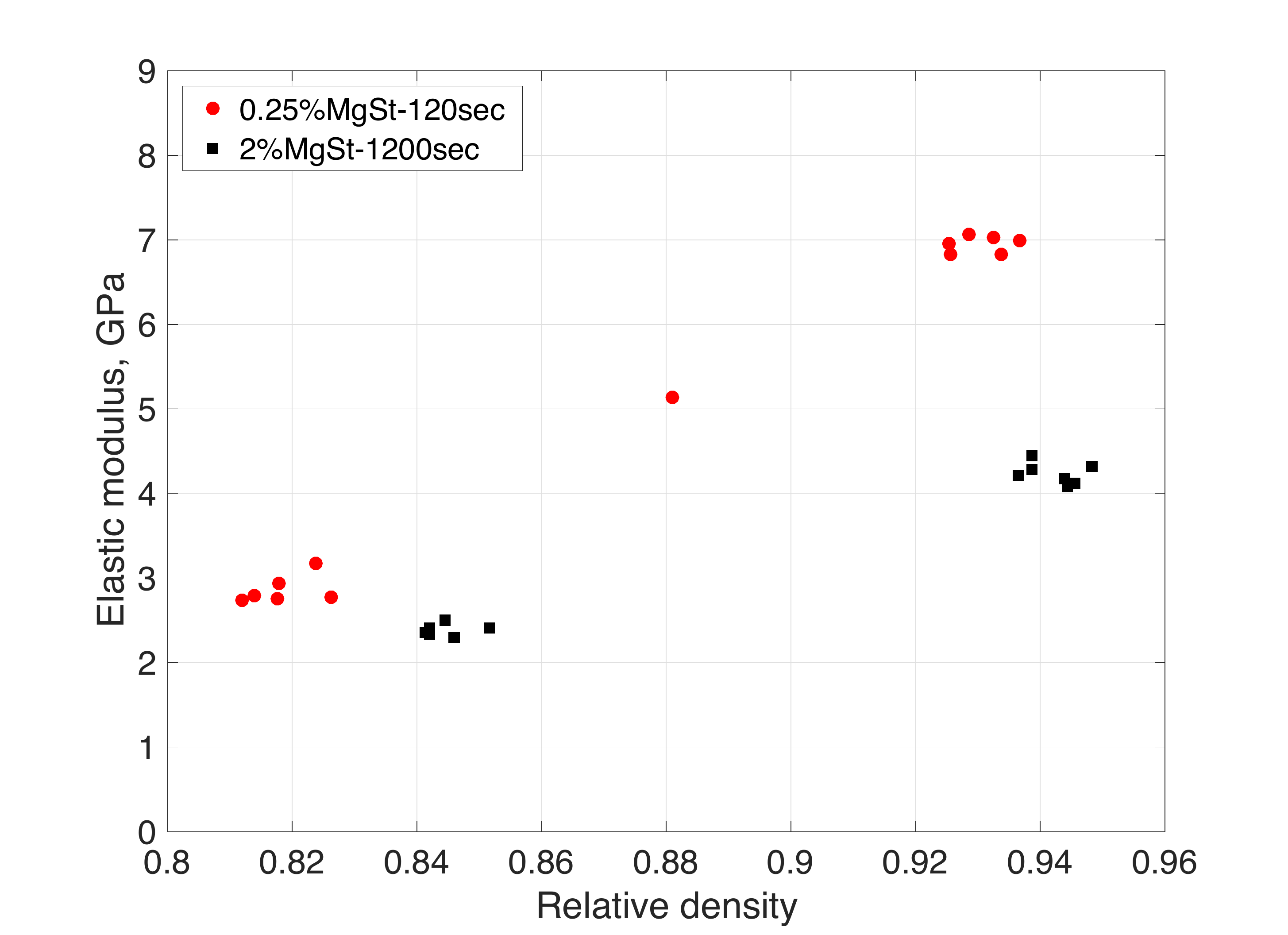}
\label{Fig:E-AR-SD}
}
\renewcommand{\thesubfigure}{(j)}
\subfigure[]
{
\hspace{-1.1cm}
\includegraphics[scale=0.27]{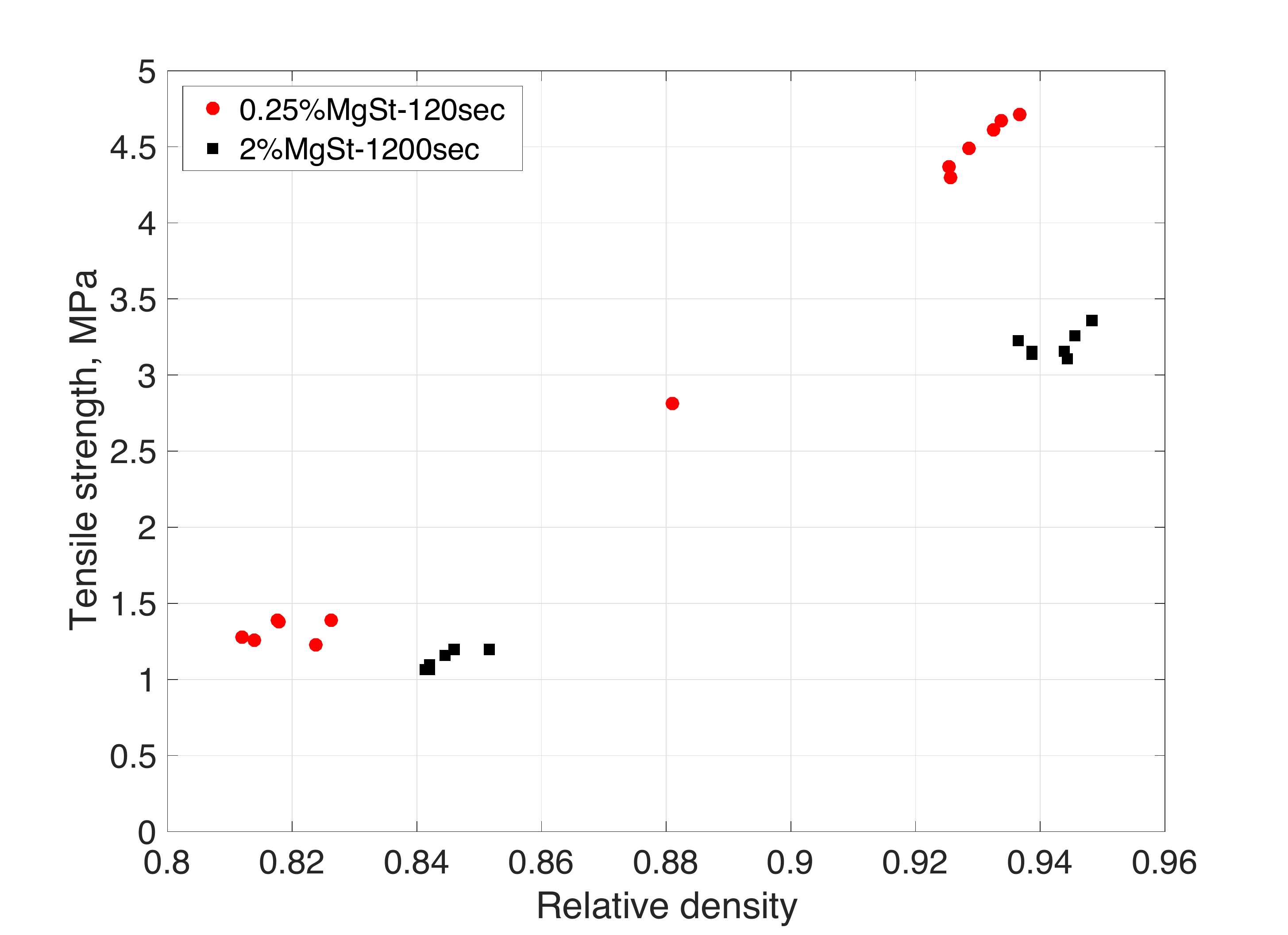}
\label{Fig:TS-AR-SD}
}
\end{tabular} 
\caption[]{Elastic modulus and tensile strength vs. out-of-die relative density for different particle size distributions of spray-dried lactose; (a,b) 0-75 $\mu$m, (c,d) 75-106 $\mu$m, (e,f) 106-150 $\mu$m, (g,h) 150-212 $\mu$m, and (i,j) as-received.}
\end{figure}}

\begin{figure}[H]
\centering
\begin{tabular}{c c}
\vspace{-0.1cm}
\subfigure[]
{
\hspace{-1cm}
\includegraphics[scale=0.23]{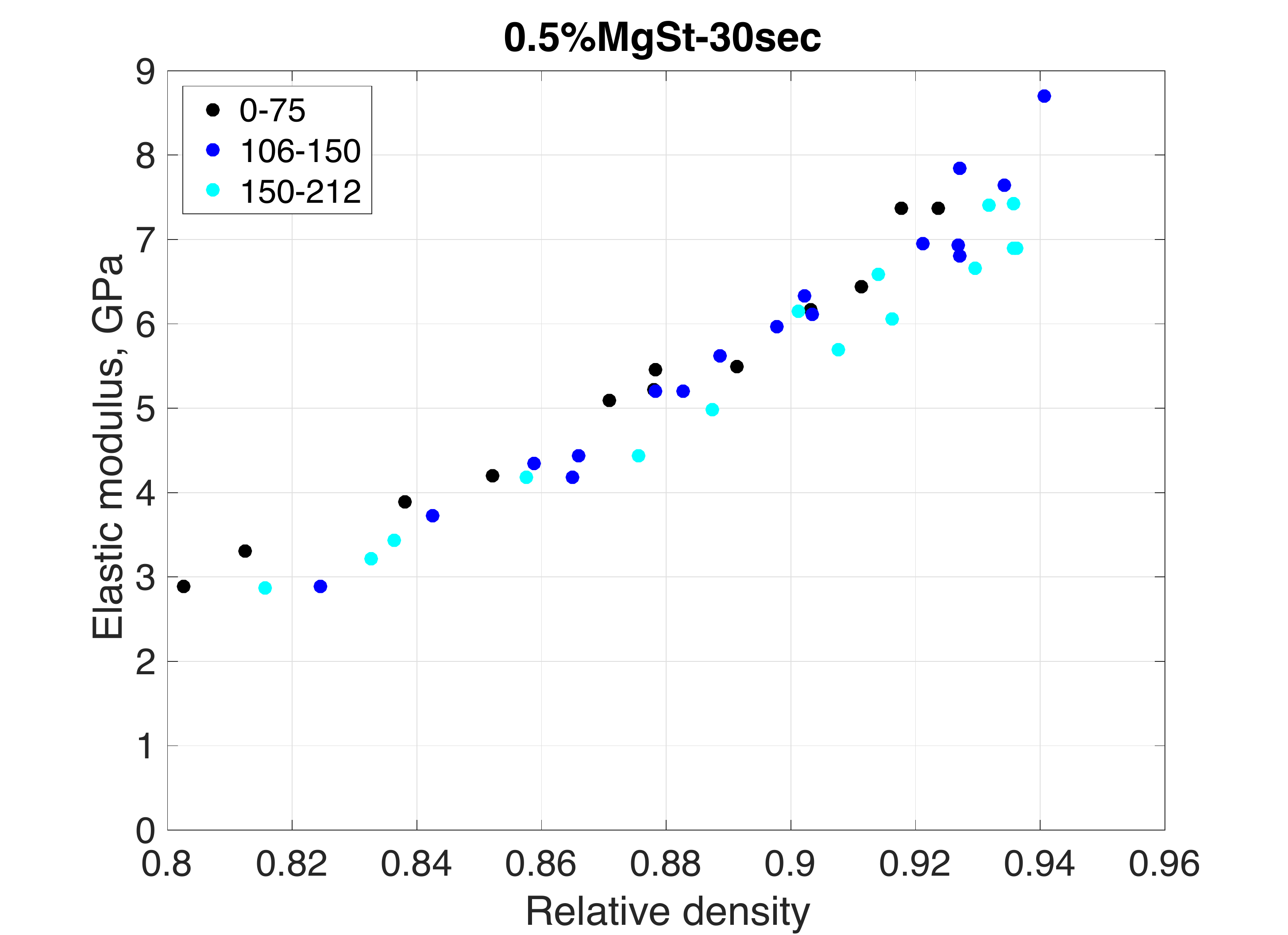}
}
\subfigure[]
{
\hspace{-1cm}
\includegraphics[scale=0.23]{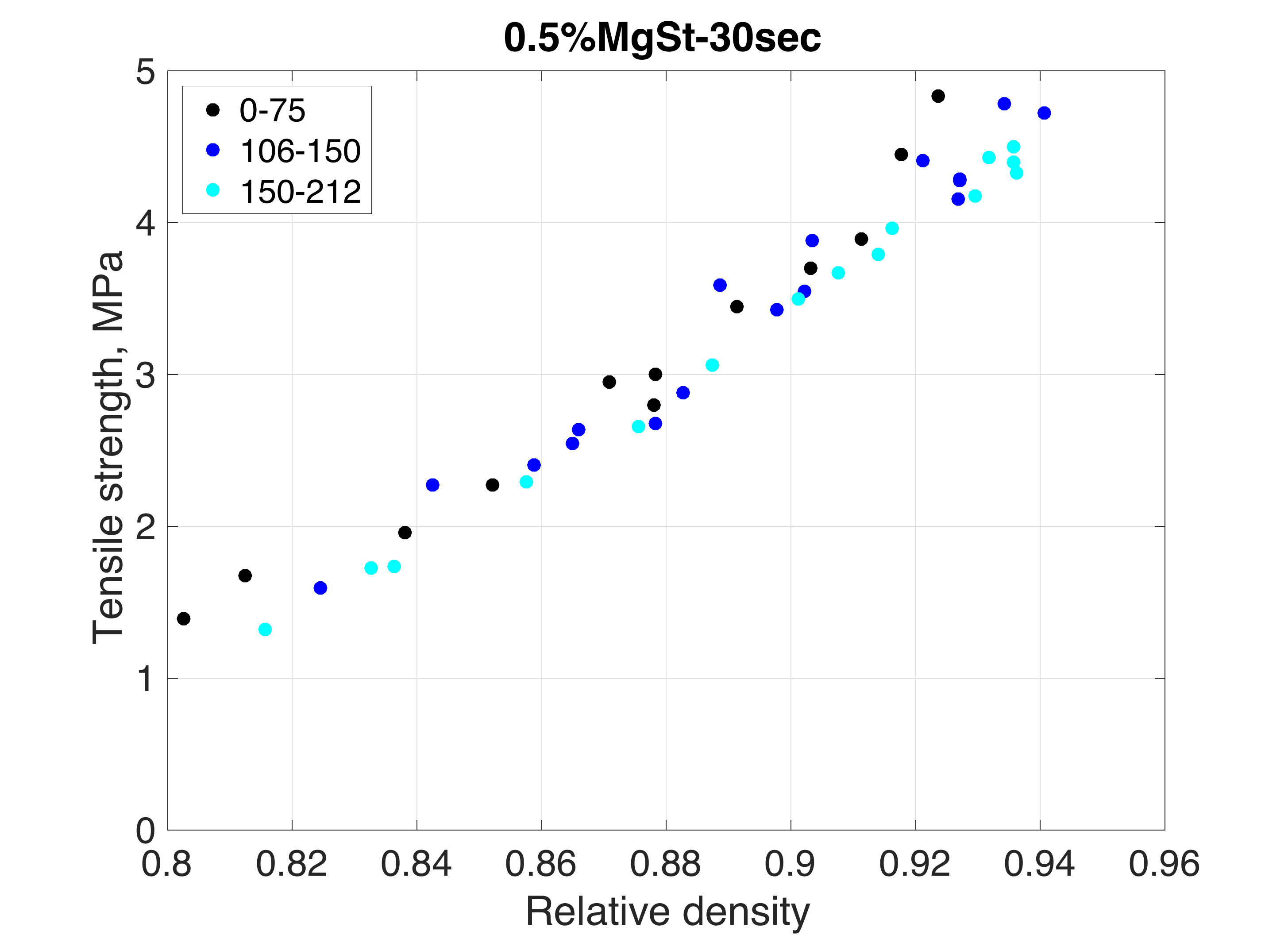}
}
\\
\vspace{-0.1cm}
\subfigure[]
{
\hspace{-1cm}
\includegraphics[scale=0.23]{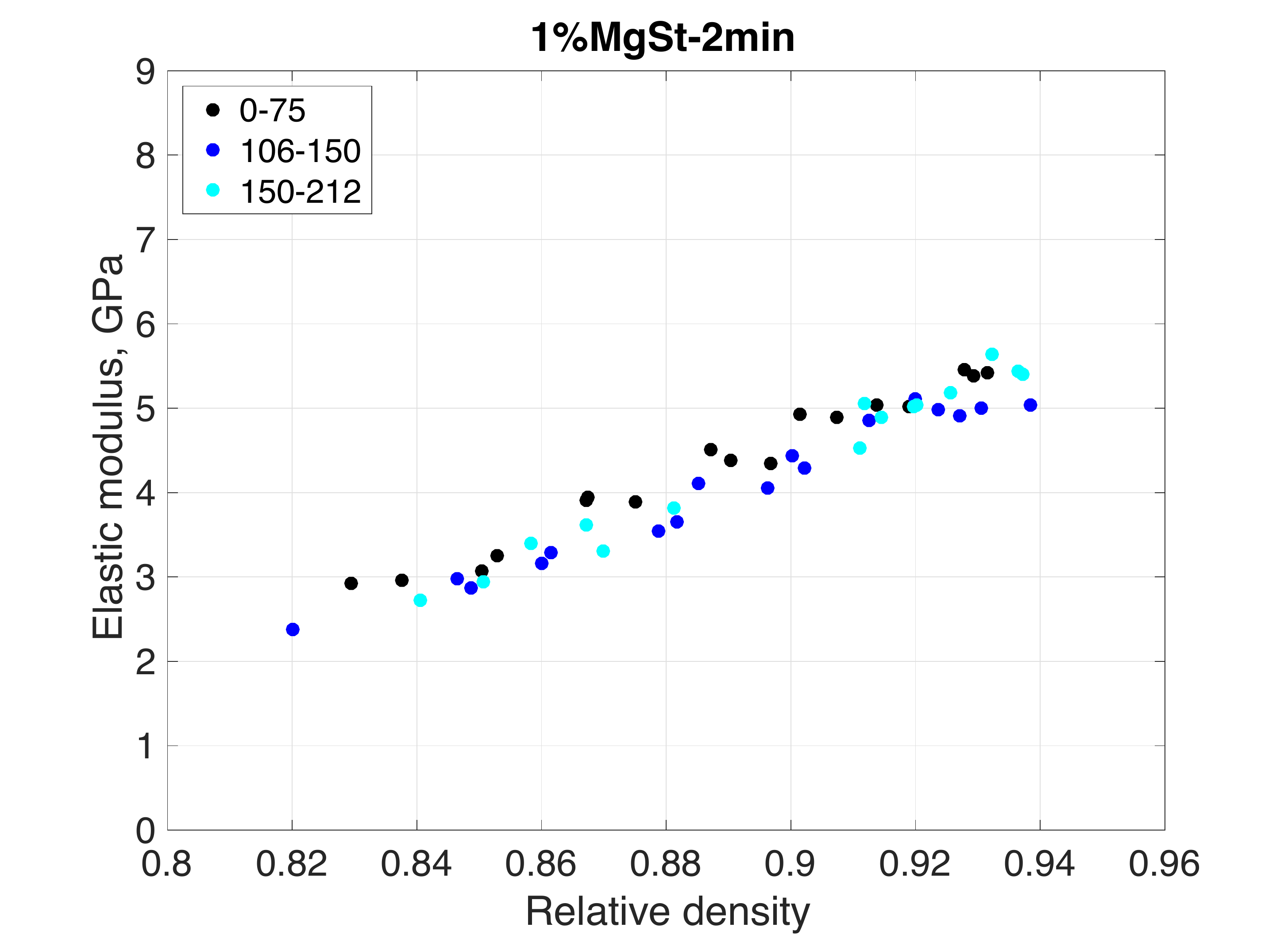}
}
\subfigure[]
{
\hspace{-1cm}
\includegraphics[scale=0.23]{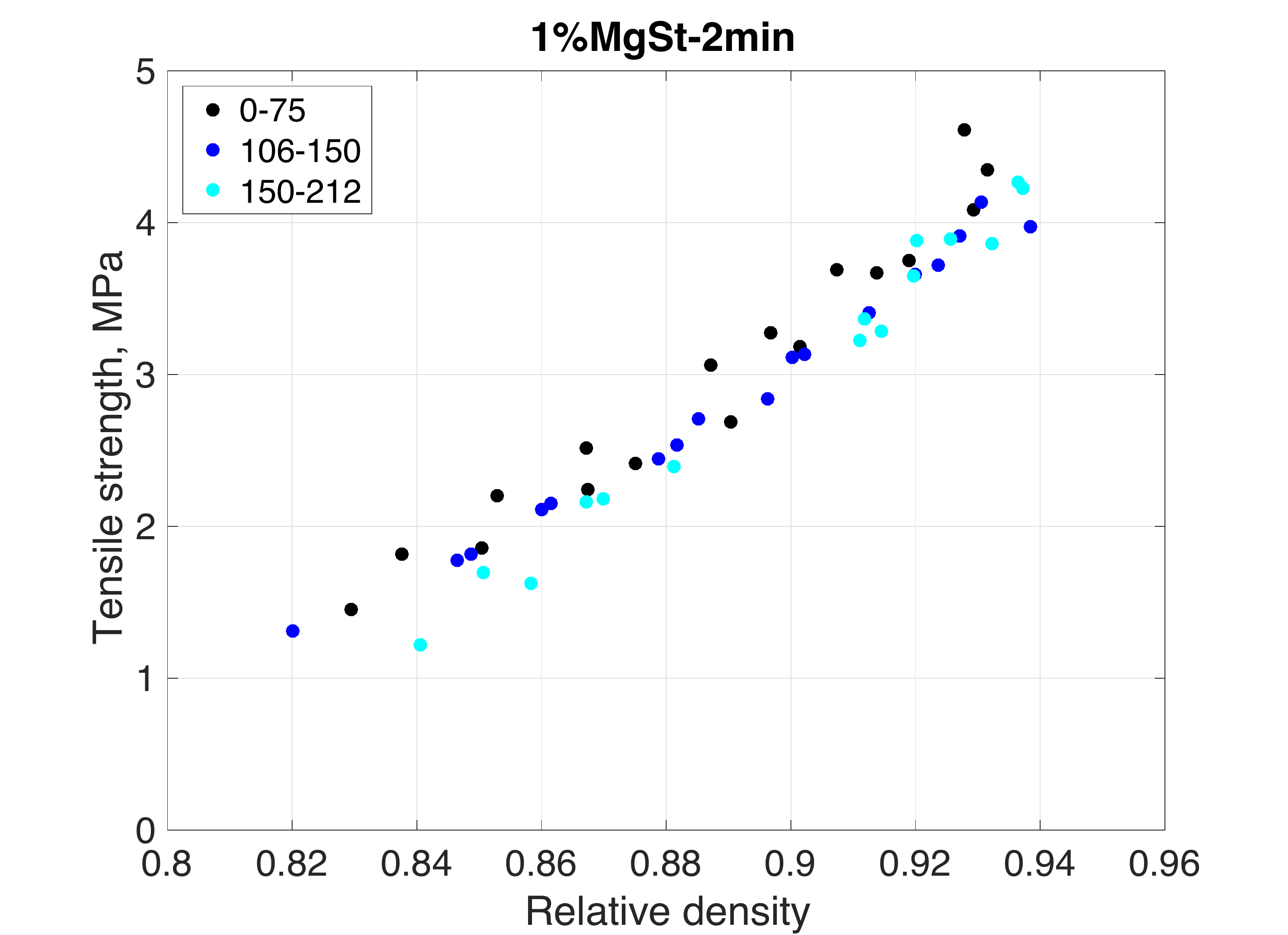}
}
\\
\vspace{-0.1cm}
\subfigure[]
{
\hspace{-1cm}
\includegraphics[scale=0.23]{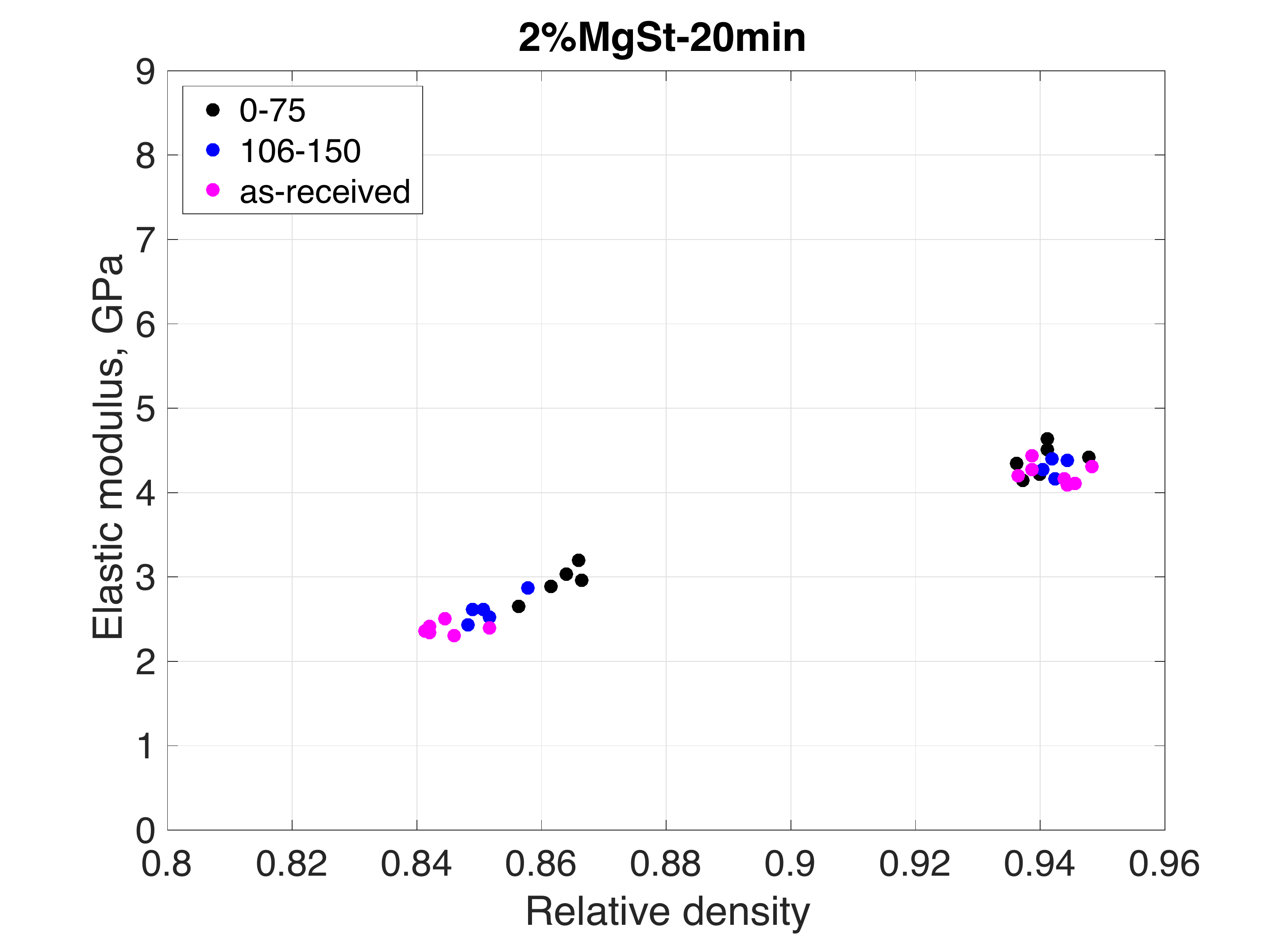}
}
\subfigure[]
{
\hspace{-1cm}
\includegraphics[scale=0.23]{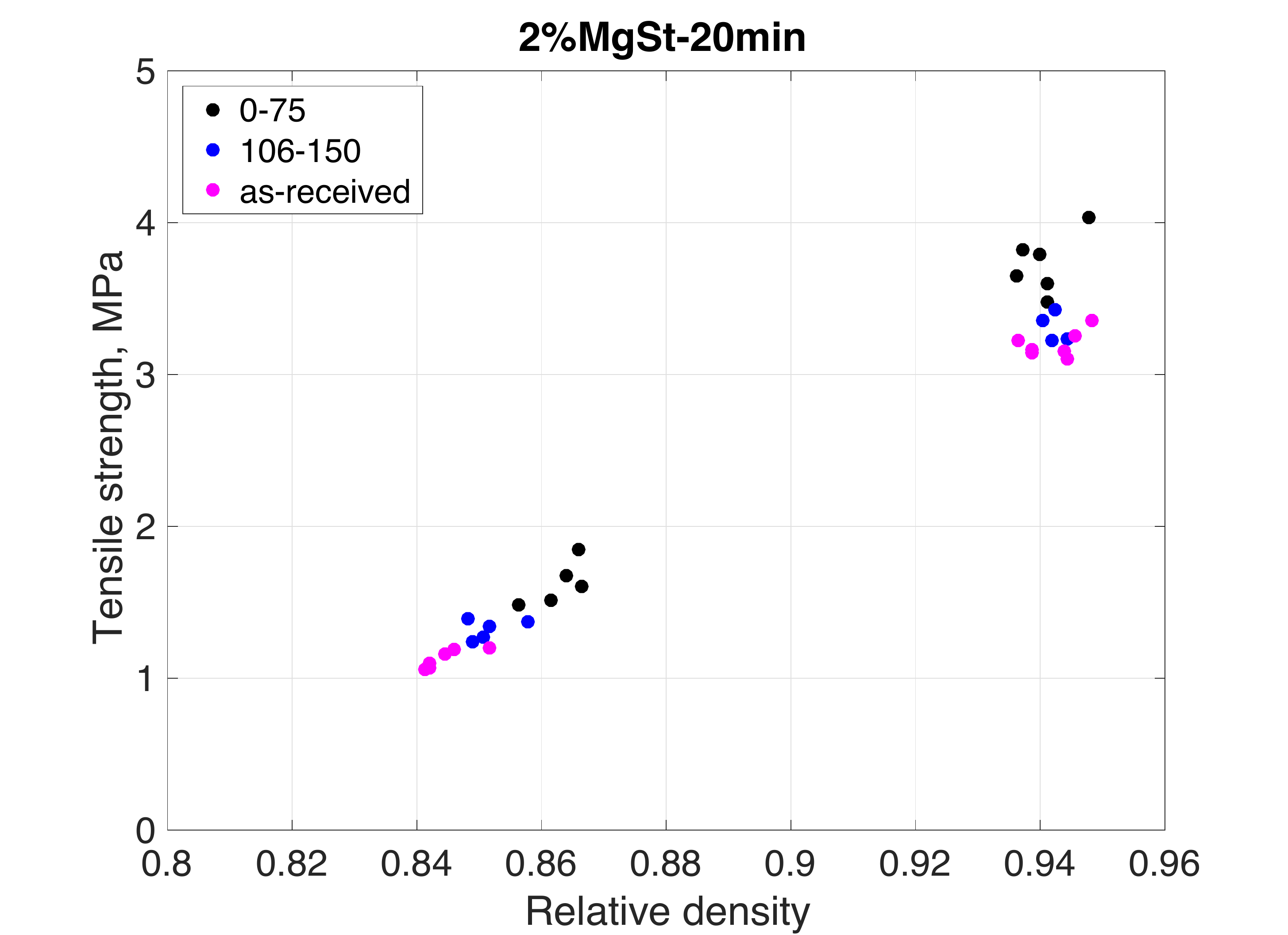}
} 
\end{tabular} 
\vspace{-0.1cm}
\caption{Particle size effect on elastic modulus and tensile strength of spray-dried lactose tablets at different lubrication conditions (a, b) 0.5\%MgSt-30sec, (c, d) 1\%MgSt-2min, and (e, f)  2\%MgSt-20min.} 
\label{Fig:comparisons-SD}
\end{figure}

\pagebreak

\begin{figure}[H]
\centering
\begin{tabular}{cc}
\subfigure[]
{
\includegraphics[scale=0.41]{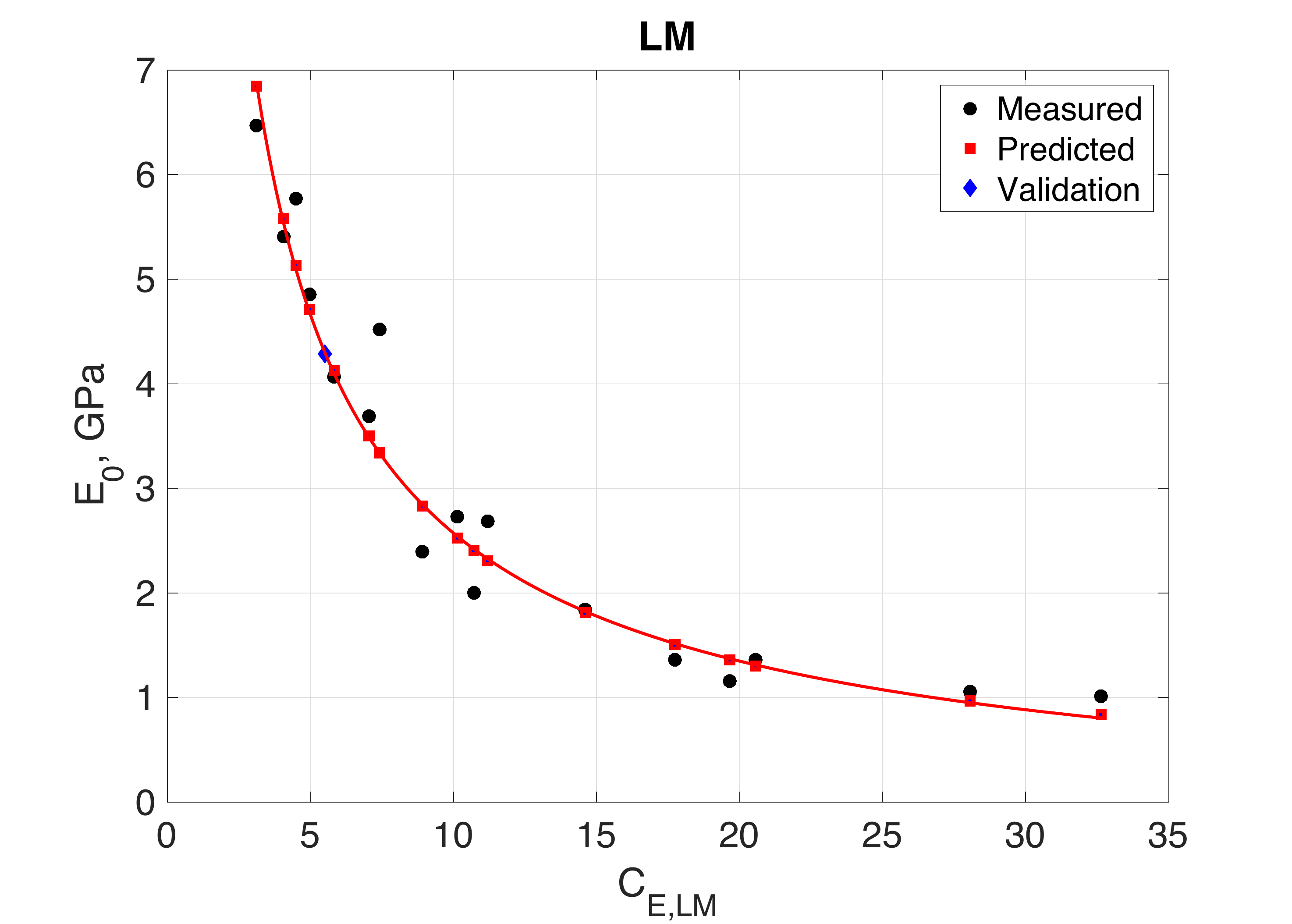}
\label{Fig:E-E}
}
\\
\subfigure[]
{
\includegraphics[scale=0.41]{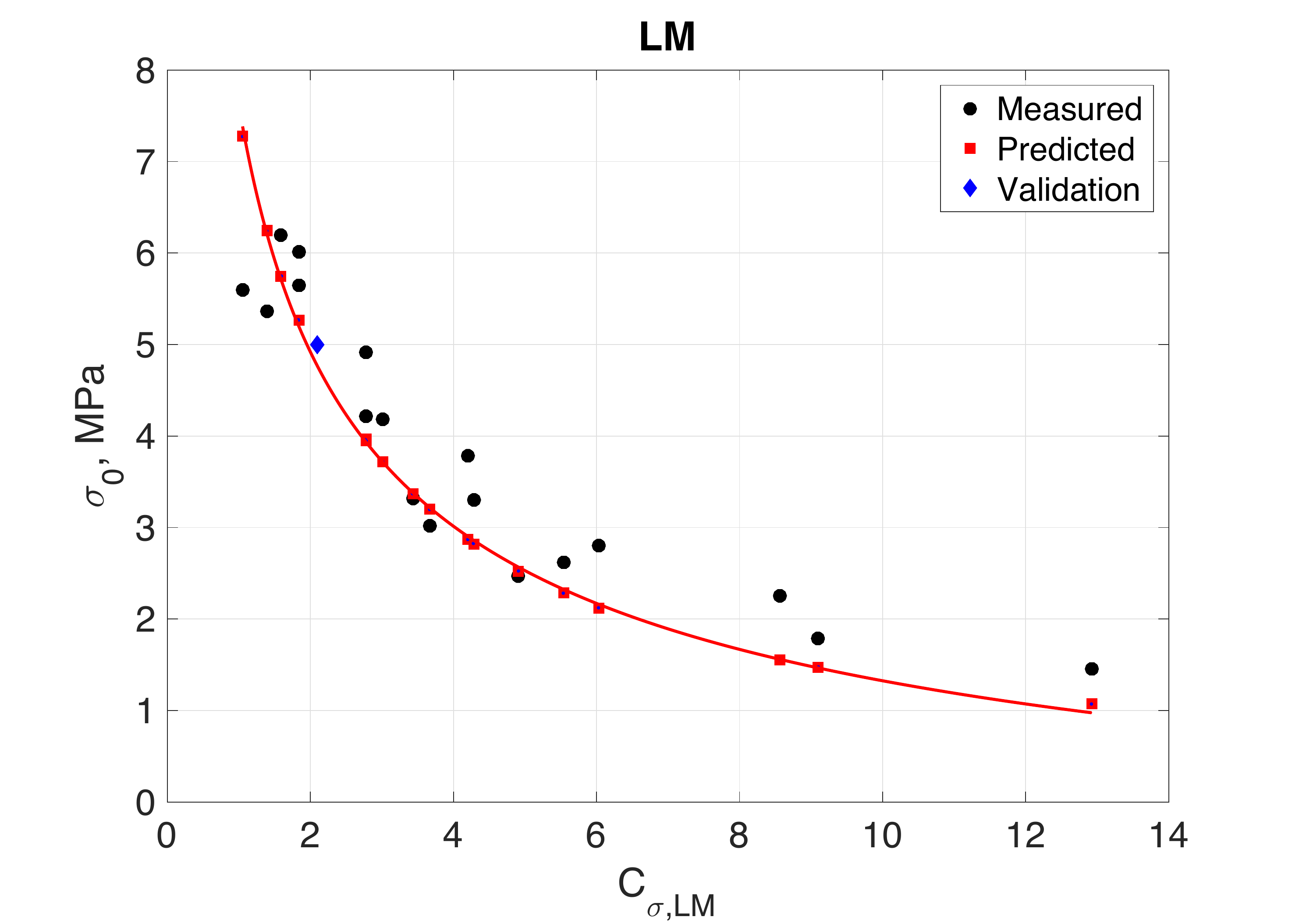}
\label{Fig:sigma-sigma}
}
\end{tabular} 
\caption{The measured and predicted values of $\text{E}_{0}$ and ${\sigma}_{0}$ as a function of parameter ``$C$" for lactose monohydrate. A validation point, in blue, is also provided.} 
\label{Fig:E0-sigma0-LM}
\end{figure}

\begin{figure}[H]
\centering
\begin{tabular}{cc}
\subfigure[]
{
\includegraphics[scale=0.41]{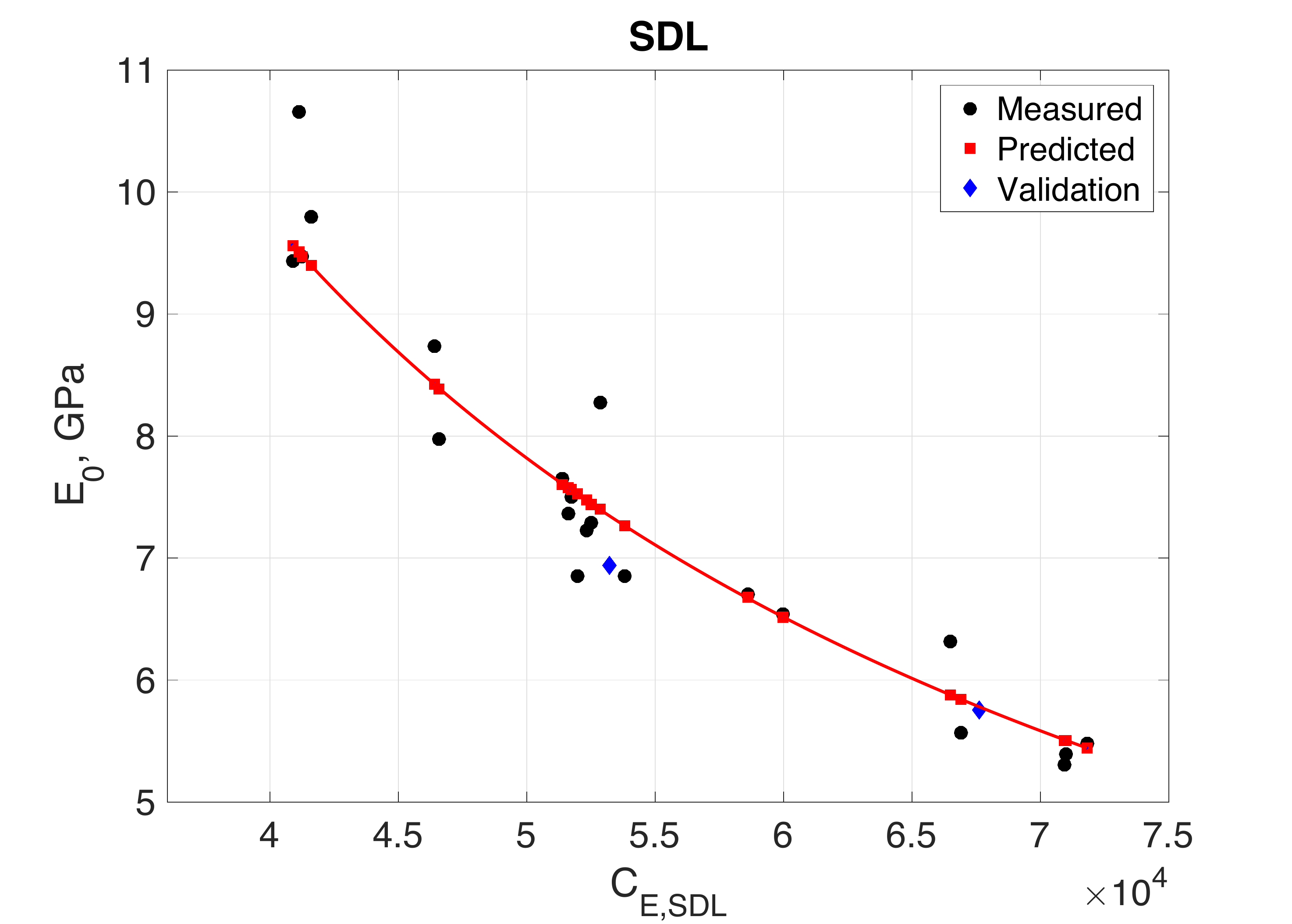}
\label{Fig:E0-C-SDL}
}
\\
\subfigure[]
{
\includegraphics[scale=0.41]{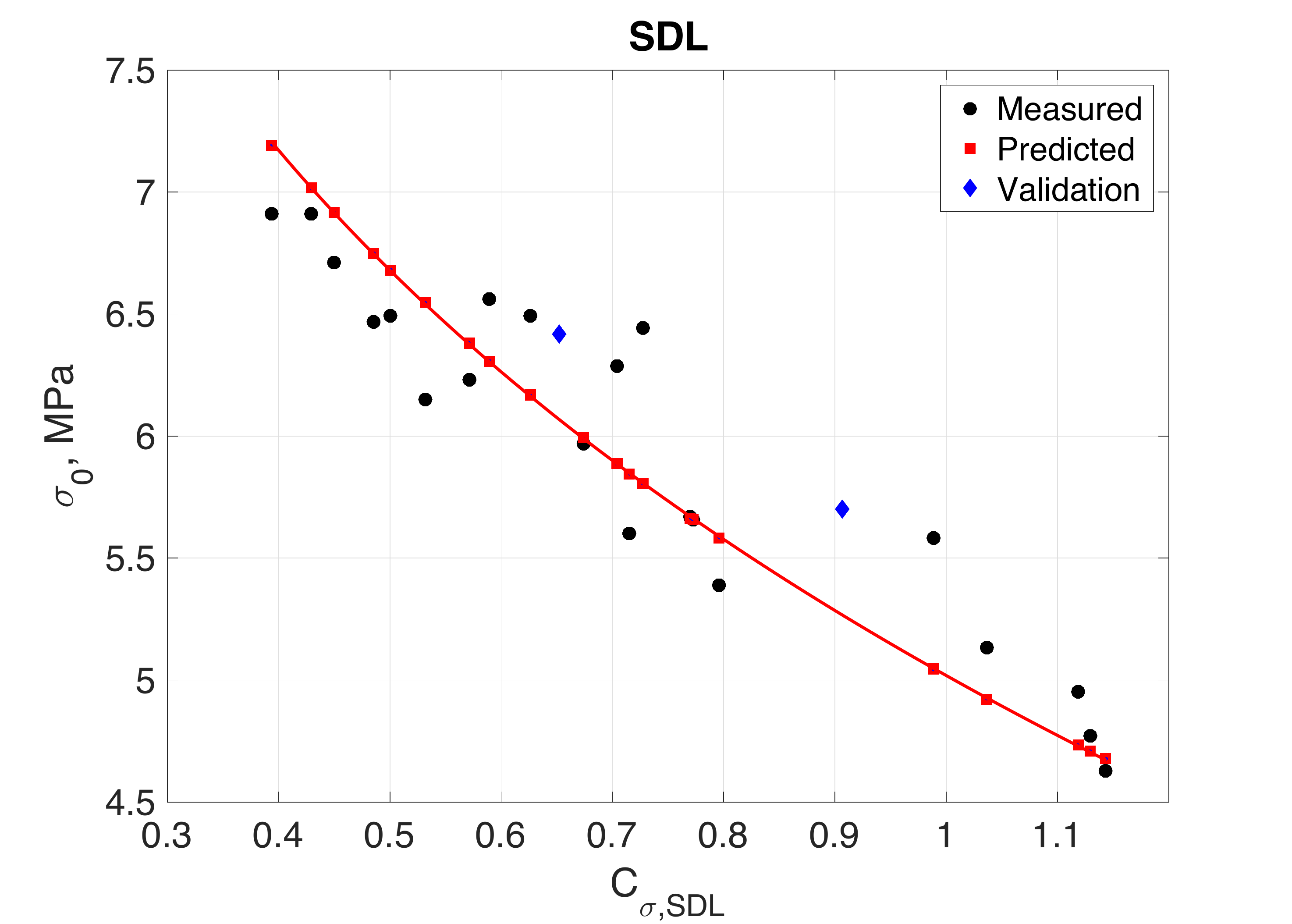}
\label{Fig:sigma0-C-SDL}
}
\end{tabular} 
\caption{The measured and predicted values of $\text{E}_{0}$ and ${\sigma}_{0}$ as a function of parameter ``$C$" for spray-dried lactose. Two validation points, in blue, are provided.} 
\label{Fig:E0-sigma0-SDL}
\end{figure}

\pagebreak

\begin{figure}[H]
\centering
\begin{tabular}{cc}
\subfigure[]
{\hspace{-0.9cm}
\includegraphics[scale=0.27]{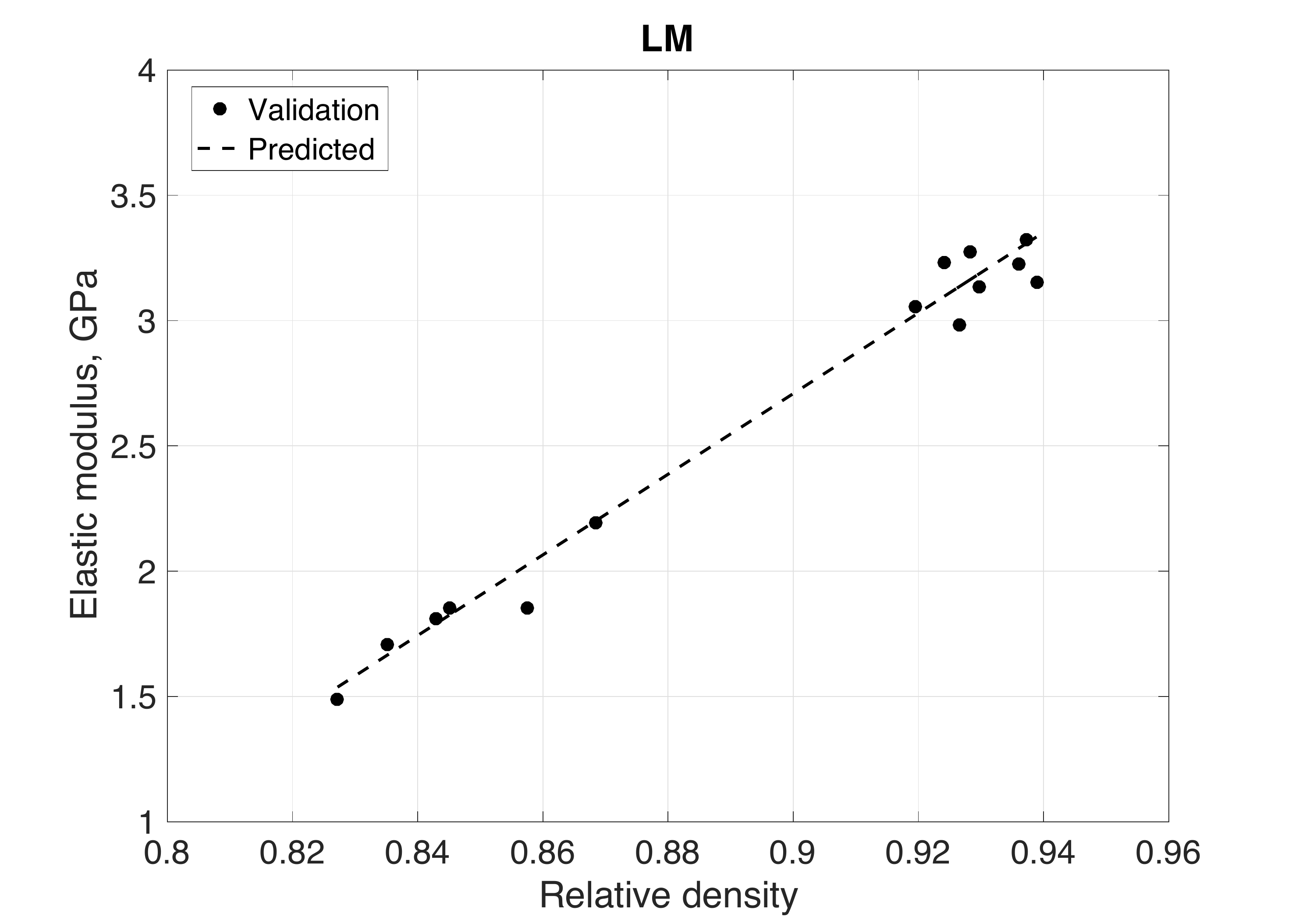}
\label{Fig:E-LM-val}
}
\subfigure[]
{\hspace{-0.9cm}
\includegraphics[scale=0.27]{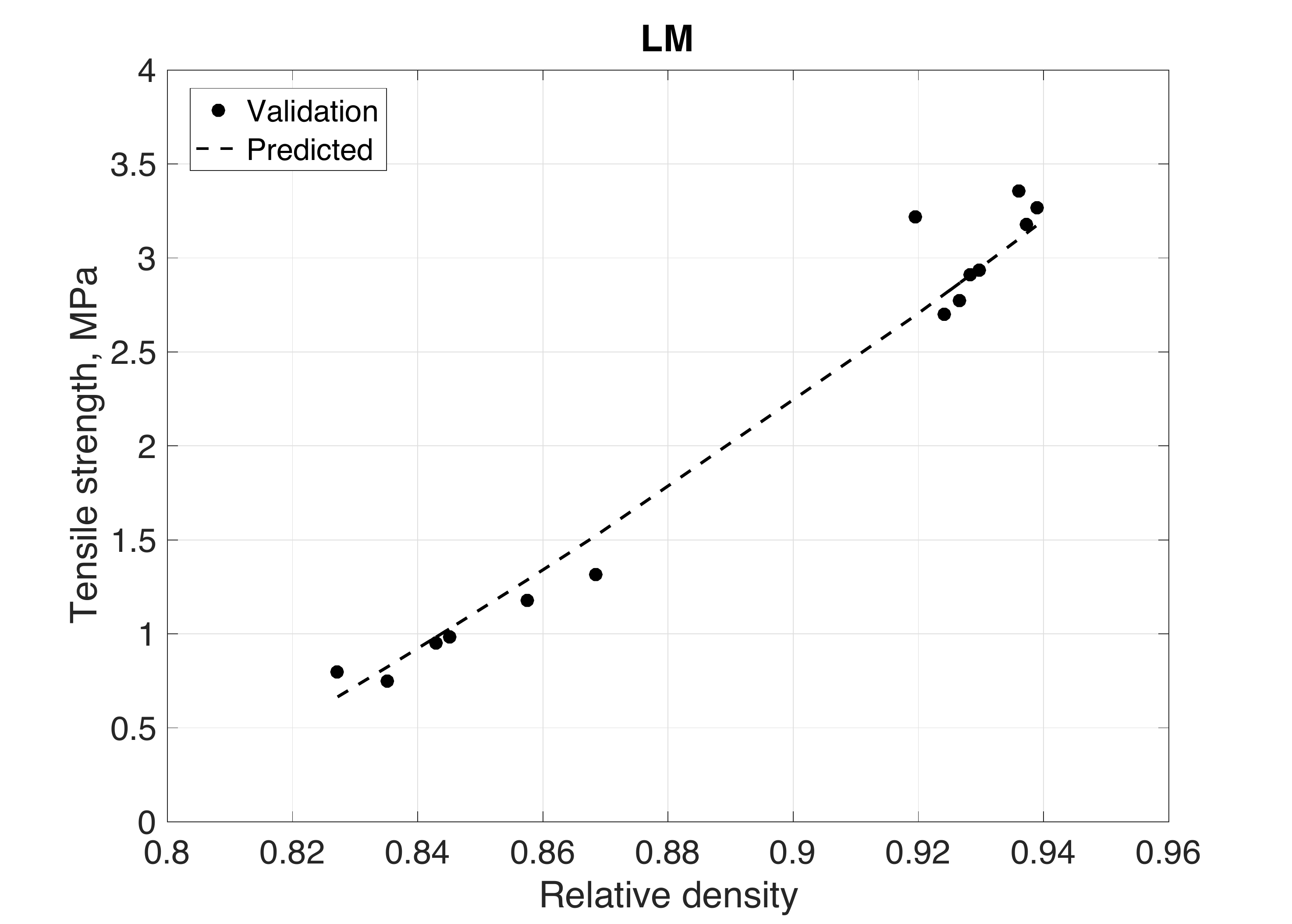}
\label{Fig:sigma-LM-val}
}
\end{tabular} 
\caption{Comparsion of the validation experiments to the model prediction for (a) elastic modulus and (b) tensile strength of lactose monohydrate tablets as a function of relative density (case $5$ in Table~\ref{Table:DOE}).} 
\label{Fig:LM-val}
\end{figure}

\begin{figure}[H]
\centering
\begin{tabular}{cc}
\subfigure[]
{\hspace{-0.9cm}
\includegraphics[scale=0.27]{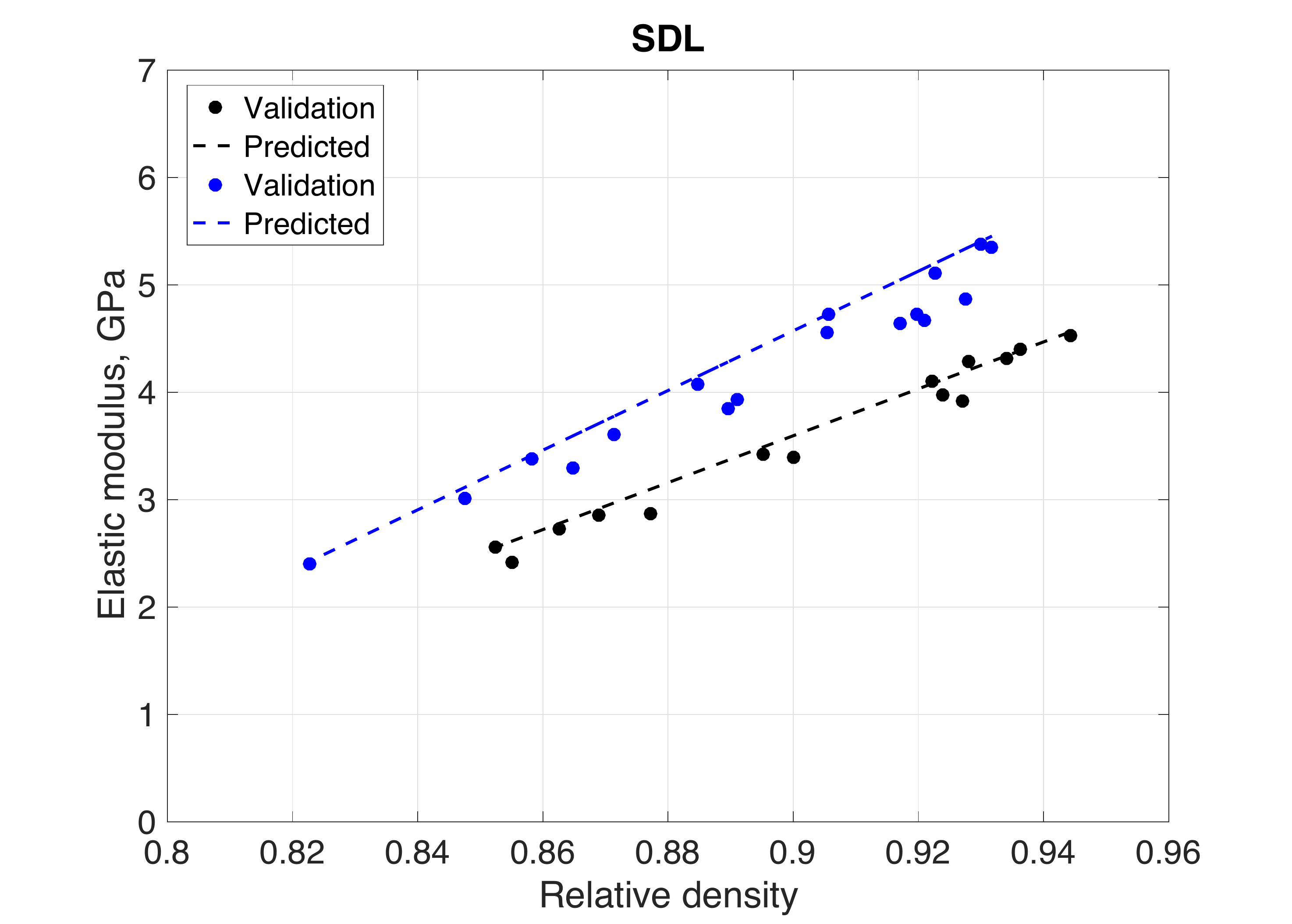}
\label{Fig:E-SD-val}
}
\subfigure[]
{\hspace{-0.9cm}
\includegraphics[scale=0.27]{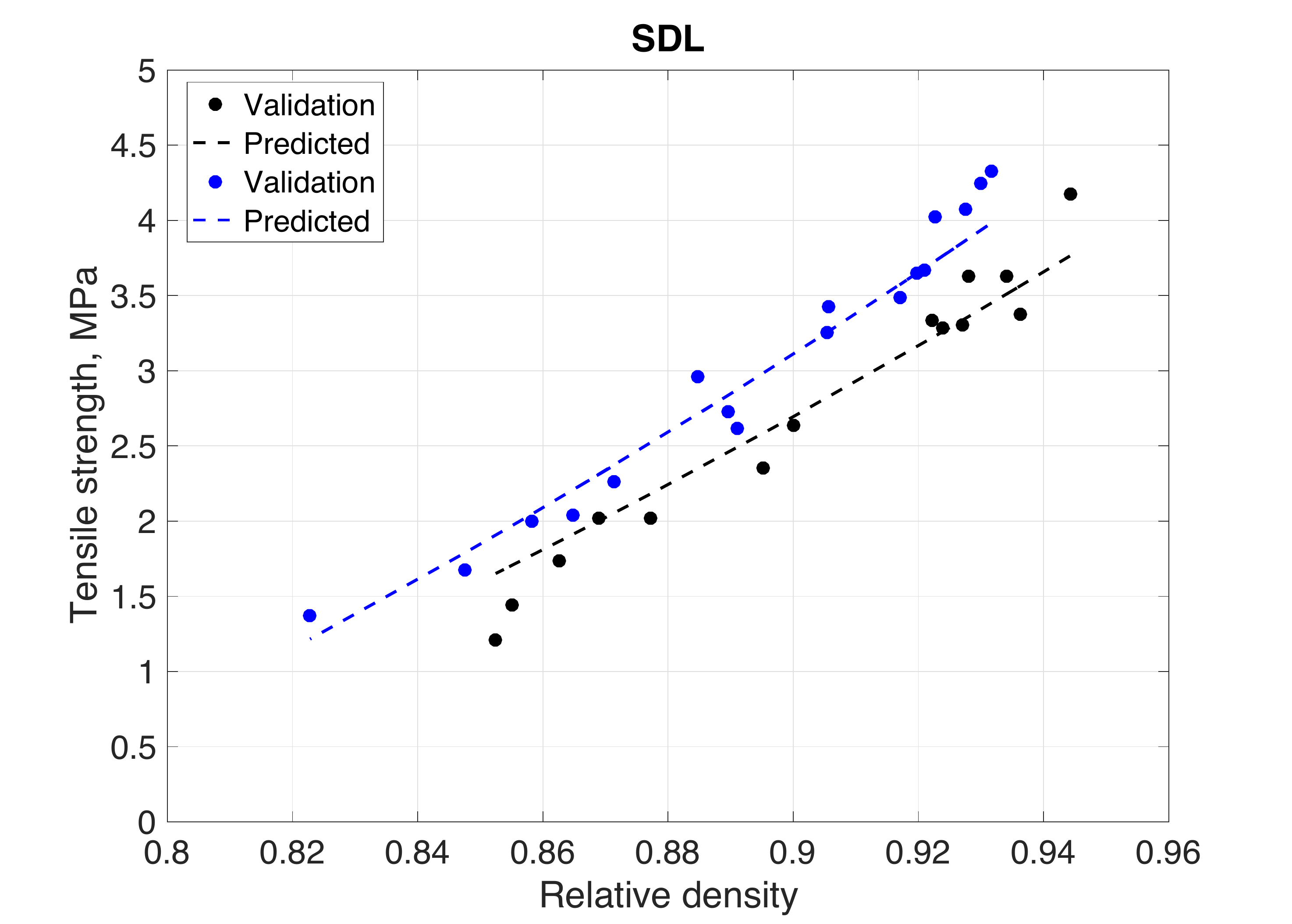}
\label{Fig:sigma-SD-val}
}
\end{tabular} 
\caption{Comparsion of the validation experiments to the model predictions for (a) elastic modulus and (b) tensile strength of spray-dried lactose tablets as a function of relative density (cases $24$ (in black) and $33$ (in blue) in Table~\ref{Table:DOE}).} 
\label{Fig:SDL-val}
\end{figure}

\begin{figure}[H]
\centering
\begin{tabular}{cc}
\subfigure[]
{
\includegraphics[scale=0.4]{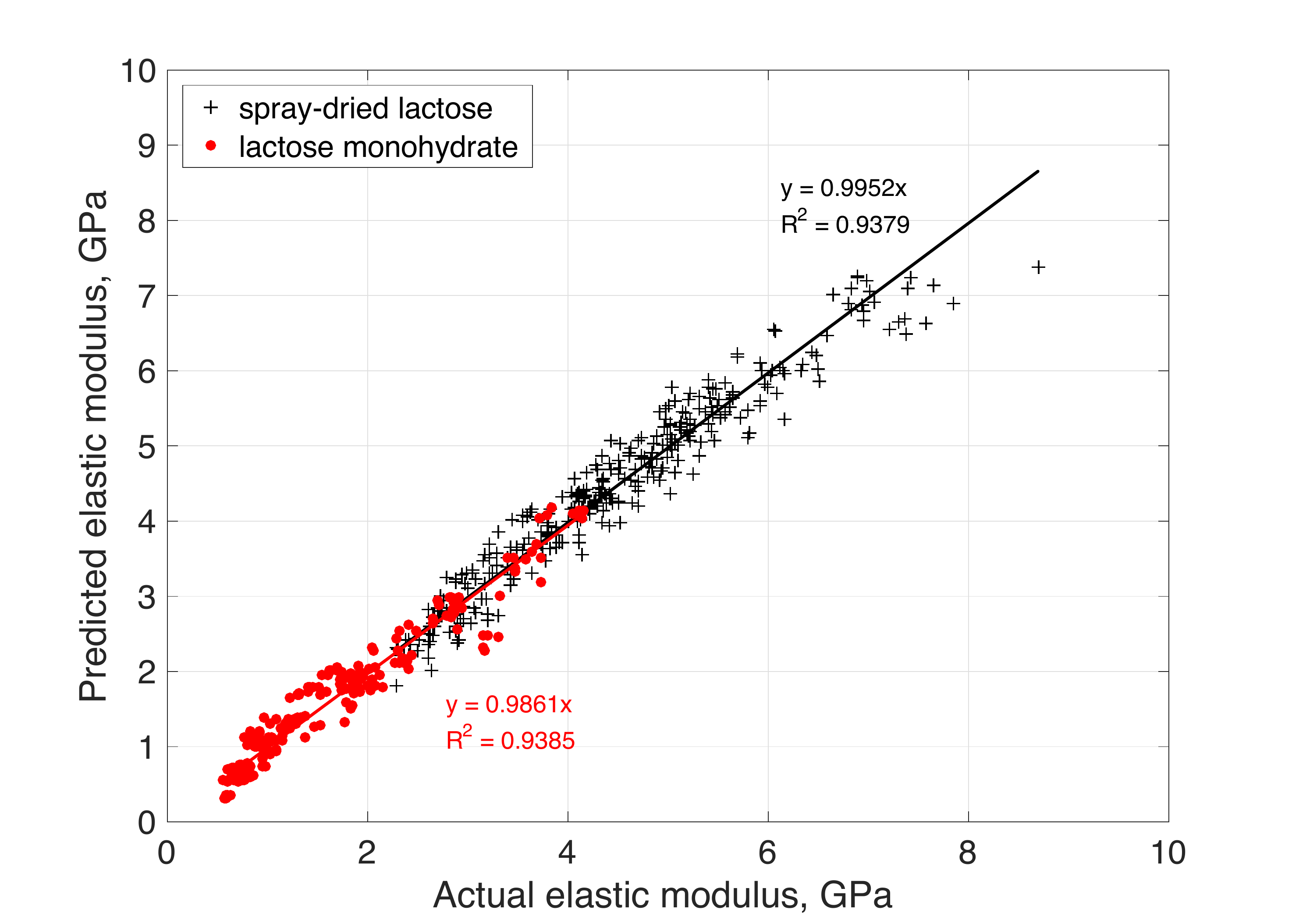}
\label{Fig:E-E}
}
\\
\subfigure[]
{
\includegraphics[scale=0.4]{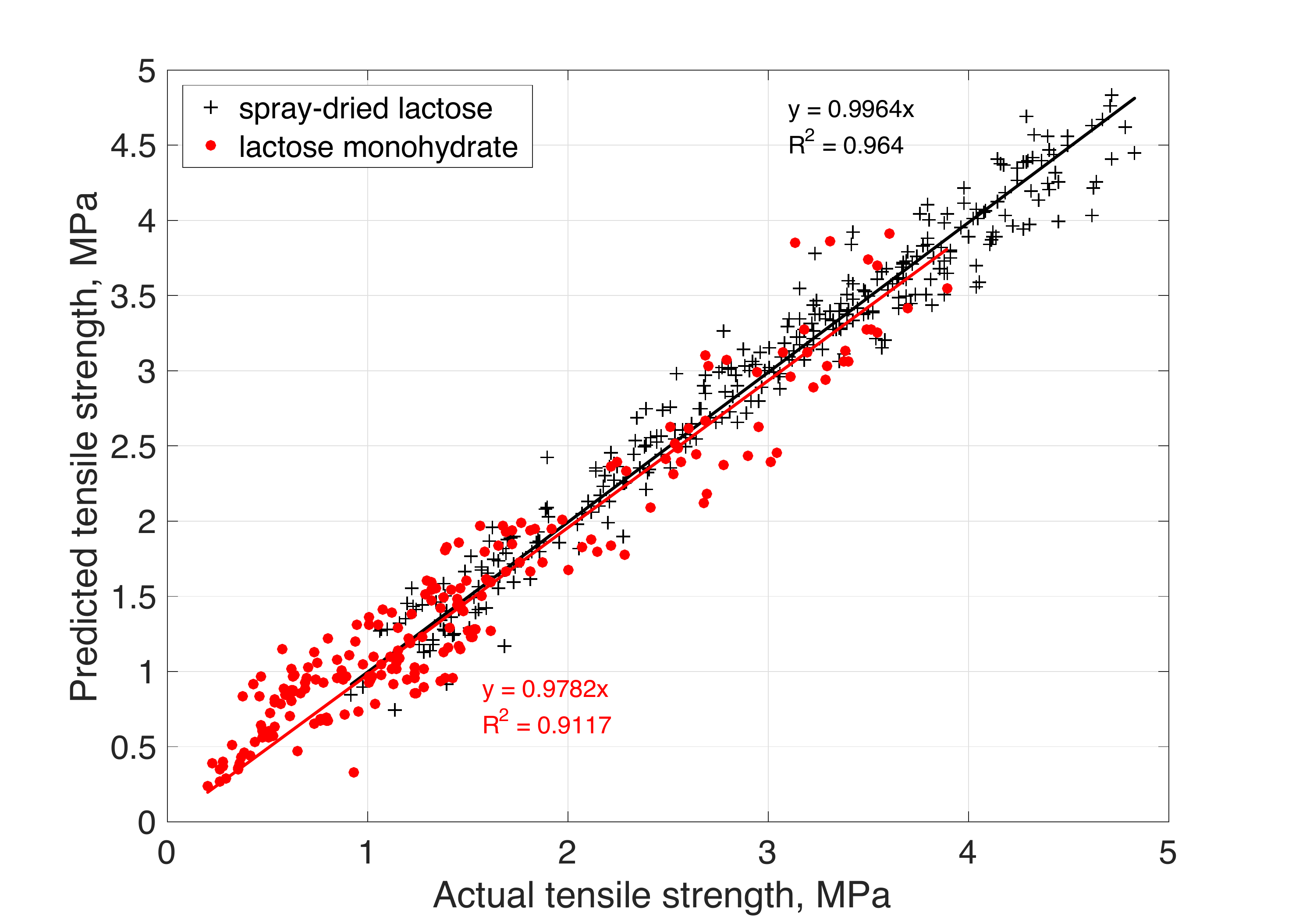}
\label{Fig:sigma-sigma}
}
\end{tabular} 
\caption{Predicted vs. actual values of (a) elastic modulus, where the $R^2$ was $0.94$ for both powders and (b) tensile strength, where $R^2$ was $0.96$ and $0.91$ for spray-dried lactose and lactose monohydrate, respectively.} 
\label{Fig:CP-AR-LM-SD}
\end{figure}

\pagebreak

\begin{figure}[H]
\centering
\begin{tabular}{cc}
\subfigure[]
{
\includegraphics[scale=0.4]{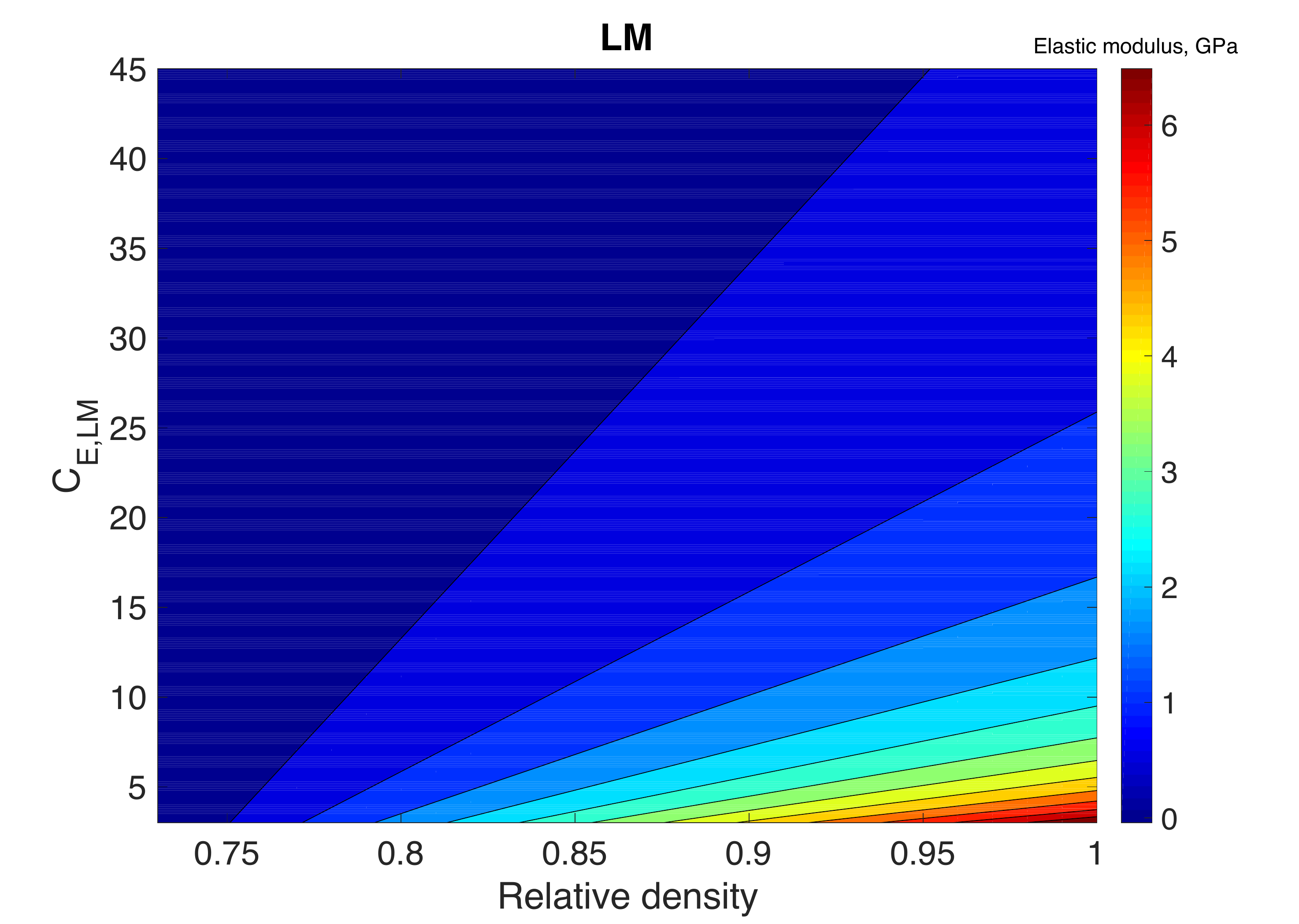}
\label{Fig:E-LM-pred}
}
\\
\subfigure[]
{
\includegraphics[scale=0.4]{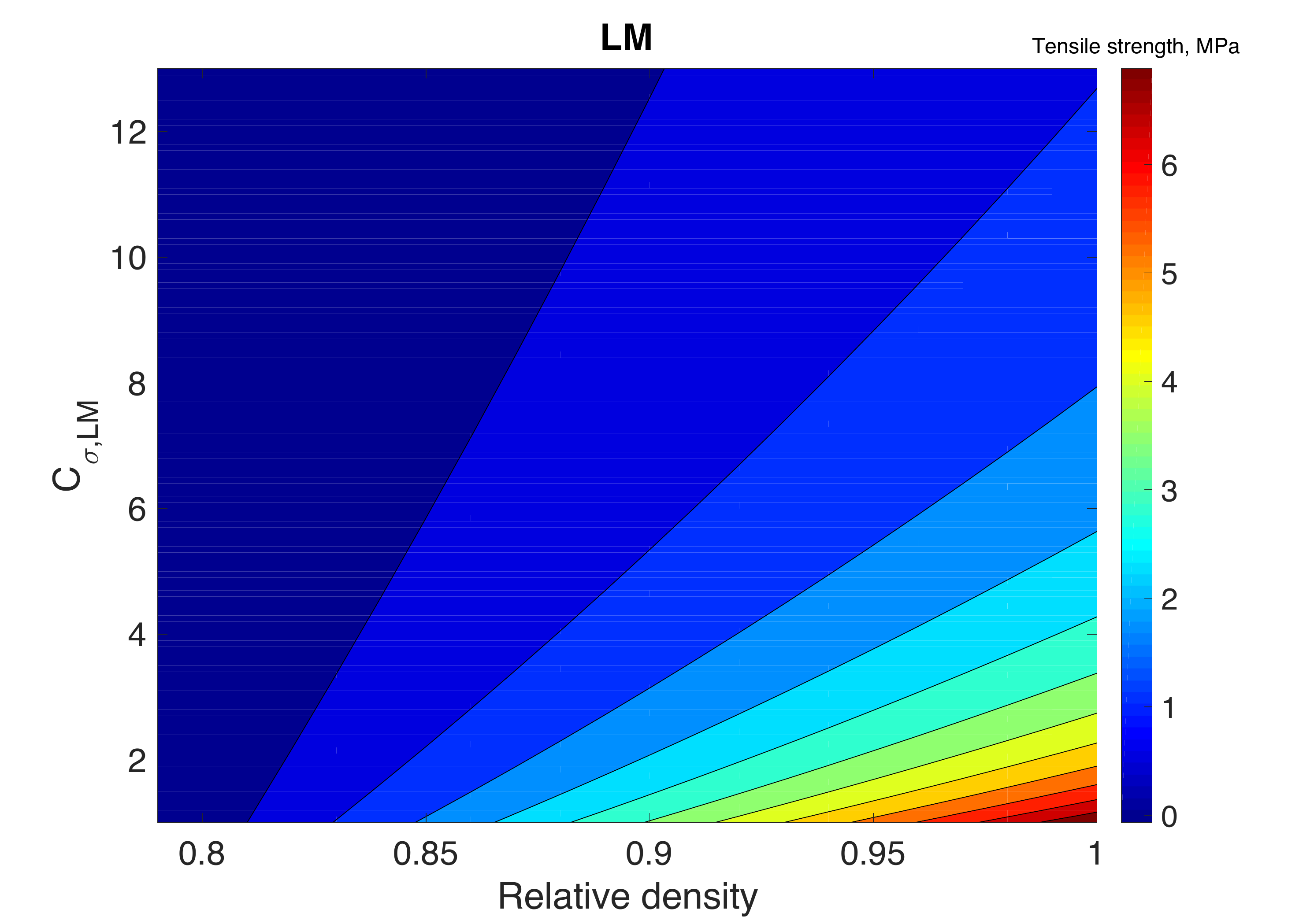}
\label{Fig:contour}
}
\end{tabular} 
\caption{Contour plots of (a) elastic modulus as a function of $C_{\text{E,LM}}$ and relative density and (b) tensile strength as a function of $C_{{\sigma,LM}}$ and relative density for lactose monohydrate.}
\label{Fig:contour-LM}
\end{figure}

\begin{figure}[H]
\centering
\begin{tabular}{cc}
\subfigure[]
{
\includegraphics[scale=0.4]{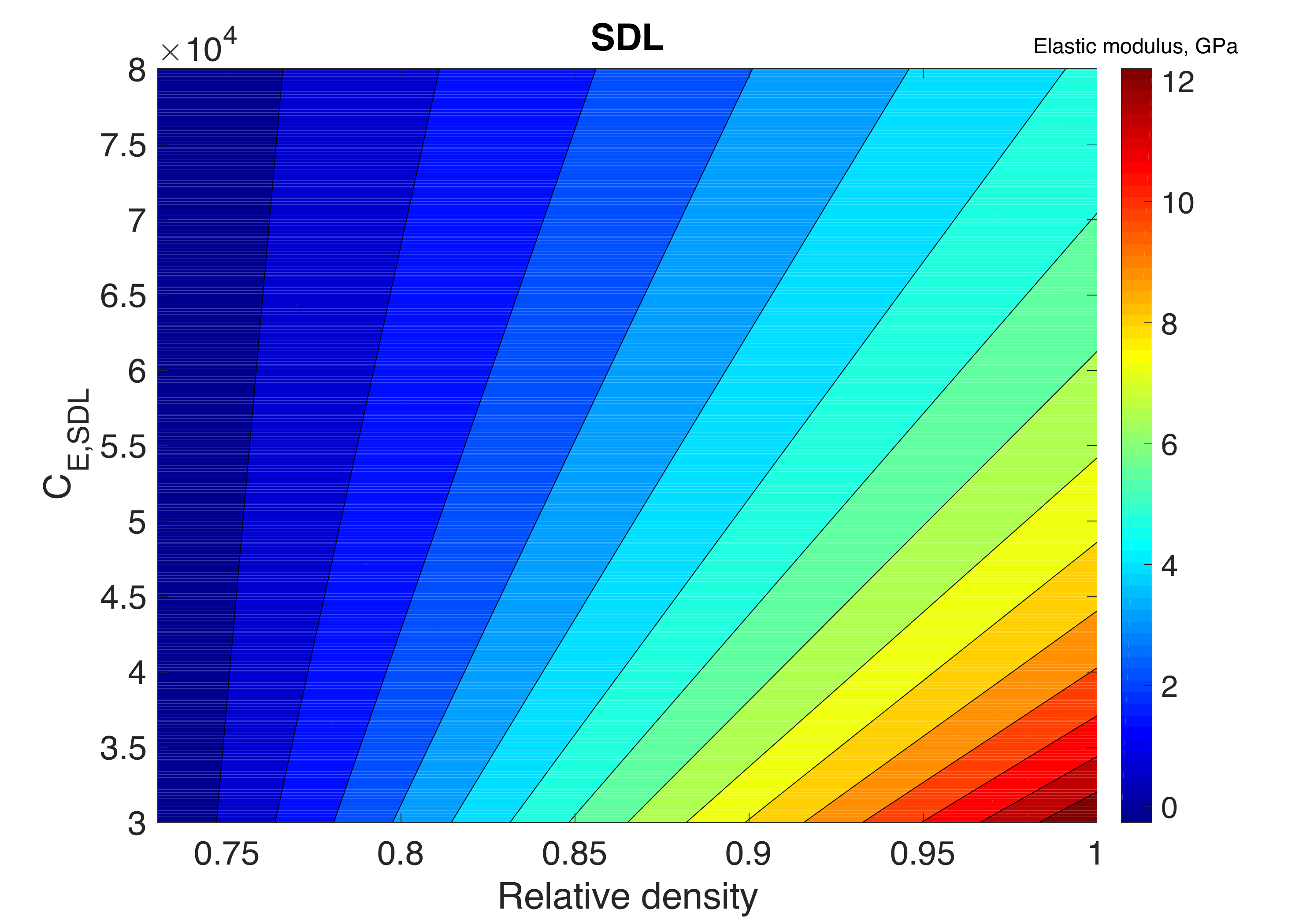}
\label{Fig:E-SD-pred}
}
\\
\subfigure[]
{
\includegraphics[scale=0.4]{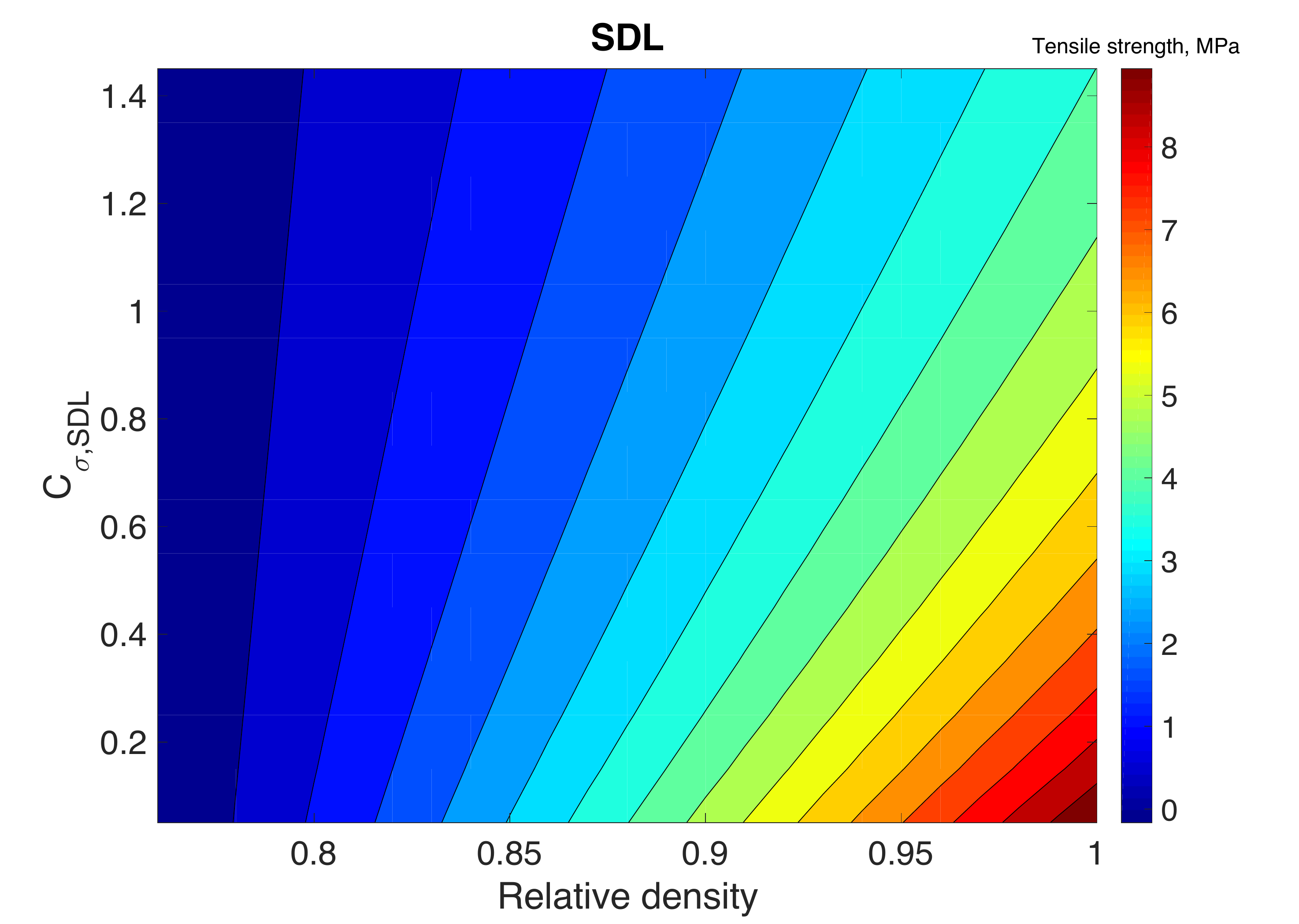}
\label{Fig:contour-SDL}
}
\end{tabular} 
\caption{Contour plots of (a) elastic modulus as a function of $C_{\text{E,SDL}}$ and relative density and (b) tensile strength as a function of $C_{{\sigma,SDL}}$ and relative density for spray-dried lactose.}
\label{Fig:contour-SDL}
\end{figure}

\begin{figure}[H]
\centering
\includegraphics[trim=1cm 0cm 1cm 1cm, clip=true, totalheight=0.42\textheight]{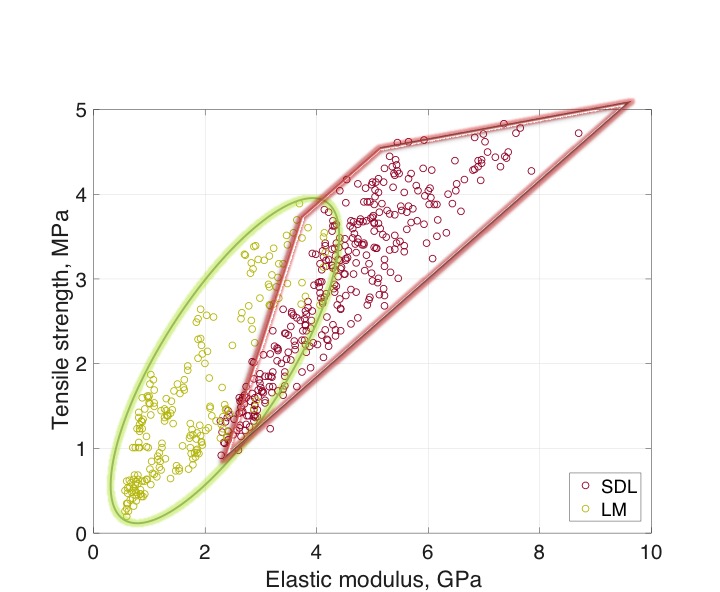}
\caption{Elastic modulus and tensile strength regime for lactose monohydrate and spray-dried lactose tablets. Relative density of tablets ranged from $0.8$ to $0.94$.} 
\label{Fig:E-TS-total}
\end{figure}

{\singlespacing
\begin{table}[H]
\centering
\caption{Mean ($\mu$) and standard deviation ($\sigma$) for each particle size distribution, MgSt concentration ($C_l$), and the mixing time ($t_m$) for all the cases studied.}
\footnotesize{
\begin{tabular}{c c|cc c|c|c} 
\hline
\hline
Powder &Cases & PSD ($\mu$m) & $\mu$ ($\mu$m) & $\sigma$ ($\mu$m) & $c_{l}$ (\%) & $t_{m}$ (sec) 
\\
\hline
\hline
Lactose monohydrate& 1& 0-75 & 62.83 & 22.87 &0.5 & 30
\\
& 2& 0-75 &62.83 & 22.87& 0.25 & 120
\\
& 3& 0-75&62.83 & 22.87 & 0.5 & 120
\\
& 4& 0-75 & 62.83 & 22.87& 1 & 120
\\
& 5 & 0-75 &62.83 & 22.87& 2 & 120
\\
&6& 0-75 &62.83 & 22.87& 2 & 300
\\
&7&0-75 &62.83 & 22.87& 0.25 & 1200
\\
&8& 0-75 & 62.83 & 22.87& 0.25 & 2400
\\
& 9 & 0-75 &62.83 & 22.87& 2 & 1200
\\
\\
& 10& 75-106 & 114&26.89&0.25 &120
\\
&11& 75-106 & 114&26.89&0.25 & 1200
\\
& 12 & 75-106 & 114&26.89&2 & 1200
\\
\\
&13 & 106-150 &149.3&25.6& 0.25 &120
\\
&14& 106-150 & 149.3&25.6&0.5 & 120
\\
& 15& 106-150 &149.3&25.6& 0.25 & 1200 
\\
&16 & 106-150 &149.3&25.6& 2 &1200
\\
\\
&17& as-received & 77.72&31.85&0.25 &120
\\
&18 & as-received &77.72&31.85& 0.25 &1200
\\ 
&19 & as-received &77.72&31.85&2 &1200
\\
\\
Spray-dried lactose & 20 & 0-75 & 65.14&17.15& 0.5 & 30
\\
& 21& 0-75 & 65.14&17.15& 2 & 30
\\
& 22& 0-75 &65.14&17.15&1 & 120
\\
& 23& 0-75 & 65.14&17.15&0.5 & 600
\\
& 24 & 0-75&65.14&17.15& 2 & 600
\\
& 25& 0-75 &65.14&17.15& 2& 1200
\\
\\
&26 &75-106 & 89.39&17.74&1 & 30
\\
&27 & 75-106 & 89.39&17.74& 0.5 & 120
\\
&28&75-106& 89.39&17.74&2 &120
\\
&29&75-106 & 89.39&17.74 &1 & 600
\\
\\
&30&106-150&120.9&27.04&0.5&30
\\
&31&106-150&120.9&27.04&2&30
\\
&32&106-150&120.9&27.04&1&120
\\
&33&106-150&120.9&27.04&0.5&600
\\
&34&106-150&120.9&27.04&2&600
\\
&35 &106-150&120.9&27.04&2&1200
\\
\\
&36&150-212&171.9&33.91&0.5&30
\\
&37&150-212&171.9&33.91&2&30
\\
&38&150-212&171.9&33.91&1&120
\\
&39&150-212&171.9&33.91&0.5&600
\\
&40&150-212&171.9&33.91&2&600
\\
\\
&41&as-received&128.3&39.5&0.25&120
\\
&42&as-received&128.3&39.5&2&1200
\\
\hline
\end{tabular}
}
\label{Table:DOE}
\end{table}
}

{\singlespacing
\begin{table}[H]
    \centering
\caption[Tensile strength and elastic modulus at zero porosity together with the critical relative density determined for each mechanical property for all cases studied.]{Tensile strength at zero porosity ($\sigma_{0}$), critical relative density ($\bar{\rho}_{c, \sigma_{t}}$), elastic modulus at zero porosity ($\text{E}_{0}$), and critical relative density ($\bar{\rho}_{c, \text{E}}$) together with their $R^{2}$ values for all the cases.}
  \footnotesize\setlength{\tabcolsep}{4pt}{
    \begin{tabular}{c   c|cc  c|cccc} 
      \hline
\hline
    Powder &Cases & $\sigma_{0}$ (MPa) & $\bar{\rho}_{c, \sigma_{t}}$ (\%)  & $R^{2} (\%)$ & $\text{E}_{0}$ (GPa) &$\bar{\rho}_{c, \text{E}}$ (\%) & $R^{2} (\%)$
    \\
\hline
\hline
Lactose monohydrate& 1& 5.6 &76.16  &98.07  &6.47 & 72.93 & 96.30
\\
& 2&5.36  &77.27 &98.57 & 5.4& 71.29& 97.62 
\\
& 3& 6.19&78.07 &98.18  & 5.77 & 73.7& 94.70
\\
& 4& 5.64 & 79.61 &98.90& 4.86 & 74.42 &98.63
\\
& 5 & 4.99 &79.7 & 97.34& 4.28& 73.14&98.14
\\
&6& 4.21 &82.47 & 99.47& 3.69 &75.3 &95.32
\\
&7&4.92 &80.08 & 98.28& 4.52 &75.12 &97.18
\\
&8& 3.31 & 81.26 & 96.98& 2.4 &75.68 &98.37
\\
& 9 & 3.79 &82.9 & 98.45& 2.73 &76.68 &96.3
\\
\\
& 10& 4.18 & 82.13 &98.40&2.69 &76.82&96.38
\\
&11& 2.81 & 81.9&99.11&1.36& 68.71&96.26
\\
& 12 & 1.78 & 83.72&99.42&1.06 & 70.27&52.66
\\
\\
&13 & 3.3 &82.78 & 99.35 &1.36&69.58 &93.13
\\
&14& 2.47 & 82.34 &97.24&1.16 & 72.02&96.73
\\
& 15& 2.26 &81.22&97.83&1.01& 61.14 & 97.31 
\\
&16 & 1.45&84.28&98.03&-& - &-
\\
\\
&17& 6.01 & 81.05&99.03&4.07&74.52 &96.60
\\
&18 & 3.02 &83.38&97.84&2& 75.6 &98.05
\\ 
&19 & 2.62 &85.1&96.90&1.84&78.37 &93.57
\\
\\
Spray-dried lactose & 20 & 6.91 & 74.65&96.21&9.8&72.78 &96.51
\\
& 21&6.49 &77.13&96.37& 7.23 &74.09&95.87
\\
& 22& 6.56 &76.89&95.70&7.29 &72.45&97.21 
\\
& 23&6.23 &76.76&95.84&6.85 &73.55&96.97
\\
& 24 & 5.7&79.32&97.42& 5.76&74.42 &98.33 
\\
& 25& 5.58&79.22&98.46& 5.48&70.42&96.24 
\\
\\
&26 &6.15& 75.09&95.24&8.74 &72.71&97.45
\\
&27 &6.49 &76.39&94.93&7.98&74.2&97.03
\\
&28&5.39& 76.86&94.93&6.7&74.06&97.19
\\
&29&5.67& 76.87&97.79 &6.54&72.75&97.32
\\
\\
&30&6.71&75.2&96.60&10.66&76.53&96.18
\\
&31&5.6&76.17&94.68&7.5&72.96&96.67
\\
&32&5.97&76.19&98.73&6.85&73.25&96.76
\\
&33&6.42&77.87&96.75&6.94&73.27&96.15
\\
&34&5.13&77.73&96.52&5.57&72.62&97.07
\\
&35 &4.77&77.87&99.22&5.39&71.23&98.72
\\
\\
&36&6.47&75.51&99.62&9.43&74.48&96.44
\\
&37&5.66&76.74&98.25&7.65&73.37&96.33
\\
&38&6.44&79.05&98.52&7.37&74.81&97.50
\\
&39&6.29&77.22&96.61&8.27&74.77&97.77
\\
&40&4.95&76.83&98.84&6.31&71.99&96.65
\\
\\
&41&6.91&76.36&99.54&9.47&73.97&99.30
\\
&42&4.63&78.08&99.53&5.3&71.68&98.06
\\
\hline
    \end{tabular}
    }
    \label{Table:TS-E-results}
\end{table}
}

\begin{table}[H]
\centering
\caption{Optimal coefficients and their residual error of elastic modulus optimization problem for lactose monohydrate and spray-dried lactose.}
\small\setlength{\tabcolsep}{4pt}{
\begin{tabular}{c c c c cc c|c }
\hline
\hline
Powder & $\text{E}_{0,\emptyset}$ (GPa) &$b_{1}$ & $b_{2}$& $b_{3}$ & $b_{4}$ &$\rho_{c, {\text{E}}}$ & E-norm\\
\hline
\hline
Lactose monohydrate &28.1221 &0.1491& 0.2645 &1.7082 &841.377& 0.7316 &3.31\\
Spray-dried lactose & 390932.897 & 0.1654 &0.0859 & -0.0179 & 0.0000266 & 0.7354 & 5.529\\
\hline
\label{Table:LM-C}
\end{tabular}}
\end{table}

\begin{table}[H]
\centering
\caption{Optimal coefficients and their residual error of tensile strength optimization problem for lactose monohydrate and spray-dried lactose.}
\small\setlength{\tabcolsep}{4pt}{
\begin{tabular}{c c c c ccc |c }
\hline
\hline
Powder &  $\sigma_{0,\emptyset}$ (MPa) &$d_{1}$&$d_{2}$ & $d_{3}$& $d_{4}$  & $\rho_{c, {\sigma_{t}}}$ &$\sigma$-norm\\
\hline
\hline
Lactose monohydrate &14.9186 &0.1982& 0.301 & 1.2984 & 498.924 &0.7924 &3.739\\
Spray-dried lactose & 10.0276 & 0.3342 & 0.1237 & 0.2154 & 7.5374&0.7653 & 3.31\\
\hline
\label{Table:SDL-C}
\end{tabular}
}
\end{table}

\end{document}